\documentclass[prx,amsmath,amssymb, notitlepage, twocolumn,
nofootinbib,
superscriptaddress,
longbibliography
]{revtex4-1}

\usepackage{amsmath}
\usepackage{tabularx,graphicx}
\usepackage{epstopdf}
\usepackage{makecell}
\usepackage{graphicx}
\usepackage{latexsym}
\usepackage{amssymb}
\usepackage{amsthm}
\usepackage{color, colortbl}
\usepackage{psfrag}
\usepackage{bbm}
\usepackage{bm}
\usepackage{titlesec}
\usepackage{dsfont}
\usepackage{slashed}
\usepackage[utf8]{inputenc}
\usepackage[symbols,nogroupskip,sort=none]{glossaries-extra}
\usepackage{nomencl}

\usepackage{multirow}
\usepackage{url}
\usepackage{xr}
\usepackage{xcite}
\usepackage{slashed}
\usepackage{microtype}
\usepackage{tikz}
\usepackage{feynmp-auto}

 \definecolor{mygreen}{RGB}{50,200,50}

\usepackage{tikz}
\usetikzlibrary{decorations.pathmorphing,decorations.markings}

\tikzset{photon/.style={decorate, draw=black, thick,
decoration={snake, amplitude=2pt, segment length=6pt}}}
\tikzset{partial ellipse/.style args={#1:#2:#3}{
    insert path={+ (#1:#3) arc (#1:#2:#3)}}}
\usepackage{float}
\makeatletter
\newcommand*{\addFileDependency}[1]{
  \typeout{(#1)}
  \@addtofilelist{#1}
  \IfFileExists{#1}{}{\typeout{No file #1.}}
}
\makeatother

\newcommand*{\myexternaldocument}[1]{%
    \externaldocument{#1}%
    \addFileDependency{#1.tex}%
    \addFileDependency{#1.aux}%
}

\myexternaldocument{main_05_12}

\newcommand{\mc}[1]{\mathcal{#1}}
\newcommand{\ew}[1]{\langle #1 \rangle}
\newcommand{\beq}{\begin{eqnarray}}
\newcommand{\eeq}{\end{eqnarray}}

\newcommand{\Tr}{{\rm Tr}}
\newcommand{\tr}{{\rm tr}}

\newcommand{\bsp}{\begin{aligned}}
\newcommand{\esp}{\end{aligned}}

\newcommand{\ie}{{i.e., }}
\newcommand{\eg}{{e.g., }}

\newcommand{\mO}{{\mathcal{O}}}

\newcommand{\lsupsc}[1]{^{#1}\hspace{-1pt}}
\newcommand{\dq}[1]{\frac{d^3 #1}{(2\pi)^3}}
\newcommand{\Trace}{{\text{Tr}}}
\newcommand{\floor}[1]{\left\lfloor #1\right\rfloor}

\usepackage{xcolor}
\definecolor{darkblue}{rgb}{0.,0.,0.4}
\definecolor{darkred}{rgb}{0.5,0.,0.}
\definecolor{BlueViolet}{RGB}{138,43,226}
\definecolor{SkyBlue}{RGB}{30,144,255}
\definecolor{DarkGreen}{RGB}{0,100,0}

\usepackage[colorlinks=true,linkcolor=darkblue,citecolor=blue,urlcolor=darkred]{hyperref}
\usepackage[normalem]{ulem}

\usepackage{mathrsfs}

\usepackage{comment}

\newcommand{\z}{\mathbb{Z}}

\makeatletter
\newcommand{\colim@}[2]{%
  \vtop{\m@th\ialign{##\cr
    \hfil$#1\operator@font colim$\hfil\cr
    \noalign{\nointerlineskip\kern1.5\ex@}#2\cr
    \noalign{\nointerlineskip\kern-\ex@}\cr}}%
}
\newcommand{\colim}{%
  \mathop{\mathpalette\colim@{\rightarrowfill@\textstyle}}\nmlimits@
}
\makeatother

\makenomenclature

\numberwithin{corollary}{section}
\numberwithin{theorem}{section}
\numberwithin{lemma}{section}
\numberwithin{definition}{section}
\numberwithin{example}{section}
\numberwithin{proposition}{section}
\numberwithin{remark}{section}

\usepackage{pst-all}

\usepackage{leftidx}

\usepackage[off]{auto-pst-pdf} 
\usepackage{booktabs}
\usepackage{yfonts}
\makeatletter

\usepackage[caption=false]{subfig}
\usepackage{enumitem}
\usepackage{tikz}

\begin{document}

\title{Topological phases and quantum criticality\\
from $SU(2)$ Chern--Simons--matter theories}  

\author{Yunchao Hao}
\affiliation{Department of Physics, National University of Singapore, 117551, Singapore}

\author{Yingcheng Li}
\affiliation{Department of Physics, National University of Singapore, 117551, Singapore}

\author{Kangle Li}
\affiliation{Department of Physics, National University of Singapore, 117551, Singapore}

\author{Liujun Zou}
\affiliation{Department of Physics, National University of Singapore, 117551, Singapore}

\begin{abstract}

Motivated by recent numerical studies where various $SU(2)$ Chern-Simons-matter theories emerge, we analytically study topological phases and quantum criticality in two-dimensional systems described by such theories. First, we classify $SU(2)_k$ topological orders in all lattice spin systems with a $p4\times SO(3)$ symmetry, where $k$ is an arbitrary nonzero integer. We find that for each odd $k$, the topological order can emerge in systems with an arbitrary Lieb-Schultz-Mattis (LSM) anomaly, and the symmetry cannot permute anyons. If the system has a nontrivial (respectively, trivial) LSM anomaly, then there is exactly one (respectively, nine) symmetry-enriched topological (SET) phases. On the other hand, $SU(2)_k$ topological order with any even $k$ can only emerge in systems with a trivial LSM anomaly. If $k\notin\{6, 10, 14, \cdots\}$, the symmetry cannot permute anyons, and there are 16 SET phases. If $k\in\{6, 10, 14, \cdots\}$, there are 4 different ways how the symmetry can permute anyons, and there are 64 SET phases. Next, we analyze the $SU(2)_k$ Chern-Simons theories coupled to $N_f$ flavors of gapless matter fields that can be either bosonic or fermionic. For both types of theories, we consider a joint large-$N_f$ and large-$k$ limit with $N_f/k$ fixed, and compute the scaling dimensions of the bilinear operators of the bosons or fermions to the order of $1/N_f$. These results sharpen our understanding of these emergent exotic topological phases and quantum criticality, and provide useful guidance to explore them further.

\end{abstract}

\maketitle

\tableofcontents

\section{Introduction}\label{sec:intro}

In two spatial dimensions, a large class of topological phases admit natural low-energy descriptions in terms of Chern-Simons (CS) gauge theories, and the continuous transitions out of such phases are often captured by CS gauge fields coupled to gapless matter. These Chern-Simons-matter theories furnish a unified framework that encodes both the topological data of the gapped phase and the universal critical properties at the transition, where strong gauge fluctuations can drive the system to an interacting conformal field theory (CFT).

Recent numerical progress has sharpened the case for revisiting non-Abelian Chern-Simons-matter theories. Refs. \cite{Zhang2024} and \cite{Luo2023} studied spin-1/2 and spin-$3/2$ square-lattice antiferromagnets, respectively, both of which have a $p4$ space group symmetry and an $SO(3)$ spin rotational symmetry. Evidence for $p4\times SO(3)$ symmetric $SU(2)_1$ and $SU(2)_3$ topological orders was reported in Refs. \cite{Zhang2024} and \cite{Luo2023}, respectively. These findings raise two questions. First, which $p4\times SO(3)$ symmetric $SU(2)_1$ and $SU(2)_3$ topological phases can emerge in such lattice systems? In other words, what is the classification of  $p4\times SO(3)$ symmetric $SU(2)_1$ and $SU(2)_3$ topological phases in this setup? Second, which of these symmetry-enriched topological phases are realized in Refs. \cite{Zhang2024} and \cite{Luo2023}?

Around the $SU(2)_1$ and $SU(2)_3$ topological phases, on the phase diagrams in Refs. \cite{Zhang2024} and \cite{Luo2023}, there are also adjacent phases with some spontaneously broken symmetries. Although the nature of the phase transitions on these phase diagrams was not fully investigated numerically, field theories that govern the continuous quantum phase transitions, if any, out of the $SU(2)_1$ or $SU(2)_3$ topological phase are naturally some $SU(2)$ Chern-Simons-matter theory with critical bosonic matter. Independently, a family of new $(2+1)$-dimensional CFTs with $USp(2N_f)$ global symmetry was uncovered using the fuzzy sphere approach,{\footnote{This global symmetry is denoted by $\mathrm{Sp}(N)$ in Ref. \cite{Zhou2025}.}} some of which are also described by critical $SU(2)$ Chern-Simons-matter theories  \cite{Zhou2025}. Together, these results motivate a systematic study of the critical properties of $SU(2)$ Chern-Simons-matter theories.

With these motivations, in this paper we develop a unified perspective on (i) the symmetry-enriched $SU(2)_k$ topological orders that can be realized on lattice spin systems with $p4\times SO(3)$ symmetry, and (ii) the critical properties of the $SU(2)$ Chern-Simons-matter theories.

We begin with classifying symmetry-enriched $SU(2)_k$ topological orders on lattice spin systems with $p4\times SO(3)$ symmetry in Sec.~\ref{sec:SET}, where $k$ can be any nonzero integer. We find that as long as $k$ is odd and the lattice system has a nontrivial Lieb-Schultz-Mattis anomaly, then there is only a single possibility for $p4\times SO(3)$ symmetric $SU(2)_k$ topological phase. In particular, this classification uniquely determines the $p4\times SO(3)$ symmetric $SU(2)_1$ and $SU(2)_3$ topological phases realized in Refs. \cite{Zhang2024} and \cite{Luo2023}. On the other hand, $SU(2)_k$ topological phases with an even $k$ can only emerge in $p4\times SO(3)$ symmetric systems that have a trivial Lieb-Schultz-Mattis anomaly, which are typically systems with integer spins. Next, we analyze the $SU(2)_k$ Chern-Simons-matter theories with $N_f$ flavors of critical matter fields. Specifically, we discuss the case with critical bosonic matter in Sec.~\ref{sec:critical_bosons} and the case with critical fermionic matter in Sec.~\ref{sec:critical_fermions}. Our focus here is to determine the scaling dimensions of various bilinear operators, in the limit where $N_f\rightarrow\infty$, $k\rightarrow\infty$ while their ratio $N_f/k$ is fixed. These scaling dimensions are given in Eqs. \eqref{eq:Delta_sigma}, \eqref{eq:Delta_A0_boson}, \eqref{eq:Delta_S_fermion} and \eqref{eq:Delta_A0_fermion}. Taken together, our results refine the characterization of the emergent $SU(2)$ Chern-Simons-matter theories and provide useful guidance to further study them.

%--------------------------------------------------------------

\section{Symmetry-enriched $SU(2)_k$ topological orders with $p4\times SO(3)$ symmetry}\label{sec:SET}

\begin{figure}
    \centering
    \includegraphics[width=\linewidth]{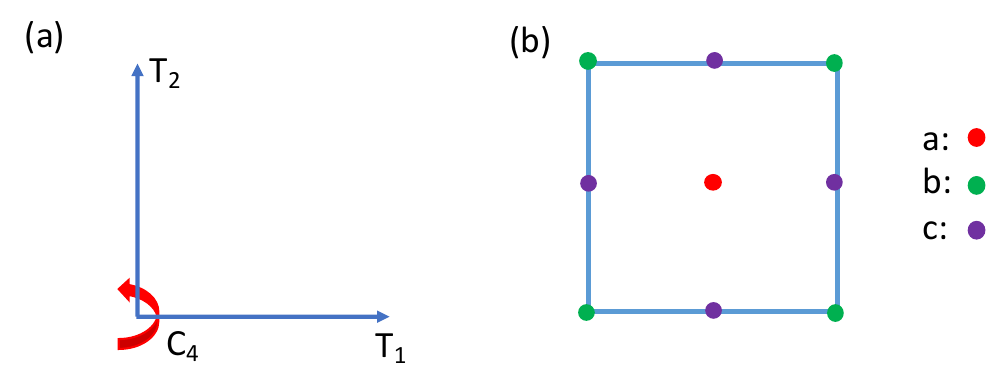}
    \caption{(a) Symmetry generators of the $p4$ space group, where $T_1$ and $T_2$ are translation vectors along two orthogonal directions, and $C_4$ is a 4-fold rotation. (b) The irreducible Wyckoff positions of a $p4$ symmetric lattice. $a$, $b$ and $c$ can be viewed as the rotation centers of $C_4$, $T_1C_4$, and $T_1C_4^2$ (also $T_2C_4^2$), respectively. Alternatively, $a$ can be viewed as the vertices of a square lattice, $b$ can be viewed as the plaquette centers of a square lattice, and $c$ can be viewed as the edge centers of a square lattice.}
    \label{fig: p4}
\end{figure}

In this section, we apply the framework of Ref.~\cite{Ye2023} to classify symmetry-enriched topological (SET) phases of the $SU(2)_k$ topological order with a microscopic symmetry group
\begin{equation}
G = p4 \times SO(3),
\end{equation}
where $p4$ is generated by translations $T_1$ and $T_2$, and a four-fold rotation $C_4$ (see Fig. \ref{fig: p4}), while $SO(3)$ denotes the spin rotation symmetry. We will only explicitly discuss the case with $k>0$, while the classifications for $k$ and $-k$ are the same. The results of our classification are summarized in Sec. \ref{subsubsec: classification result}.

\subsection{Brief review of the characterization and classification of SET phases}

We start by briefly reviewing how to characterize and classify SET phases, and refer the readers to Refs. \cite{Barkeshli2019, Ye2023, Li2026} for more details about this general theory.

Without taking symmetries into account, a topological order is characterized by a set of anyons with certain fusion and braiding properties. For the $SU(2)_k$ topological orders, these properties are summarized in Appendix \ref{app:SU2k_data}.

Given a topological order and a symmetry, there can be different symmetric topological orders that are organized into different SET phases. Each SET phase is characterized by the following 3 pieces of data \cite{Barkeshli2019}:{\footnote{More precisely speaking, there is an additional layer of data that describes what type of $p4\times SO(3)$ symmetry-protected topological phase is stacked on the system. As this data does not affect the properties related to anyon, in  this paper we will disregard it.}}
\begin{enumerate}

    \item How each anyon is permuted by each symmetry, as described by a set of invertible maps $\rho_g$ between the anyons, where $g$ is an element of the symmetry group $G$.

    \item How the fusion vertex basis states transform under the symmetries, as described by a set of $U$-symbols.

    \item What fractional quantum number under the symmetry is carried by each anyon, as described by a set of $\eta$-symbols.
    
\end{enumerate}

Let us unpack the meaning of the above data. The first piece of data should be self explanatory, and we denote by $^ga$ the anyon obtained by acting with the symmetry corresponding to $g\in G$ on the anyon $a$, \ie $\rho_g(a)= \lsupsc{g}a$. For the second piece of data, suppose that anyons $a$ and $b$ can fuse into anyon $c$ with the fusion multiplicity being $N_{ab}^c$. Namely, for a state with three well separated anyons $a$, $b$ and $\bar c$ (\ie the anti-particle of $c$) at positions $x_1$, $x_2$ and $x_3$, respectively, there is an $N_{ab}^c$-dimensional degenerate subspace of locally indistinguishable states. We denote such a state by $|a_1, b_2, \bar c_3; \mu\rangle$, where $\mu$ is the index of the fusion vertex basis states, \ie an orthonormal basis of the degenerate subspace. Denote the microscopic symmetry action for an element $g\in G$ by $R_g$. The symmetry localization hypothesis states that
\beq
\begin{split}
&R_g|a_1, b_2, \bar c_3; \mu\rangle\propto V_g^{a}(^gx_1)V_g^{b}(^gx_2)V_g^{\bar c}(^gx_3)\\
&\cdot\sum_{\nu=1}^{N_{ab}^c}[U_g(^ga, ^gb; ^gc)]_{\mu\nu}|^ga_{^g1}, ^gb_{^g2}, ^g\bar c_{^g3}; \nu\rangle,
\end{split}
\eeq
where $V_g^a(^gx_1)$, $V_g^b(^gx_2)$ and $V_g^{\bar c}(^gx_3)$ are unitary operators supported near $^gx_1$, $^gx_2$ and $^gx_3$, respectively ($^gx_{1,2,3}$ are the position obtained when $x_{1,2,3}$ are transformed by the symmetry $R_g$), and $U_g(^ga, ^gb; ^gc)$, also known as the $U$-symbol, is an $N_{ab}^c$-by-$N_{ab}^c$ unitary matrix describing how the fusion vertex basis transforms under the symmetry. For the third piece of data, denote by $|a_1, \bar a_2\rangle$ a state with anyons $a$ and $\bar a$ at well separated positions $x_1$ and $x_2$, respectively. The $\eta$-symbol, which is a $U(1)$ phase factor, is defined by
\beq \label{eq: eta definition}
\begin{split}
&\eta_a(g, h)\\
=&\langle a_1, \bar a_2|V_{gh}^{\lsupsc{\overline{gh}}a}(x_1) R_gV_h^{\lsupsc{\overline{gh}}a}(^{\bar g}x_1)R_g^{-1}V_g^{^{\bar g}a}(x_1)|a_1, \bar a_2\rangle,
\end{split}
\eeq
for each pair of group elements $g,h\in G$ and each type of anyon $a$, where $\bar g=g^{-1}$. The $\eta$-symbols measure the difference between the action of $gh$ and the action of $h$ followed by the action of $g$ on anyons, which are similar to projective representations in quantum mechanics, so they capture the fractional quantum numbers carried by the anyons.

The above symmetry-related data (\ie the maps $\rho_g$, the $U$- and $\eta$-symbols), along with the symmetry-independent data that characterizes the fusion and braiding of anyons (\ie anyon types, fusion rules, $F$- and $R$-symbols \cite{Kawagoe2019}), can be extracted microscopically from a topological state, and they satisfy a set of nontrivial equations and have some ``gauge freedom'' \cite{Li2026}.

For the purpose of the present work, from now on we will use a crystalline equivalence principle to convert SET phases with symmetry group $G=p4\times SO(3)$, where $p4$ is a lattice symmetry and $SO(3)$ is an internal symmetry, to SET phases with a purely internal symmetry $G=p4\times SO(3)$ \cite{Thorngren2016}. Then classifying the SET phases of the $SU(2)_k$ topological order amounts to finding all possible anyon permutation patterns $\rho_g$, $U$- and $\eta$-symbols in the present symmetry setting. To distinguish different lattice systems with the same symmetry group (\eg square lattice spin-$3/2$ systems, checkerboard lattice spin-1 systems, etc), an anomaly-matching analysis is needed when carrying out this classification \cite{Ye2023}. In the rest of this section, we perform this classification.

\subsection{Classification of $p4\times SO(3)$ symmetric $SU(2)_k$ topological orders}

As reviewed in Appendix \ref{app:SU2k_data}, the $SU(2)_k$ topological order has $k+1$ types of anyons, labeled by
\begin{equation}
j \in \left\{0,\frac12,1,\dots,\frac{k}{2}\right\},
\end{equation}
with $j=0$ the trivial anyon. The fusion rule is
\begin{equation}
j_1 \times j_2
=
\sum_{j_3 = |j_1-j_2|}^{\min(j_1+j_2,\;k-j_1-j_2)} j_3,
\end{equation}
where $j_3$ increases in integer steps.

An important property we need below is that there are only two Abelian anyons, $j=0$ and $j=k/2$, and their fusion rule is that $\frac{k}{2}\times\frac{k}{2}=0$. That is, the fusion of the Abelian anyons is described by a group $\mc{A}=\z_2$.

\subsubsection{Anyon permutation patterns and $U$-symbols} \label{subsubsec: anyon permutation}

We start by classifying all anyon permutation patterns. As discussed in Appendix \ref{subapp: topological symmetry}, the symmetries can permute anyons only if $k=4n+2$ with some integer $n\geqslant 1$, \ie $k\in\{6, 10, 14, \cdots\}$. In this case, the $SU(2)_k$ topological order has an ``emergent'' $G_{\rm top}=\z_2$ topological symmetry. Denote by $\tilde{\mathbf{1}}$ and $\tilde g$ the identity and generator of $G_{\rm top}$, respectively, and denote by $\tilde\rho_{\tilde g}(j)$ the anyon obtained from acting $\tilde g$ on anyon $j$. This topological symmetry permutes the anyons as follows:
\beq \label{eq: topological permutation main}
\tilde\rho_{\tilde{\mathbf{1}}}(j)=j,
\quad
\tilde\rho_{\tilde g}(j)=
\begin{cases}
    j, & {\text{if }} j\in\z,\\
    \frac{k}{2}-j, & {\text{if }} j\in\z+\frac{1}{2}.
\end{cases}
\eeq
When $j_3$ is a fusion outcome of $j_1$ and $j_2$, the $U$-symbol of this topological symmetry is
\beq \label{eq: topological U symbol main}
\begin{split}
\tilde U_{\tilde{\mathbf{1}}}(j_1, j_2; j_3)=1,
\quad
\tilde U_{\tilde g}(j_1, j_2; j_3)=(-1)^{x(j_1, j_2, j_3)},
\end{split}
\eeq
with
\beq
\begin{split}
x(j_1, j_2, j_3)
=&\floor{j_1}+\floor{j_2}+\floor{j_3}\\
	&+
	p(j_1)
	\left(
	\floor{j_1}+\floor{j_2}+\floor{j_3}
	\right)\\
    &+
	p(j_2)\floor{j_1},
\end{split}
\eeq
where $\floor{j}=j$ if $j\in\z$ and $\floor{j}=j-\frac{1}{2}$ if $j\in\z+\frac{1}{2}$, and $p(j)=0$ if $j\in\z$ and $p(j)=1$ if $j\in\z+\frac{1}{2}$. A valid set of $\eta$-symbols for this topological symmetry is simply $\tilde\eta_{j}(\tilde g, \tilde h)=1$ for all anyon $j$ and all elements $\tilde g, \tilde h\in G_{\rm top}$.

Below we discuss the different cases for how the microscopic $G=p4\times SO(3)$ symmetry can permute the anyons.

\begin{itemize}
    
    \item $k\neq 4n+2$ with an integer $n\geqslant 1$.

    In this case, since no anyon is permuted, the $U$-symbols can be taken as $U_g(j_1, j_2; j_3)=1$ for all $g\in G$ when $j_3$ is a fusion outcome of $j_1$ and $j_2$.

    \item $k=4n+2$ with an integer $n\geqslant 1$.

    In this case, the anyon permutation pattern is characterized by a group homomorphism from the microscopic symmetry $G=p4\times SO(3)$ to the topological symmetry $G_{\rm top}=\z_2$,
    \beq
        \varphi: G\rightarrow G_{\rm top},
    \eeq
    such that
    \beq \label{eq: anyon permutation and U}
    \begin{split}
    \rho_g(j)&=\tilde\rho_{\varphi(g)}(j),\\
    U_g(j_1, j_2; j_3)&=\tilde U_{\varphi(g)}(j_1, j_2; j_3),
    \end{split}
    \eeq
    with $\tilde\rho$ and $\tilde U$ given by Eqs. \eqref{eq: topological permutation main} and \eqref{eq: topological U symbol main}, respectively.

    There are 4 different homomorphisms $\varphi$, determined by which generators of $p4$ are mapped into the generator $\tilde g$ of the $G_{\rm top}$ under $\varphi$. The first case is where no generator of $p4$ is mapped to $\tilde g$, the second case is where the $C_4$ rotation generator is mapped to $\tilde g$ but the translation generators $T_{1,2}$ are not, the third case is where the translation generators $T_{1,2}$ are mapped to $\tilde g$ but the $C_4$ generator is not, and the last case is where both the $C_4$ rotation generator and $T_{1,2}$ are mapped to $\tilde g$.
    
\end{itemize}

The above describes all anyon permutation patterns and their corresponding $U$-symbols for all $SU(2)_k$ topological orders.

To finish the classification, what remains is to find all possible $\eta$-symbols. We carry out this procedure in two steps. We first find all possible sets of $\eta$-symbols compatible with the $G=p4\times SO(3)$ symmetry. In this step, we do not specify the type of the lattice system. Namely, some of these sets of $\eta$-symbols can only appear in square lattice spin-3/2 systems, but some other can only appear in checkerboard lattice spin-1 systems, etc. In the second step, we use anomaly matching to distinguish different types of lattice systems and refine the classification.

\subsubsection{Classification of fractional quantum numbers}

Although there are cases where symmetries can permute anyons, the classification of the possible $\eta$-symbols is simplified by two facts. First, in all cases, the two Abelian anyons $j=0$ and $j=\frac{k}{2}$ will not be permuted by any symmetry, even though the other anyons can be permuted. Second, no matter whether any anyon is permuted by the symmetry, a valid set of $\eta$-symbols is $\eta_j(g, h)=1$ for all anyon $j$ and all $g, h\in G$, because the topological symmetry has a trivial set of $\eta$-symbols. Because of these two facts, different sets of $\eta$-symbols are classified by $H^2(G, \mc{A})$, where $\mc{A}=\z_2$ is the group that describes the fusion of the Abelian anyons in the topological order \cite{Barkeshli2019}. Hence the $\eta$-symbols are classified by $H^2(G, \mc{A})=H^2(p4\times SO(3), \z_2)=\z_2^4$, \ie there are $2^4=16$ types of $\eta$-symbols for each anyon permutation pattern.

To write down the $\eta$-symbols explicitly, for each anyon $j$, we can write
\begin{equation}\label{eq:eta_from_t}
\eta_j(g,h) = M_{j,\,t(g,h)},
\end{equation}
where $M_{j,a}=\theta_{j\times a}/(\theta_j\theta_a)$ is the mutual braiding phase between anyon $j$ and the Abelian anyon $a=t(g,h)$. (see Eq. (176) of Ref. \cite{Barkeshli2019}). From Appendix \ref{app:SU2k_data}, the relevant mutual braiding phases are
\beq \label{eq: M matrix}
M_{j,0}=1,
\quad
M_{j, k/2}=(-1)^{2j}.
\eeq
In the above, the Abelian anyon $t(g, h)$ is regarded as a cocycle that represents an element in $H^2(G, \mc{A})$. Namely, we can parameterize the Abelian-anyon-valued cocycle as
\begin{equation}
t(g,h)=\frac{k}{2}\,w(g,h),
\end{equation}
where $w(g,h)\in \{0,1\}\cong\mathbb Z_2$ is a representative cocycle in $H^2(G, \z_2)$ that takes value in $\{0, 1\}$. Then Eq. \eqref{eq:eta_from_t} becomes
\begin{equation}\label{eq:eta_general}
\eta_j(g,h)=(-1)^{2j\,w(g,h)}.
\end{equation}

According to Eq. (35) of Ref. \cite{Ye2023}, $w$ can be written as
\begin{equation}\label{eq:w_parameterization}
w
=
P(n_1 B_{xy}
+
n_2 B_{c^2}
+
n_3 A_{x+y}^2
+
n_4 w_2),
\end{equation}
where $n_1,n_2,n_3,n_4\in\{0,1\}$, and $P(x)=1$ for odd $x$ and $P(x)=0$ for even $x$. The ``spatial generators'' $B_{xy}$, $B_{c^2}$ and $A_{x+y}^2$ generate $H^2(p4,\mathbb{Z}_2)$, while $w_2$ is the generator of $H^2(SO(3),\mathbb{Z}_2)\cong\mathbb{Z}_2$, i.e., the second Stiefel-Whitney class. To write down these cocycles explicitly, we parameterize a general element of $p4$ as
\begin{equation}
g=T_1^xT_2^yC_4^c,
\qquad
x,y\in\mathbb Z,\quad c\in\{0,1,2,3\}.
\end{equation}
The representative 2-cocycles for the spatial generators are (see Eqs. (B22-B24) in Ref. \cite{Ye2023})
\begin{align}
B_{xy}(g_1,g_2)
&=
P\left(P_c(c_1)y_1x_2+P(c_1)y_2(y_1+x_2)\right),\label{eq:Bxy}\\
B_{c^2}(g_1,g_2)
&=
P\left(\frac{[c_1]_4+[c_2]_4-[c_1+c_2]_4}{4}\right),\label{eq:Bc2}\\
A_{x+y}^2(g_1,g_2)
&=
P\left((x_1+y_1)(x_2+y_2)\right),\label{eq:Axy2}
\end{align}
where $[c]_4$ denotes $c \bmod 4$, and $P_c(c)=1-P(c)$. For the $SO(3)$ sector, it is convenient to restrict to a
$\mathbb Z_2\times \mathbb Z_2$ subgroup generated by $U_\pi$ and $U_\pi'$, \ie $\pi$ spin rotations around two orthogonal axes, and
parameterize an element as $U_\pi^{i_1}(U_\pi')^{i_2}$
with $i_1,i_2\in\{0,1\}$. On this subgroup, a representative
cocycle for the second Stiefel-Whitney class is (see Eq. (C13) in Ref. \cite{Ye2023})
\begin{equation}\label{eq:w2_cocycle}
w_2(g_1,g_2)=P(i_1j_1+i_2j_2+i_1j_2),
\end{equation}
where $g_1 = U_\pi^{i_1}(U_\pi')^{i_2}$ and
$g_2 = U_\pi^{j_1}(U_\pi')^{j_2}$.

Put together, for each anyon permutation pattern, all possible sets of $\eta$-symbols are classified by $\z_2^4$, with explicit expressions given by Eq. \eqref{eq:eta_general}, where the cocycle $w$ is parametrized by $n_{1,2,3,4}\in\{0, 1\}$ via Eq. \eqref{eq:w_parameterization}.

What is the physical meaning of the parameters $n_{1,2,3,4}$? Since Eq. \eqref{eq:eta_general} implies that $
\eta_{\frac{1}{2}}(g, h)=(-1)^{w(g, h)}$, $n_{1,2,3,4}$ are linked to the fractional quantum numbers carried by the anyon $j=1/2$ through Eq. \eqref{eq: eta definition}, which is the same as the fractional quantum number carried by the anyon $j=k/2$ if $k$ is odd. Intuitively speaking, $n_4=1$ ($n_4=0$) means that the anyon $1/2$ carries a projective (linear) representation under $SO(3)$. The meaning of $n_{1,2,3}$ can be read off from Eq. (198) from Ref. \cite{Ye2022}, and their values tell us whether a $1$ or $-1$ is obtained when acting the $C_2$ rotation, $T_1T_2C_2$ rotation, and $T_1C_2$ rotation twice on the anyon $j=1/2$, as shown in Table \ref{tab: fractional quantum numbers}.

\begin{table}[htbp]
\centering
\caption{The meaning of the fractional quantum numbers $n_{1,2,3}$. For each combination of $n_{1,2,3}$, we list whether the three types of 180-degree rotations give a $-1$ when the corresponding rotations act twice on the anyon $j=1/2$.}
\label{tab: fractional quantum numbers}
\begin{tabular}{c|c|c|c}
\hline\hline
$(n_1, n_2, n_3)$ & $C_2^2=-1$ & $(T_1T_2C_2)^2=-1$ & $(T_1C_2)^2=-1$ \\
\hline
$(0, 0, 0)$ & no & no & no \\
$(0, 0, 1)$ & no & no & yes\\
$(0, 1, 0)$ & yes & yes & yes\\
$(0, 1, 1)$ & yes & yes & no\\
$(1, 0, 0)$ & no & yes & no\\
$(1, 0, 1)$ & no & yes & yes\\
$(1, 1, 0)$ & yes & no & yes\\
$(1, 1, 1)$ & yes & no & no
\\
\hline\hline
\end{tabular}
\end{table}

%--------------------------------------------------------------

\subsubsection{Anomaly matching}\label{subsec:anomaly_indicators}

In the above classification of fractional quantum numbers, we have included all possible fractional quantum numbers relevant to $G=p4\times SO(3)$ symmetric $SU(2)_k$ topological orders, without specifying the types of lattice systems. However, given a type of lattice systems, only some of these possibilities can be realized. Now we specialize to different types of lattices and classify the fractional quantum numbers for each type.

At the first glance, there appears to be infinitely many types of lattice systems, such as square lattice spin-1/2 and spin-3/2 systems studied in Refs. \cite{Zhang2024} and \cite{Luo2023}, checkerboard lattice spin-1 systems, Lieb lattice spin-1/2 systems, etc. It turns out that all these lattice systems can be organized into a finite set of equivalence classes, known as the lattice homotopy classes \cite{Po2017, ElseThorngren2020}. For a review of the general notion of lattice homotopy, we refer the reader to Sec. II. A of Ref. \cite{Ye2022} or Sec. IV of Ref. \cite{Ye2023}. For the purpose of the current paper, we just need the fact that there are 8 different lattice homotopy classes for $p4\times SO(3)$ symmetric lattice spin systems, and each of them has a simple representative system, where half-odd-integer spins are put at some of the irreducible Wyckoff positions in Fig. \ref{fig: p4}. These 8 lattice homotopy classes can be labeled as
\beq
0,\ a,\ b,\ c,\ a+b,\ a+c,\ b+c,\ a+b+c,
\eeq
where, for example, $0$ means that no half-odd-integer spin is put at any irreducible Wyckoff position, $a$ means that half-odd-integer spins are put at the irreducible Wyckoff position $a$, and $a+c$ means that half-odd-integer spins are put at irreducible Wyckoff positions $a$ and $c$ together. So a system with only integer spins belongs to class 0, the square lattice spin-1/2 and spin-3/2 systems studied in Refs. \cite{Zhang2024} and \cite{Luo2023} both belong to class $a$, and a Lieb lattice spin-1/2 system belongs to class $a+c$.

Our goal is to determine the valid fractional quantum numbers, or $\eta$-symbols, for each lattice homotopy class, and the strategy is to perform anomaly matching. In essence, from the microscopic perspective the anomaly captures the interplay between the microscopic degrees of freedom and the symmetries, and from the perspective of the low-energy quantum phase of matter it can be derived using the universal data of this phase \cite{Zou2026}. For two dimensional lattice spin systems, the anomaly of each lattice homotopy class, known as the Lieb-Schultz-Mattis anomaly, is derived in Ref. \cite{Ye2022}, and for an SET phase it can also be expressed via data such as the $U$- and $\eta$-symbols \cite{Ye2023SciPost}. The matching between these two perspectives gives stringent constraints on which SET phases can emerge in systems in a given lattice homotopy class.

Each lattice homotopy class is uniquely specified by its anomaly, as indicated in Table \ref{tab:p4_indicators} (see Table II of Ref. \cite{Ye2023}). Here the anomaly is completely characterized by the values of the following three anomaly indicators (see Eq. (8) of Ref. \cite{Ye2023}):
\begin{align}
\mathsf I_1 &= \mathcal I_3(C_2U_\pi,\; C_2U_\pi'),\label{eq:I1}\\
\mathsf I_2 &= \mathcal I_3(T_1T_2C_2U_\pi,\; T_1T_2C_2U_\pi'),\label{eq:I2}\\
\mathsf I_3 &= \mathcal I_3(T_1C_2U_\pi,\; T_1C_2U_\pi'),\label{eq:I3}
\end{align}
Recall that $C_2=C_4^2$ is the 2-fold lattice rotation, $T_{1,2}$ are the lattice translations, and  $U_\pi$ and $U_\pi'$ are $\pi$ spin rotations around two orthogonal axes, respectively. Note that each argument in the above anomaly indicators, such as $C_2U_\pi$ and $T_1T_2C_2U_\pi'$, is an order-2 element, \ie its square is identity. In the above, $\mathcal I_3(g_1, g_2)$ is defined for a $\z_2\times\z_2$ group generated by $g_1$ and $g_2$:
\beq\label{eq:indicator_with_z2z2 main}
\begin{aligned}
&\mathcal{I}_3\left(g_1, g_2\right)\\ 
=& \frac{1}{D^2}\sum_{\substack{a,b,x,u,\\
^{g_1}a = a,\\
a \times b \times ^{g_1}b \rightarrow ^{g_2}a}}
{d_b}\frac{\theta_x}{\theta_a}
R_u^{b, ^{g_1}b} F_{{^{g_2}}a}^{a,b,{^{g_1}}b}
F_{{^{g_2}}a}^{a,{^{g_1}}b,b}\\
&\times U_{g_1}^{-1}(a,b;x)U_{g_1}^{-1}(x,^{g_1}b;^{g_2}a)\\
&\times \frac{1}{\eta_b(g_1, g_1)}\frac{\eta_a(g_2, g_1)}{\eta_a(g_1, g_2)}
\end{aligned}
\eeq
where the total quantum dimension $D$, topological spin $\theta_a$, $F$- and $R$-symbols are symmetry-independent topological properties of the $SU(2)_k$ topological order, which can be found in Appendix \ref{app:SU2k_data}. In the summation, the symbol ``$a\times b\times ^{g_1}b\rightarrow ^{g_2}a$'' means that $^{g_2}a$ is a fusion outcome of $a$, $b$ and $^{g_1}b$. Eq. \eqref{eq:indicator_with_z2z2 main} is a simplified version of Eq. (6) of Ref. \cite{Ye2023}, where we have used the fact that all $F$- and $U$-symbols are real numbers. Eq. \eqref{eq:indicator_with_z2z2 main} takes values in $\{\pm 1\}$ and is a partition function of a $(3{+}1)$-dimensional $\mathbb{Z}_2\times\mathbb{Z}_2$ symmetry-protected topological phase whose boundary realizes the given SET order~\cite{Ye2023SciPost}.

\begin{table}[htbp]
\centering
\caption{Values of the anomaly indicators $(\mathsf{I}_1,\mathsf{I}_2,\mathsf{I}_3)$ for the eight lattice homotopy classes with symmetry group $p4 \times SO(3)$.}
\label{tab:p4_indicators}
\begin{tabular}{l|cccccccc}
\hline\hline
 & 0 & $a$ & $b$ & $c$ & $a{+}b$ & $a{+}c$ & $b{+}c$ & $a{+}b{+}c$ \\
\hline
$\mathsf I_1$ & 1 & $-1$ & 1 & 1 & $-1$ & $-1$ & 1 & $-1$ \\
$\mathsf I_2$ & 1 & 1 & $-1$ & 1 & $-1$ & 1 & $-1$ & $-1$ \\
$\mathsf I_3$ & 1 & 1 & 1 & $-1$ & 1 & $-1$ & $-1$ & $-1$ \\
\hline\hline
\end{tabular}
\end{table}

Substituting into Eqs. \eqref{eq:I1}, \eqref{eq:I2} and \eqref{eq:I3} the $U$-symbol in Eqs. \eqref{eq: anyon permutation and U} and \eqref{eq: topological U symbol main}, the $\eta$-symbols in Eqs. \eqref{eq:eta_general} and \eqref{eq:w_parameterization}, and the symmetry-independent topological properties of the $SU(2)_k$ topological order collected in Appendix \ref{app:SU2k_data}, we obtain the anomaly indicators (see Appendix \ref{app: calculating anomalies}):
\beq \label{eq: calculated anomaly indicators}
\begin{split}
\mathsf I_1 &= (-1)^{kn_2 n_4},\\
\mathsf I_2 &= (-1)^{k(n_1+n_2)n_4},\\
\mathsf I_3 &= (-1)^{k(n_2+n_3)n_4}.
\end{split}
\eeq
The above results apply to all $k$ and all anyon permutation patterns. It shows that only the parity, but not the precise value, of $k$ is important to determine the anomaly.

\subsubsection{Classification results} \label{subsubsec: classification result}

Finally, matching the anomalies Eq. \eqref{eq: calculated anomaly indicators} to Table \ref{tab:p4_indicators} yields the following classification.

\paragraph{Even $k$.}
When $k$ is even, the anomaly is always trivial, \ie $\mathsf{I}_{1,2,3}=1$, so an $SU(2)_k$ topological order can be realized \emph{only in the trivial lattice homotopy class $0$}, such as a lattice system with only integer spins. Conversely, on a system in the lattice homotopy class 0, if $k\neq 4n+2$ with an integer $n\geqslant 1$, all $2^4=16$ combinations of $(n_1, n_2, n_3, n_4)$ in Eq. \eqref{eq:w_parameterization} give rise to valid fractional quantum numbers via Eq. \eqref{eq:eta_general}, \ie there are 16 SET phases of such $SU(2)_k$ topological order in these systems. If $k=4n+2$ with an integer $n\geqslant 1$, then there are $16\times 4=64$ SET phases, where the additional factor of 4 counts the 4 anyon permutation patterns described in Sec. \ref{subsubsec: anyon permutation}.

\paragraph{Odd $k$.}
When $k$ is odd, the anomaly indicators depend nontrivially on the fractionalization data $(n_1,n_2,n_3,n_4)$. Different choices can match different lattice homotopy classes. Specifically, to realize $SU(2)_k$ on a square-lattice spin-1/2 or spin-$3/2$ system, which belongs to lattice homotopy class $a$ and is studied in Refs. \cite{Zhang2024} and \cite{Luo2023}, since $(\mathsf{I}_1,\mathsf{I}_2,\mathsf{I}_3) = (-1,1,1)$ in this case, we need
\begin{equation}
n_2 n_4 = 1, \ (n_1+n_2) n_4 = 0, \ (n_2+n_3) n_4 = 0 \pmod{2},
\end{equation}
which uniquely fixes $(n_1,n_2,n_3,n_4) = (1,1,1,1)$.

More generally, Table~\ref{tab:classification_SU2k} summarizes the anomaly-compatible fractional quantum numbers for each lattice homotopy class. We can see that if the lattice homotopy class comes with a nontrivial Lieb-Schultz-Mattis anomaly, then there is exactly one SET phase of $SU(2)_k$ topological order, as long as $k$ is odd. If the lattice homotopy class comes with no anomaly, there are 9 SET phases of the $SU(2)_k$ topological order, where either $n_4=0$ or $n_1=n_2=n_3=0$.

\begin{table}[htbp]
\centering
\caption{Classification of symmetry-enriched $SU(2)_k$ topological 
orders in lattice systems with $p4\times SO(3)$ symmetry, assuming that $k$ is odd. The fractional quantum numbers are parameterized by $(n_1,n_2,n_3,n_4)\in\{0,1\}^{\otimes 4}$ through Eqs. \eqref{eq:eta_general} and \eqref{eq:w_parameterization}.}
\label{tab:classification_SU2k}
\begin{tabular}{l|c}
\hline\hline
Lattice homotopy class & Fractional quantum numbers \\
\hline
$0$       & $(n_1, n_2, n_3, 0)$, or $(0,0,0,1)$ \\
$a$       & $(1,1,1,1)$ \\
$b$       & $(1,0,0,1)$ \\
$c$       & $(0,0,1,1)$ \\
$a+b$     & $(0,1,1,1)$ \\
$a+c$     & $(1,1,0,1)$ \\
$b+c$     & $(1,0,1,1)$ \\
$a+b+c$   & $(0,1,0,1)$ \\
\hline\hline
\end{tabular}
\end{table}

In this way, anomaly matching refines the cohomology classification $H^2(G, \mc{A})=\mathbb Z_2^4$ into the subset of SET phases that are actually compatible with a given $p4\times SO(3)$ symmetric lattice system. Together with the anyon permutation patterns and their corresponding $U$-symbols discussed in Sec. \ref{subsubsec: anyon permutation}, this result gives a complete classification of $p4\times SO(3)$ symmetric $SU(2)_k$ topological phases.

In passing, we remark that our classification implies that, with appropriate models, the $p4\times SO(3)$ symmetric $SU(2)_1$ ($SU(2)_3$) topological phase with $(n_1, n_2, n_3, n_4)=(1,1,1,1)$ can emerge in a square lattice spin-3/2 (spin-1/2) system, in addition to the square lattice spin-1/2 (spin-3/2) system studied in Ref. \cite{Zhang2024} (Ref. \cite{Luo2023}). However, constructing such models and verifying the emergence of these phases is beyond our current scope.

%--------------------------------------------------------------
\section{$SU(2)$ Chern--Simons theory with critical bosons}
\label{sec:critical_bosons}

%---------------------------------------------------------------------
 
%---------------------------------------------------------------------
 
After classifying the $p4\times SO(3)$ symmetric $SU(2)_k$ topological phases and pinning down the SET phases realized in Refs. \cite{Zhang2024} and \cite{Luo2023}, we now turn to the $SU(2)_k$ Chern-Simons theory coupled to $N_f$ flavors of critical bosons that are in the fundamental representation of the $SU(2)$ gauge group. These theories are natural candidates that describe continuous quantum phase transitions between the $SU(2)_k$ topological phases and various phases with spontaneously broken symmetries, and some CFTs identified in Ref. \cite{Zhou2025} are also potentially described by such theories.

We tune the theory to a fixed point that will be described in more detail below. The main purpose of this section is to classify the gauge invariant bilinear operators of the bosonic matter and calculate their scaling dimensions, which can be used to diagnose the critical point. The logic flow is as follows. First, because the fundamental representation
of $SU(2)$ is pseudoreal, the flavor symmetry group of the theory is enhanced from a naive $SU(N_f)$ group to a larger $USp(2N_f)$ group. This enlarged global symmetry strongly constrains the allowed quartic interactions between the bosons and fixes
how gauge invariant bilinear operators of bosons organize into
its representations. In particular, these bilinears split into a singlet representation and an anti-symmetric traceless representation that will be defined later. The singlet operator is special, and
at the fixed point it is more naturally represented by
a Hubbard-Stratonovich field $\sigma$. After establishing this operator basis, we compute the scaling dimensions
of $\sigma$ and the anti-symmetric traceless bilinear operators to the order of $1/N_f$, where the results are given in Eqs. \eqref{eq:Delta_sigma} and \eqref{eq:Delta_A0_boson}. The full technical
details are collected in Appendix~\ref{app:boson_details}.

%---------------------------------------------------------------------
\subsection{Action}
\label{subsec:boson_setup}
%---------------------------------------------------------------------

We start by describing the Chern-Simons-boson theory.

Denote the $SU(2)$ gauge field by $A^{a}T^{a}$, where the generators of $SU(2)$ are $T^{a}=\sigma^{a}/2$ for $a=1,2,3$, with $\sigma^a$ the standard Pauli matrices. The matter fields consist of $N_f$ flavors of complex bosons $\Phi_{\alpha,i}$,
where $\alpha=1,2$ is an $SU(2)$ color index and
$i=1,\dots,N_f$ is a flavor index.
The Euclidean action is
$S=S_{\mathrm{CS}}+S_{\mathrm{GF}}+S_{\mathrm{gh}}+S_{\mathrm{matter}}$,
with
\begin{widetext}
\begin{align}
S_{\mathrm{CS}}
&= \frac{k}{4\pi}\!\int\!d^3x\left(
  -\frac{i}{2}\varepsilon_{\mu\nu\lambda}A_\mu^{a}\partial_\nu A_\lambda^{a}
  -\frac{i}{6}\varepsilon_{\mu\nu\lambda}f^{abc}
  A_\mu^a A_\nu^b A_\lambda^c\right),\label{eq: CS action bosonic matter}\\
S_{\mathrm{GF}}
&= -\frac{N_f}{32\xi}\int d^3x d^3y\frac{\partial_\mu A^a_\mu(x)\partial_\nu A^a_\nu(y)}{2\pi^2|x-y|^2},\ \xi\leqslant0 \label{eq: gauge fixing action bosonic matter}\\
S_{\mathrm{gh}}
&= \int\!d^3x\left(
  \partial_\mu\bar c^{a}\partial^\mu c^{a}
  + f^{abc}\partial_\mu\bar c^{a} A_\mu^{b}c^{c}\right), \label{eq: ghost action bosonic matter}\\
S_{\mathrm{matter}}
&= \int\!d^3x\left[
(D_\mu\Phi_{\alpha, i})^\dagger(D^\mu\Phi_{\alpha, i})
  + \lambda\left(\Phi_{\alpha,i}^\dagger\Phi_{\alpha, i}\right)^2
  \right],\label{eq: original bosonic matter action}
\end{align}
\end{widetext}
where $D_\mu\Phi_{\alpha, i}=(\partial_\mu\delta_{\alpha\beta}-iA_\mu^{a}T^{a}_{\alpha\beta})\Phi_{\beta, i}$. Throughout this paper, repeated indices are summed over unless otherwise stated, and tensors with superscripts and subscripts are identical. In the above, $S_{\mathrm{CS}}$ is the $SU(2)_k$ Chern-Simons action, $S_{\mathrm{GF}}$ is a gauge-fixing term with $\xi=0$ corresponding to the Landau gauge where $\partial_\mu A^a_\mu=0$ ($\xi\leqslant0$ to ensure the boundedness of action), $S_{\mathrm{gh}}$ is the action for the Faddeev-Popov ghost $c$, and $S_{\mathrm{matter}}$ describes the bosonic matter and its coupling to the gauge field.

%---------------------------------------------------------------------
\subsection{Flavor symmetry and the single quartic coupling}
\label{subsec:boson_symmetry}
%---------------------------------------------------------------------

Next, we determine the flavor symmetry, since it provides a natural
classification of operators.

Define the symplectic conjugate
\begin{equation}
\tilde\Phi_{\alpha,i}
\equiv
\varepsilon_{\alpha\beta}\Phi^*_{\beta,i}.
\end{equation}
It is important to note that $\tilde\Phi$ transforms in the same way as $\Phi$ under gauge transformations. Now we can combine $\Phi$ and $\tilde\Phi$ into a single $2\times 2N_f$ matrix
\begin{equation}
X=[\Phi\,|\,\tilde\Phi].
\end{equation}
Then the $SU(2)$ gauge group acts on $X$ from the left, while the flavor symmetry acts from the right. The covariant derivative becomes
\begin{equation}
D_\mu X
=
\partial_\mu X-iA_\mu^aT^aX,
\end{equation}
and the kinetic term of the bosonic matter can be written as
\begin{equation} \label{eq: bosonic kinetic term}
\mathcal L_{\mathrm b,kin}
=
\frac12\,
\operatorname{Tr}
\left[
(D_\mu X)^\dagger(D^\mu X)
\right],
\end{equation}
where the trace is over both the $SU(2)$ color index and the
$2N_f$ flavor indices. The factor of $1/2$ compensates for the
doubling of fields in $X=[\Phi\,|\,\tilde\Phi]$, since the
$\tilde\Phi$ columns are not independent degrees of freedom.

The matrix $X$ obeys the pseudoreality constraint
\begin{equation}
X^*=\varepsilon X\Omega,
\quad
\Omega=
\begin{pmatrix}
0 & \mathbb I_{N_f}\\
-\mathbb I_{N_f} & 0
\end{pmatrix}.
\label{eq:reality}
\end{equation}
Without violating this constraint, the kinetic term Eq. \eqref{eq: bosonic kinetic term} is invariant under
\begin{equation} \label{eq: USp transformation}
X\rightarrow XU,
\quad
U^\dagger U=\mathbb I_{2N_f},
\quad
U^T\Omega U=\Omega.
\end{equation}
These two conditions imply that $U$ is a $USp(2N_f)$ matrix. Therefore, the manifest $SU(N_f)$ flavor symmetry is enhanced to
\begin{equation}
G_{\mathrm{flavor}}
=
USp(2N_f),
\end{equation}
with the $USp(2N_f)$ transformation given in Eq. \eqref{eq: USp transformation}.{\footnote{Since $X$ is not gauge invariant, the faithful flavor symmetry group is actually  $USp(2N_f)$ with its center modded out. However, it is unnecessary to make this distinction for our purpose.}} This way of identifying enlarged flavor symmetries
from pseudoreality is standard in discussions of gauge theories with $USp$ gauge groups, where our gauge group is $SU(2)=USp(2)$
\cite{WangNahumMetlitskiXuSenthil2017, Zou2021}.

Identifying this enlarged symmetry is
important because the operator spectrum should be organized into
representations of $USp(2N_f)$ rather than only of the manifest
$SU(N_f)$ subgroup. The same symmetry also constrains the allowed quartic interactions. A
systematic classification of gauge invariant and $USp(2N_f)$-invariant
quartic operators of the bosons, as reviewed in Appendix~\ref{app:quartic_classification},
shows that the only gauge invariant and $USp(2N_f)$-invariant quartic interaction of the bosons is proportional to $\left(\Phi_{\alpha, i}^\dagger\Phi_{\alpha, i}\right)^2$. This justifies the use of a single quartic coupling $\lambda$ in
$S_{\mathrm{matter}}$.

\subsection{Fixed point action}

Now we can describe the fixed point we are interested in.

To do so, it is convenient to introduce a Hubbard-Stratonovich field $\sigma$ and eliminate the quartic interaction of the bosons. Concretely, we replace $S_{\rm matter}$ in Eq. \eqref{eq: original bosonic matter action} by
\beq \label{eq: matter action with HS field}
\bsp
S_{\rm matter}
=
\int\!d^3x\Bigg[
&(D_\mu\Phi_{\alpha, i})^\dagger(D^\mu\Phi_{\alpha, i})\\
  -&
  \sigma\left(\Phi_{\alpha,i}^\dagger\Phi_{\alpha, i}\right) +\eta\sigma^2
  \Bigg],
\esp
\eeq
where $\eta=-1/(4\lambda)$ and $\sigma$ is an imaginary dynamical field that should be integrated over in the path integral. One can check that integrating $\sigma$ out indeed converts this new $S_{\rm matter}$ to its original form in Eq. \eqref{eq: original bosonic matter action}. 

From now on, we take the total action to be $S=S_{\mathrm{CS}}+S_{\mathrm{GF}}+S_{\mathrm{gh}}+S_{\rm matter}$, with $S_{\mathrm{CS}}$, $S_{\mathrm{GF}}$, $S_{\mathrm{gh}}$ and $S_{\rm matter}$ given by Eqs. \eqref{eq: CS action bosonic matter}, \eqref{eq: gauge fixing action bosonic matter}, \eqref{eq: ghost action bosonic matter} and \eqref{eq: matter action with HS field}, respectively. The fixed point we are interested in is achieved by setting 
(the renormalized version of) $\eta=0$, where the field $\sigma$ is gapless. We will see that such a fixed point can be accessed by a $1/N_f$ expansion with respect to a joint large-$N_f$ and large-$k$ limit, where
\beq
N_f\rightarrow\infty,\quad  k\rightarrow\infty,\quad {\rm with\ } \frac{N_f}{k}\sim\mc{O}(1) {\rm \ fixed.}
\eeq

In passing, we note that this fixed point is not expected to be the only one that can be achieved by the original action defined in Eqs. \eqref{eq: CS action bosonic matter}, \eqref{eq: gauge fixing action bosonic matter}, \eqref{eq: ghost action bosonic matter} and \eqref{eq: original bosonic matter action}. There can be another fixed point with $\eta\rightarrow -\infty$, where the field $\sigma$ has a large gap and can be ignored in the low-energy physics. This fixed point is analogous to the tricritical point in an Ising transition, since both of them have a vanishing quartic interaction between the bosons. However, such tricritical-type fixed points are expected to be more unstable than the critical fixed points with $\eta=0$ (\ie having more symmetric relevant local perturbations), so we will focus on the case with $\eta=0$.

%---------------------------------------------------------------------
\subsection{Bilinear operators}
\label{subsec:boson_bilinears}
%---------------------------------------------------------------------

Having identified the enlarged flavor symmetry and specified the fixed point of our interest, now we organize the gauge invariant bilinear operators of the bosons as representations of the $USp(2N_f)$ flavor symmetry group. 

Let
$I,J=1,\ldots,2N_f$ denote the $USp(2N_f)$ fundamental indices and
write the scalar field as $X_{\alpha I}$, where $\alpha=1,2$ is the
$SU(2)$ gauge index. The basic gauge invariant bilinear operator of the bosons is
\begin{equation}
\mathcal O_{IJ}
=
\varepsilon^{\alpha\beta}X_{\alpha I}X_{\beta J}.
\end{equation}
Because the bosonic fields commute while $\varepsilon^{\alpha\beta}$ is
anti-symmetric, $\mathcal O_{IJ}$ is antisymmetric in the
$USp(2N_f)$ indices:
\begin{equation}
\mathcal O_{IJ}=-\mathcal O_{JI}.
\end{equation}
Therefore, the bilinear operators transform in
\begin{equation}
\wedge^2\mathbf{2N_f}
=
\mathbf 1\oplus \mathbf A_0,
\end{equation}
where $\mathbf 1$ is the singlet representation and $\mathbf A_0$ is the anti-symmetric traceless representation of
$USp(2N_f)$. Explicitly, any bilinear operator $\mc{O}_{IJ}$ can be decomposed into a singlet part $\mc{O}_{\rm sing}$ and an anti-symmetric traceless part $\mc{O}^{A_0}_{IJ}$, where
\begin{align}
\mathcal O_{\mathrm{sing}}
\propto
\Omega^{IJ}\mathcal O_{IJ},
\quad
\mathcal O^{A_0}_{IJ}
=
\mathcal O_{IJ}
-
\frac{1}{2N_f}\Omega_{IJ}
\Omega^{KL}\mathcal O_{KL}.
\end{align}
This is the full $USp(2N_f)$ classification of
bilinear operators of bosons, and $O^{A_0}_{IJ}$ is called ``traceless'' because $O^{A_0}_{IJ}\Omega_{IJ}=0$.

The singlet component is proportional to the naive mass operator
\begin{equation}
|\Phi|^2_{\mathrm{sing}}
=
\frac{1}{\sqrt{N_f}}\sum_i\Phi_i^\dagger\Phi_i.
\end{equation}
However, the equation of motion of $\sigma$ in Eq. \eqref{eq: matter action with HS field} sets this operator to zero, so we should identify the scaling dimension of the most relevant operator in this representation as the scaling dimension of $\sigma$ \cite{Chester2016, BenvenutiKhachatryan2019}.

For the explicit calculation of the scaling dimensions of bilinear operators in the anti-symmetric traceless representation, it is
convenient to choose an $SU(N_f)\subset USp(2N_f)$ subgroup
that acts on the original fields $\Phi_{\alpha, i}$. In this notation, the
number-conserving block of $\mathcal O^{A_0}_{IJ}$ contains the familiar
traceless bilinear
\begin{equation} \label{eq: boson adjoint mass}
|\Phi|^2_{\mathrm{adj}}
=
\Phi_i^\dagger\Phi_j
-
\frac{\delta_{ij}}{N_f}
\sum_l\Phi_l^\dagger\Phi_l.
\end{equation}
Here the subscript ``adj'' means the adjoint representation of this
chosen $SU(N_f)$ subgroup, not the adjoint representation of
$USp(2N_f)$. Such operators are representative components of the full $\mathbf A_0$ multiplet, and other components of the same
$USp(2N_f)$ multiplet include the pairing operators
\begin{equation}
\varepsilon^{\alpha\beta}\Phi_{\alpha i}\Phi_{\beta j},
\quad
\varepsilon^{\alpha\beta}
\tilde\Phi_{\alpha i}\tilde\Phi_{\beta j},
\end{equation}
All these components have the same scaling dimension at a
$USp(2N_f)$-invariant fixed point, and it suffices to calculate for one of them.

In summary, the two bilinear operators whose dimensions we will compute are the Hubbard-Stratonovich field $\sigma$ and $|\Phi|^2_{\mathrm{adj}}$ introduced in Eq. \eqref{eq: boson adjoint mass}.

%---------------------------------------------------------------------
\subsection{Scaling dimensions at $\mathcal O(1/N_f)$}
\label{subsec:boson_scaling_dimensions}
%---------------------------------------------------------------------

Now we are ready to calculate the scaling dimensions of the bilinear operators of the bosons, which mean the scaling dimensions of $\sigma$ and $|\Phi|^2_{\mathrm{adj}}$ in practice.

For any operator $O$, its scaling dimension $\Delta_O$ is defined through the behavior of its two-point correlation function via
\beq
\langle O^\dag(x)O(0)\rangle\sim\frac{1}{|x|^{2\Delta_O}}
\eeq
for large distance $|x|$, or
\beq \label{eq: scaling dimension from correlator}
\langle O^\dag(p)O(p)\rangle\sim |p|^{2\Delta_O-3}
\eeq
for small momentum $p$. For our purpose, we work in the joint large-$N_f$,
large-$k$ limit, where
\begin{equation}
\label{eq:large_Nf_limit_boson}
N_f\to\infty,\quad k\to\infty,\quad
\frac{N_f}{k}=\mathcal O(1)\;\text{fixed}.
\end{equation}
We will directly calculate the two-point correlation functions of $\sigma$ and $|\Phi|^2_{\mathrm{adj}}$ to the order of $1/N_f$.

Here we sketch the workflow of this calculation, while leaving its details to Appendix \ref{app:boson_details}. Starting from the action $S=S_{\rm CS}+S_{\rm GF}+S_{\rm gh}+S_{\rm matter}$, with $S_{\rm CS}$, $S_{\rm GF}$, $S_{\rm gh}$ and $S_{\rm matter}$ given in Eqs. \eqref{eq: CS action bosonic matter}, \eqref{eq: gauge fixing action bosonic matter}, \eqref{eq: ghost action bosonic matter} and \eqref{eq: matter action with HS field}, respectively, we first integrate out the fields $\Phi_{\alpha, i}$ to get an effective action. This effective action modifies the propagators of the gauge field $A^a_\mu$ and the field $\sigma$, and its structure naturally allows a $1/N_f$ expansion, with each term in the expansion having some diagrammatic interpretation. Using this effective action, we calculate the relevant two-point correlation functions, from which we can read off the scaling dimensions:
\beq \label{eq:Delta_sigma}
\bsp
\Delta_\sigma=
&2+
\frac{1}{N_f}
\Bigg\{
\frac{N_f^2}{32\pi^2\Delta_b}-\frac{8}{3\pi^2}\\
&-\frac{3N_f^2}{32\pi^2\Delta_b^2}\left[\left(\frac{N_f}{32}\right)^2-\left(\frac{k}{4\pi}\right)^2\right]
\Bigg\}
+\mc{O}(1/N_f^2)
\esp
\eeq
and
\beq \label{eq:Delta_A0_boson}
\bsp
\Delta_{|\Phi|^2_{\rm adj}}
=1+\frac{1}{N_f}
\left(
\frac{8}{3\pi^2}
-\frac{N_f^2}{32\pi^2\Delta_b}
\right)
+\mc{O}(1/N_f^2),
\esp
\eeq
with
\beq
\Delta_b=\left(\frac{N_f}{32}\right)^2+\left(\frac{k}{4\pi}\right)^2.
\eeq

%--------------------------------------------------------------
 
\section{$SU(2)$ Chern--Simons theory with critical fermions}
\label{sec:critical_fermions}

The quantum critical theory in Sec. \ref{sec:critical_bosons} naturally describes a quantum phase transition between the $SU(2)_k$ topological phase and a phase with a spontaneously broken symmetry. In this section, we study the $SU(2)_k$ Chern-Simons theory coupled to $N_f$ flavors of gapless two-component Dirac fermions, which naturally describes a quantum phase transition between two topological phases captured by the $SU(2)_{k+\frac{N_f}{2}}$ Chern-Simons theory and the $SU(2)_{k-\frac{N_f}{2}}$ Chern-Simons theory, respectively{\footnote{Note that the Chern-Simons theory contains fermionic matter, so its topological properties are different from those of the standard $SU(2)$ Chern-Simons theory.}}. Across such a transition, no spontaneous symmetry breaking is expected. For the consistency of the theory, we demand that $k+\frac{N_f}{2}\in\z$.

As in Sec. \ref{sec:critical_bosons}, our goal here is also to classify the gauge invariant mass and pairing operators of the fermions and calculate their scaling dimensions, and these scaling dimensions can be used to diagnose the critical point. Again, we work in the limit
\beq
N_f\to\infty,\quad k\to\infty,\quad
\frac{N_f}{k}=\mathcal O(1)\;\text{fixed}.
\eeq
The analysis below shares many similarities with that in Sec. \ref{sec:critical_bosons}. For example, the flavor symmetry group here is also $USp(2N_f)$, and the mass and pairing operators fall into singlet and anti-symmetric traceless representations of the $USp(2N_f)$ symmetry. The scaling dimensions of these two types of operators are given by Eqs. \eqref{eq:Delta_S_fermion} and \eqref{eq:Delta_A0_fermion}, respectively.

\subsection{Action}

As in Sec. \ref{sec:critical_bosons}, we first introduce the Chern-Simons-fermion theory.

We still denote the $SU(2)$ gauge field by $A^aT^a$, where the generators of $SU(2)$ are $T^a=\sigma^a/2$ for $a=1,2,3$, with $\sigma^a$ the Pauli matrices. The matter fields consist of $N_f$ flavors of gapless two-component Dirac fermions $\psi_{\alpha, i}$, where $\alpha=1,2$ is the $SU(2)$ color index and $i=1, 2, \cdots, N_f$ is a flavor index. The Euclidean action is $S=S_{\rm CS}+S_{\rm GF}+S_{\rm gh}+S_{\rm matter}$, with{\footnote{In addition to this action, this theory also contains a gravitational Chern-Simons term (see, for example, Ref. \cite{Zou2018}). However, for the purpose of this work that term is unimportant.}}
\begin{widetext}
\label{eq:action_pieces}
\begin{align}
  S_{\mathrm{CS}}
  &=
  \frac{k}{4\pi}\int d^{3}x\,
  \biggl(
    -\frac{i}{2}\,\epsilon_{\mu\nu\lambda}\,
      A_\mu^a\,\partial_\nu A_\lambda^a
    -\frac{i}{6}\,\epsilon_{\mu\nu\lambda}\,
      f^{abc}\,A_\mu^a A_\nu^b A_\lambda^c
  \biggr), \\[4pt]
  S_{\mathrm{GF}}
  &=-\frac{N_f}{32\xi}\int d^3x d^3y\frac{\partial_\mu A^a_\mu(x)\partial_\nu A^a_\nu(y)}{2\pi^2|x-y|^2},\ \xi\leqslant 0 \\[4pt]
  S_{\mathrm{gh}}
  &=
  \int d^{3}x\,
  \bigl(
    \partial_\mu \bar{c}^{\,a}\,\partial^\mu c^a
    +
    f^{abc}\,\partial_\mu \bar{c}^{\,a}\,A_\mu^b\,c^c
  \bigr), \\[4pt]
  S_{\mathrm{matter}}
  &=
  \int d^{3}x\,
  \sum_{i=1}^{N_f}
    \bar\psi_i \,\gamma^\mu D_\mu \psi_i,
\end{align}
\end{widetext}
where $D_\mu\psi=\bigl(\partial_\mu -\,iA_\mu^a T^a\bigr)\psi$ and the Euclidean gamma matrices satisfy $\{\gamma^\mu, \gamma^\nu\}=2\delta^{\mu\nu}$. As we can see, this action differs from the action in Sec. \ref{sec:critical_bosons} only by $S_{\rm matter}$.

\subsection{Flavor symmetry}

As in Sec. \ref{sec:critical_bosons}, we now analyze the flavor symmetry, because it provides a way to organize the mass and pairing operators in its irreducible representations.

Similar to the case with bosonic matter, here we will also define a conjugate of the fermionic fields that transforms in the same way as the original fermionic fields under both the gauge and Euclidean transformations. Then we can express the kinetic term using the original fermionic fields and these conjugate fields, and the relation between these fields yields the correct flavor symmetry group, which turns out to be $USp(2N_f)$, rather than a naive $SU(N_f)$.

Concretely, we first make the spinor index of the Dirac fermions explicit, \ie each fermion field $\psi_{\alpha i,s}$ carries an $SU(2)$ color index
$\alpha=1,2$, a flavor index $i=1,\ldots,N_f$, and a two-component spinor index $s=1,2$. Now we define
\beq
\tilde\psi_{\alpha i, s}=(\varepsilon_c)_{\alpha\beta}(C)_{st}\bar\psi_{\beta i, t},
\eeq
where $\varepsilon_c$ and $C$ act on the color and spinor indices, respectively. Here $C$ is a charge conjugation operation, which is defined such that $\tilde\psi$ transforms in the same way as $\psi$ under the Euclidean transformations. One can also check that $\tilde\psi$ transforms in the same way as $\psi$ under gauge transformations due to the presence of $\varepsilon_c$.

Next, we package $\psi_{\alpha i}$ and $\tilde\psi_{\alpha i}$ into a 2-by-$2N_f$ matrix
\beq
X=[\psi | \tilde\psi],
\eeq
where each entry is a two-component spinor. The gauge and Euclidean transformations act on $X$ simply by left multiplication. This matrix satisfies a reality condition
\beq \label{eq: reality fermion}
X^*=(\varepsilon_c\otimes\varepsilon_E) X \Omega,
\quad
\Omega=
\begin{pmatrix}
0 & \mathbb I_{N_f}\\
-\mathbb I_{N_f} & 0
\end{pmatrix}.
\eeq

In terms of $X$, the kinetic Lagrangian in the action can be written as
\beq
\mc{L}_{\rm matter}=\frac{1}{2}{\rm Tr}(\bar X \gamma^\mu D_\mu X),
\eeq
where the trace is over the $2N_f$ flavor indices.  Taking the reality constraint Eq. \eqref{eq: reality fermion} into account, this Lagrangian is invariant under
\beq
X\rightarrow XU,
\quad
U^\dagger U=\mathbb I_{2N_f},
\quad
U^T\Omega U=\Omega.
\eeq
Therefore, the flavor symmetry, as given above, is still
\beq
G_{\rm flavor}=USp(2N_f).
\eeq

From the above analysis, we see that the enhancement of the flavor symmetry group from the naive $SU(N_f)$ to the actual $USp(2N_f)$ is due to the pseudo-reality of the fundamental representation of the $SU(2)$ gauge group and the spinor representation of the Euclidean group, similar to the enhancement of the flavor symmetry group in Sec. \ref{sec:critical_bosons}.

%--------------------------------------------------------------
\subsection{Bilinear operators}
\label{subsec:bilinear_operators}
%--------------------------------------------------------------

We next classify the gauge-invariant and Euclidean-invariant fermion bilinear operators
using the enlarged $USp(2N_f)$ global symmetry. These are the usual mass and pairing operators of the fermions.

In terms of $X$, a general gauge-invariant and Euclidean-invariant fermion bilinear operator can be expanded using
\beq
O_{IJ}=(\varepsilon_c)_{\alpha\beta}(\varepsilon_E)_{st}X_{\alpha I, s}X_{\beta J, t}.
\eeq
The anti-commutation relation between the fermionic fields and the anti-symmetries of $\varepsilon_c$ and $\varepsilon_E$ imply that the indices $I$ and $J$ should be anti-symmetrized, \ie
\beq
O_{IJ}=-O_{JI}.
\eeq
Therefore, these
bilinear operators transform in the antisymmetric tensor product of two
$USp(2N_f)$ fundamental representations,
\begin{equation}
\wedge^2\mathbf{2N_f}
=
\mathbf{1}\oplus\mathbf{A}_0 .
\end{equation}
Here $\mathbf A_0$ denotes the antisymmetric traceless representation of
$USp(2N_f)$. It should not be confused with the adjoint
representation of $USp(2N_f)$, which is the symmetric square of
the fundamental.

The singlet component is the flavor-symmetric fermion mass operator
\begin{equation} \label{eq: fermion singlet mass}
\mathcal O_S
=
\frac{1}{\sqrt{N_f}}\sum_{i=1}^{N_f}\bar\psi_i\psi_i .
\end{equation}
The remaining bilinears belong to the non-singlet channel
$\mathbf A_0$. For explicit calculations later, it is again convenient
to choose an $SU(N_f)\subset USp(2N_f)$ subgroup acting on the
original complex fermions and use the number-conserving representative
\begin{equation}
\mathcal O_{A_0}
\sim
\bar\psi_i\psi_j
-
\frac{\delta_{ij}}{N_f}
\sum_{l=1}^{N_f}\bar\psi_l\psi_l .
\end{equation}
Equivalently, one may use an off-diagonal representative such as
\begin{equation} \label{eq: fermion adjoint mass}
\mathcal O_{A_0}
=
\bar\psi_1\psi_2 .
\end{equation}
This operator is an $SU(N_f)$-adjoint representative of the full
$USp(2N_f)$ multiplet $\mathbf A_0$. Other components are pairing operators of the fermions. By the
$USp(2N_f)$ symmetry, all components of this multiplet have the
same scaling dimension at the fixed point.

Thus the two channels whose scaling dimensions we compute
are
\begin{equation}
\mathcal O_S\in \mathbf 1,
\qquad
\mathcal O_{A_0}\in \mathbf A_0,
\end{equation}
with $\mc{O}_S$ and $\mc{O}_{A_0}$ given by Eqs. \eqref{eq: fermion singlet mass} and \eqref{eq: fermion adjoint mass}, respectively.
 
%--------------------------------------------------------------
\subsection{Scaling dimensions at $\mc{O}(1/N_f)$}
\label{subsec:scaling_dimensions}
%--------------------------------------------------------------

Finally, we calculate the scaling dimensions of $\mc{O}_S$ and $\mc{O}_{A_0}$ by calculating their correlation functions and using Eq. \eqref{eq: scaling dimension from correlator}. We will work in the limit
\beq
N_f\to\infty,\quad k\to\infty,\quad
\frac{N_f}{k}=\mathcal O(1)\;\text{fixed},
\eeq
and keep the result up to the order of $\mc{O}(1/N_f)$.

The framework of doing this calculation is similar to that in Sec. \ref{subsec:boson_scaling_dimensions}, which will be discussed in more detail in Appendix \ref{app: fermion analysis}. Roughly speaking, we integrate out the fermions and obtain an effective action, which allows us to carry out an $1/N_f$ expansion of the correlation functions and thus extract the scaling dimensions using Eq. \eqref{eq: scaling dimension from correlator}.

The final result is that, up to the order $\mc{O}(1/N_f)$, the scaling dimension of the singlet mass $\mc{O}_S$ is
\begin{equation}
\label{eq:Delta_S_fermion}
\Delta_S
=
2
-
\frac{32\,N_f}{64\,k^2+N_f^2\pi^2}
+
\frac{
96\,N_f\left(N_f^2\pi^2-64\,k^2\right)
}{
\left(64\,k^2+N_f^2\pi^2\right)^2
}.
\end{equation}
and the scaling dimension of the multiplet $\mc{O}_{A_0}$ is
\begin{equation}
\label{eq:Delta_A0_fermion}
\Delta_{A_0}
=
2
-
\frac{32\,N_f}{64\,k^2+N_f^2\pi^2}.
\end{equation}

\section{Discussion}

In this work, we have carried out a systematic classification of $SU(2)_k$ topological orders on all $p4\times SO(3)$ symmetric lattice spin systems, where $k$ is an arbitrary nonzero integer. We have also performed large-$N_f$ analyses of $SU(2)_k$
Chern-Simons-matter theory with either $N_f$ flavors of gapless bosons or $N_f$ flavors of gapless two-component Dirac fermions. When the matter is bosonic, the theory naturally describes quantum phase transitions between a topological phase and a non-topological phase with some spontaneously broken symmetries. When the matter is fermionic, the theory naturally describes quantum phase transitions between two topological phases, and no spontaneous symmetry breaking is expected across this transition.

We remark on the power of our framework for classifying the topological phases. Previously, the most commonly used approach for such problems is the parton-based approach \cite{Wen2001}, where one decomposes the fundamental microscopic degrees of freedom into some fractionalized degrees of freedom, and studies the symmetries and Hamiltonians of these new degrees of freedom. As commented by many papers (see, for example, Ref. \cite{Ye2023}), this approach is neither conceptually direct nor complete, and compared to our framework of classification it is rather inefficient. If one uses this approach, one would perform some parton analysis for each $k$ and each type of lattice systems, which is formidable, especially when the symmetry can permute anyons. In contrast, we can systematically deal with all level $k$, all lattice spin systems and all patterns where the symmetry permutes anyons.

Given the classification of these symmetry-enriched topological phases, it is interesting to find concrete models to realize all of them. In particular, it is interesting to find models to show that different systems in a given lattice homotopy class, such as square lattice spin-1/2 and spin-3/2 systems, can realize, say, $SU(2)_k$ with different values of $k$. We leave this for future work.

For the quantum criticality, we note that our analysis should only be regarded as the first step towards understanding the quantum phase transitions in, say, the phase diagrams of Refs. \cite{Zhang2024} and \cite{Luo2023}. Indeed, the nature of those quantum phase transitions was not discussed in detail in Refs. \cite{Zhang2024} and \cite{Luo2023}, and currently we do not know what $N_f$ should be for those transitions. Without this information, we also cannot determine how the microscopic operators are related to the field theoretic operators, and what perturbations to the field theories we study are allowed by the microscopic symmetries, in a way similar to that in, say, Ref. \cite{Zou2018}. Once $N_f$ and how microscopic observables are related to the field theoretic operators are known, if there is further numerical result for the critical exponents, we can compare those exponents with the scaling dimensions we obtain, which serves as a way to diagnose the nature of the critical theory. We leave the more detailed studies of these quantum phase transitions to future work.

Our results of the critical $SU(2)$ Chern-Simons-matter theory are also relevant to the models numerically studied in Ref. \cite{Zhou2025}. Although which field theories are realized by the models in Ref. \cite{Zhou2025} is currently unknown, some suggestions were given therein. Some of the candidate theories there are also covered by us, including the following types.

\begin{enumerate}
    
    \item $SU(2)_1$ with $N_f$ flavors of gapless bosons in the fundamental representation, where $N_f\in\{2, 3, 4\}$. For these theories, the numerical values of Eqs. \eqref{eq:Delta_sigma} and \eqref{eq:Delta_A0_boson} are in the second and third columns below, respectively. If these theories are realized in Ref. \cite{Zhou2025}, these scaling dimensions are given by the fourth and fifth columns, respectively.

\begin{table}[h!]
\begin{tabular}{l|cc|cc}
\hline\hline
$N_f$ & $\Delta_\sigma$ & $\Delta_{|\Phi|^2_{\rm adj}}$ & $\Delta_\sigma$ & $\Delta_{|\Phi|^2_{\rm adj}}$ \\
\hline
2 & 2.92 & 0.52 & 2.797 & 0.702 \\
3 & 2.23 & 0.46 & 2.979 & 0.770\\
4 & 1.78 & 0.49 & 2.987 & 0.795\\
\hline\hline
\end{tabular}
\end{table}

\item $SU(2)_{\frac{1}{2}}$ with 3 flavors of two-component Dirac fermions in the fundamental representation. For this theory, Eqs. \eqref{eq:Delta_S_fermion} and \eqref{eq:Delta_A0_fermion} give
\beq
\Delta_S=2.99,
\quad
\Delta_{A_0}=1.08.
\eeq
If this theory is realized in Ref. \cite{Zhou2025}, the numerical values of these scaling dimensions are
\beq
\Delta_S=2.979,
\quad
\Delta_{A_0}=0.770.
\eeq

\item $SU(2)_1$ with $4$ flavors of two-component Dirac fermions in the fundamental representation. For this theory, Eqs. \eqref{eq:Delta_S_fermion} and \eqref{eq:Delta_A0_fermion} give
\beq
\Delta_S=2.16,
\quad
\Delta_{A_0}=1.42.
\eeq
If this theory is realized in Ref. \cite{Zhou2025}, the numerical values of these scaling dimensions are
\beq
\Delta_S=2.987,
\quad
\Delta_{A_0}=0.795.
\eeq
    
\end{enumerate}

In all cases, there are notable differences between our large-$N_f$ results and the numerical results in Ref. \cite{Zhou2025}, which may result from higher-order $1/N_f$ corrections on our side and finite size effects in Ref. \cite{Zhou2025}. We leave improvements of these results and the identification of the field theories realized in Ref. \cite{Zhou2025} to future work.

\begin{acknowledgments}

We thank Wei Zhu for useful discussions. We also thank various AI tools, including AI-Know and ChatGPT for helping with the analysis. LZ is
supported by the National University of Singapore start-up grants A-0009991-00-00 and A0009991-01-00.

\end{acknowledgments}

\clearpage
\onecolumngrid
\appendix

\section{Topological properties of the $SU(2)_k$ topological order}
\label{app:SU2k_data}

To systematically compute the anomaly indicators for the symmetry-enriched $SU(2)_k$ topological order in Sec.~\ref{sec:SET}, we need the explicit algebraic data that describes the topological properties of this topological order. Specifically, to evaluate Eq. \eqref{eq:indicator_with_z2z2 main}, we need the quantum dimensions, topological spins, the $F$- and $R$-symbols and the topological symmetry of this topological order. In this appendix, we collect the relevant data. All this data except the topological symmetry can be found in, for example, Sec. 5.4 of Ref. \cite{Bonderson2007}. We find the topological symmetry, especially its $U$-symbols, with the help of ChatGPT and AI-Know.

\subsection{Anyon types and fusion rules}

In the $SU(2)_k$ topological order, there are $k+1$ types of anyons, labeled by
\begin{equation}
j \in \left\{0,\frac12,1,\dots,\frac{k}{2}\right\},
\end{equation}
where $j=0$ corresponds to the trivial anyon. The fusion rules of these anyons are
\begin{equation} \label{eq: fusion rule app}
j_1 \times j_2
=
\sum_{j_3 = |j_1-j_2|}^{\min(j_1+j_2,\;k-j_1-j_2)} j_3,
\end{equation}
where $j_3$ increases in integer steps.

Because the fusion rules of $SU(2)_k$ are strictly multiplicity-free (i.e., the fusion coefficients $N_{ab}^c \in \{0, 1\}$), the $R$-symbols and $F$-symbols reduce from generic unitary matrices to simple scalar phases and real numbers, respectively. This simplification makes the $SU(2)_k$ theory particularly tractable for explicit computations of the anomaly indicators.

\subsection{Quantum integers and quantum dimensions}

It is most convenient to express all quantities using $q$-deformed numbers (quantum integers). For $SU(2)_k$, the deformation parameter is evaluated at the root of unity
\begin{equation}
q = \exp\!\left(\frac{2\pi i}{k+2}\right).
\end{equation}
The quantum integer $[n]$ is defined as
\begin{equation}\label{eq:quantum_integer}
[n] = \frac{\sin\!\left(\frac{\pi n}{k+2}\right)}{\sin\!\left(\frac{\pi}{k+2}\right)},
\end{equation}
with the corresponding quantum factorial
\begin{equation}
[n]! = [n][n-1]\cdots[1], \qquad [0]! = 1.
\end{equation}
These definitions ensure that $[n]$ reduces to the ordinary integer $n$ in the classical limit $k\to\infty$.

Using this notation, the quantum dimension of an anyon with generalized spin $j \in \{0, 1/2, \dots, k/2\}$ is simply
\begin{equation}\label{eq:quantum_dim}
d_j = [2j+1] = \frac{\sin\left(\frac{(2j+1)\pi}{k+2}\right)}{\sin\left(\frac{\pi}{k+2}\right)}.
\end{equation}
The total quantum dimension of the theory is
\begin{equation}
D = \sqrt{\sum_{j=0}^{k/2} d_j^2} = \sqrt{\frac{k+2}{2}} \cdot \frac{1}{\sin\!\left(\frac{\pi}{k+2}\right)}.
\end{equation}

\subsection{Modular $S$-matrix}

The modular $S$-matrix of $SU(2)_k$, which encodes the mutual braiding statistics and implements modular transformations on the torus, takes the form
\begin{equation}\label{eq:S_matrix}
S_{j\ell}
=
\sqrt{\frac{2}{k+2}}\,
\sin\!\left(\frac{\pi(2j+1)(2\ell+1)}{k+2}\right),
\end{equation}
where $j,\ell\in\left\{0,\frac12,1,\dots,\frac{k}{2}\right\}$. This matrix is real, symmetric, and unitary, and it satisfies $S^2 = C$, where $C$ is the charge conjugation matrix $(C)_{j\ell} = \delta_{j,\bar\ell}$. Since every $SU(2)_k$ anyon is self-conjugate ($\bar{j} = j$), $C = \mathbf{1}$ and hence $S^2 = \mathbf{1}$.

The quantum dimension is recovered from the $S$-matrix via $d_j = S_{j0}/S_{00}$, consistent with Eq.~\eqref{eq:quantum_dim}. The mutual braiding phase between anyons $j$ and $\ell$ is
\begin{equation}
M_{j,\ell} = \frac{S_{j\ell}\, S_{00}}{S_{j0}\, S_{\ell 0}}.
\end{equation}
In particular, we have Eq. \eqref{eq: M matrix}:
\beq
M_{j, 0}=1,
\quad
M_{j, \frac{k}{2}}=(-1)^{2j}.
\eeq

\subsection{$R$-symbols}

Intuitively speaking, the $R$-symbols encode the phase accumulated when two anyons $j_1$ and $j_2$ undergo a counter-clockwise half-braid to fuse into the channel $j_3$. For the $SU(2)_k$ theory, the $R$-symbol is a pure $U(1)$ phase (since the fusion is multiplicity-free):
\begin{equation}\label{eq:R_symbol}
R_{j_3}^{j_1, j_2} = (-1)^{j_3 - j_1 - j_2} \exp\!\left[ \frac{i \pi \left(j_3(j_3+1) - j_1(j_1+1) - j_2(j_2+1)\right)}{k+2} \right].
\end{equation}
The first factor $(-1)^{j_3-j_1-j_2}$ can be recognized as the standard $SU(2)$ Clebsch--Gordan sign, while the exponential factor encodes the $q$-deformation.

The topological spin $\theta_j$, which represents the phase acquired
by a $2\pi$ self-rotation of a single anyon $j$, is related to the
$R$-symbol via $\theta_j = \varkappa_j\,(R_0^{j,j})^{-1}$, where
$\varkappa_j = (-1)^{2j}$ is the Frobenius--Schur indicator. Since
in $SU(2)_k$ every representation is self-conjugate, this gives
\begin{equation}\label{eq:topological_spin}
\theta_j = \exp\!\left[ \frac{i\, 2\pi\, j(j+1)}{k+2} \right].
\end{equation}
This quantity also determines the chiral central charge $c_-$ of the theory modulo 8 via $\exp(2\pi i\, c_-/8) = D^{-1}\sum_j d_j^2 \theta_j$.

\subsection{$F$-symbols}

The $F$-symbols represent the unitary basis transformation between different tree-level fusion orderings of three anyons, implementing the associativity of the fusion algebra at the level of state spaces. Diagrammatically, the $F$-symbol relates two different orderings of splitting an anyon $j_4$ into three anyons $j_1$, $j_2$, $j_3$, with intermediate fusion channels $j_5$ and $j_6$:
\begin{equation}\label{eq:F_symbol}
\left( F_{j_4}^{j_1, j_2, j_3} \right)_{j_5, j_6} = (-1)^{j_1 + j_2 + j_3 + j_4} \sqrt{[2j_5+1][2j_6+1]} \begin{Bmatrix} j_1 & j_2 & j_5 \\ j_3 & j_4 & j_6 \end{Bmatrix}_q.
\end{equation}
Here, the curly brackets denote the quantum $6j$-symbol of the quantum group $U_q(\mathfrak{su}(2))$.

To evaluate the quantum $6j$-symbol, we first define the triangle coefficient $\Delta(a,b,c)$, which encodes the admissibility of the fusion vertex $a \times b \rightarrow c$:
\begin{equation}\label{eq:triangle}
\Delta(a,b,c) = \sqrt{\frac{[a+b-c]! \, [a-b+c]! \, [-a+b+c]!}{[a+b+c+1]!}}.
\end{equation}
The triangle inequality $|a-b|\leqslant c \leqslant \min(a+b, k-a-b)$ and the condition $a+b+c\in\z$ must be satisfied for $\Delta(a,b,c)$ to be nonzero, which is equivalent to the fusion coefficient $N_{ab}^c \ne 0$.

The quantum $6j$-symbol is then expressed as the Racah--Wigner sum:
\begin{equation}\label{eq:6j_symbol}
\begin{Bmatrix} j_1 & j_2 & j_5 \\ j_3 & j_4 & j_6 \end{Bmatrix}_q = \Delta(j_1, j_2, j_5)\, \Delta(j_1, j_4, j_6)\, \Delta(j_3, j_2, j_6)\, \Delta(j_3, j_4, j_5) \sum_z \frac{(-1)^z [z+1]!}{D(z)},
\end{equation}
where the denominator $D(z)$ is the product of quantum factorials:
\begin{align}\label{eq:D_of_z}
D(z) =\, &[z - j_1 - j_2 - j_5]! \, [z - j_1 - j_4 - j_6]! \, [z - j_3 - j_2 - j_6]! \, [z - j_3 - j_4 - j_5]! \nonumber \\
&\times [j_1 + j_2 + j_3 + j_4 - z]! \, [j_1 + j_3 + j_5 + j_6 - z]! \, [j_2 + j_4 + j_5 + j_6 - z]!.
\end{align}
The summation index $z$ runs over all integers such that the arguments of all quantum factorials in $D(z)$ are non-negative. In practice, this means
\begin{align}
z_{\min} &= \max(j_1{+}j_2{+}j_5,\, j_1{+}j_4{+}j_6,\, j_3{+}j_2{+}j_6,\, j_3{+}j_4{+}j_5), \nonumber \\
z_{\max} &= \min(j_1{+}j_2{+}j_3{+}j_4,\, j_1{+}j_3{+}j_5{+}j_6,\, j_2{+}j_4{+}j_5{+}j_6),
\end{align}
and $z$ ranges over $\{z_{\min}, z_{\min}+1, \ldots, z_{\max}\}$.

\subsection{Topological symmetry} \label{subapp: topological symmetry}

The topological symmetry is the emergent symmetry of a topological order, where anyons can be permuted. We will see that the $SU(2)_k$ topological order has a trivial topological symmetry unless $k=4n+2$ with an integer $n\geqslant 1$. When $k=4n+2$ with an integer $n\geqslant 1$, the topological symmetries form a $\z_2$ group. The action of this $\z_2$ group on the anyons is given by Eq. \eqref{eq: topological permutation}, the corresponding $U$-symbols are given by Eq. \eqref{eq: topological U symbol}, and a set of $\eta$-symbols can be taken to be Eq. \eqref{eq: topological eta symbol}.

Since the $SU(2)_k$ topological order is chiral, its topological symmetry must be unitary and preserve the quantum dimension and topological spin of the anyons. According to Eqs. \eqref{eq:quantum_dim} and \eqref{eq:topological_spin}, in order for two anyons $j_1$ and $j_2$ to have the same quantum dimension and topological spin, we need either $j_1=j_2$, or $j_2=\frac{k}{2}-j_1$ if $\frac{k}{4}-j\in\z$. Only the second case involves some nontrivial anyon permutation, where $j\rightarrow\frac{k}{2}-j$. The requirement $\frac{k}{4}-j\in\z$ implies that $k$ must be even in this case. Suppose $k=4n$ with $n\in\z$, then the anyon $j$ that will be permuted satisfies $j\in\z$. In particular, the trivial anyon with $j=0$ will become $j=\frac{k}{2}$, which is a nontrivial anyon and shows that this map is invalid. On the other hand, if $k=4n+2$ with $n\in\z$, then the anyon $j$ that will be permuted satisfies $j\in\z+\frac{1}{2}$, and there is no analogous contradiction.

To determine the topological symmetry, it is enough to check how the anyon $j=1/2$ can be permuted, as the permutations of all other anyons can be read off from it using the fusion rule Eq. \eqref{eq: fusion rule app}. In fact, when $k=4n+2$ with an integer $n\geqslant 1$ (if $n=0$ the following map does not change the anyon type), there is indeed a nontrivial $\z_2$ topological symmetry that can be characterized precisely. Denote the generator of this $\z_2$ by $g$, and denote the anyon obtained from acting $g$ on $j$ by $\rho_g(j)$. Then
\beq \label{eq: topological permutation}
\rho_g(j)=
\begin{cases}
    j, & {\text{if }} j\in\z,\\
    \frac{k}{2}-j, & {\text{if }} j\in\z+\frac{1}{2}.
\end{cases}
\eeq
When $j_3$ is a fusion outcome of $j_1$ and $j_2$, the $U$-symbol for $g$ is
\beq \label{eq: topological U symbol}
U_g(j_1, j_2; j_3)=(-1)^{x(j_1, j_2, j_3)}
\eeq
with
\beq
x(j_1, j_2, j_3)=\floor{j_1}+\floor{j_2}+\floor{j_3}
	+
	p(j_1)
	\left(
	\floor{j_1}+\floor{j_2}+\floor{j_3}
	\right)
	+
	p(j_2)\floor{j_1},
\eeq
where $\floor{j}=j$ if $j\in\z$ and $\floor{j}=j-\frac{1}{2}$ if $j\in\z+\frac{1}{2}$, and $p(j)=0$ if $j\in\z$ and $p(j)=1$ if $j\in\z+\frac{1}{2}$. A valid set of $\eta$-symbols can be taken to be trivial, \ie
\beq \label{eq: topological eta symbol}
\eta_j(g_1, g_2)=1
\eeq
for all $j$ and all $g_1, g_2\in\z_2$.  One can check that these $U$- and $\eta$-symbols satisfy the correct consistency equations (see, for example, Eqs. (19), (23) and (24) of Ref. \cite{Ye2023SciPost} for a summary of these equations).

\subsection{Key properties}

Several properties of this algebraic data are worth emphasizing, as they simplify the anomaly-indicator computations.

\begin{enumerate}
\item \textbf{Reality.} Because all factors in the $F$-symbol expression Eq. \eqref{eq:F_symbol} involve only real quantum integers and real square roots, the $F$-symbols of $SU(2)_k$ are strictly real-valued.

\item \textbf{Multiplicity-freeness.} Since $N_{ab}^c \in \{0,1\}$ for all anyons in $SU(2)_k$, the $F$- and $R$-symbols are scalars rather than matrices, and the summations in the anomaly indicator run only over anyon types, with no additional multiplicity indices.

\item \textbf{Self-conjugacy.} Every anyon $j$ in $SU(2)_k$ is its own antiparticle, $\bar{j}=j$, since all representations of $SU(2)$ are (pseudo-)real. This implies that the Frobenius--Schur indicator is $\varkappa_j = (-1)^{2j}$.

\item \textbf{Topological symmetry.} Only for $k=4n+2$ with an integer $n\geqslant 1$ will $SU(2)_k$ have a nontrivial $\z_2$ topological symmetry. For all other $k$, there is no topological symmetry that permutes the anyons.

\end{enumerate}

This algebraic data provides the rigorous foundation for evaluating the LSM anomaly indicators, and can be used as input for any systematic classification of symmetry-enriched $SU(2)_k$ topological orders following the framework of Ref.~\cite{Ye2023}.

\section{Details of the calculations of the anomaly} \label{app: calculating anomalies}

In this appendix, we derive Eq. \eqref{eq: calculated anomaly indicators}. For a given symmetry-enriched $SU(2)_k$ topological phase, because all relevant data needed for calculating the anomaly indicators in Eqs. \eqref{eq:I1}, \eqref{eq:I2} and \eqref{eq:I3} are given in Sec. \ref{sec:SET} and Appendix \ref{app:SU2k_data}, one can calculate these anomaly indicators directly. We had initially verified that Eq. \eqref{eq: calculated anomaly indicators} holds for all $k\leqslant 60$ using Mathematica and Python, for the case with no anyon permutation. Then ChatGPT helped us get an analytic proof of Eq. \eqref{eq: calculated anomaly indicators} for all $k$ and all anyon permutation patterns. In the following, we present this proof.

The strategy is to separate the calculation of the anomaly into two parts. The first part is to get a general result for the relative anomaly between two different symmetry-enriched $SU(2)_k$ topological phases with the same anyon permutation patterns, \ie the difference between the anomalies of these two SET phases. It turns out that this relative anomaly takes a particularly simple form, given by Eq. \eqref{eq: relative anomaly}. The second part is to directly calculate the anomaly indicators for a simple reference symmetry-enriched $SU(2)_k$ topological phase, for each anyon permutation pattern. These reference anomalies are given by Eq. \eqref{eq: reference anomaly app}. Combining the results of these two parts, we get Eq. \eqref{eq: calculated anomaly indicators}.

As we will see, the reason why this strategy simplifies the calculation of the anomaly indicators is because 1) although Eq. \eqref{eq:indicator_with_z2z2 main} is rather complicated for a general symmetry-enriched $SU(2)_k$ topological phase, the relative anomaly is much simpler and only takes as input the data related to the Abelian anyons, and 2) there are some simple reference SET phases whose anomalies are easy to calculate.

For the readers' convenience, we copy the definitions of the anomaly indicators Eqs. \eqref{eq:I1}, \eqref{eq:I2} and \eqref{eq:I3} here:
\beq \label{eq: 3 anomaly indicators app}
\begin{split}
\mathsf I_1 &= \mathcal I_3(C_2U_\pi,\; C_2U_\pi'),\\
\mathsf I_2 &= \mathcal I_3(T_1T_2C_2U_\pi,\; T_1T_2C_2U_\pi'),\\
\mathsf I_3 &= \mathcal I_3(T_1C_2U_\pi,\; T_1C_2U_\pi'),
\end{split}
\eeq
where $\mc{I}_3(g_1, g_2)$ is defined for a $\z_2\times\z_2$ symmetry generated by $g_1$ and $g_2$:
\beq \label{eq: master anomaly indicator app}
\begin{split}
\mathcal{I}_3\left(g_1, g_2\right) 
=& \frac{1}{D^2}\sum_{\substack{a,b,x,u,\\ 
^{g_1}a = a\\
a \times b \times ^{g_1}b \rightarrow ^{g_2}a}}
{d_b}\frac{\theta_x}{\theta_a}
R_u^{b, ^{g_1}b} F_{{^{g_2}}a}^{a,b,{^{g_1}}b}
F_{{^{g_2}}a}^{a,{^{g_1}}b,b}
U_{g_1}^{-1}(a,b;x)U_{g_1}^{-1}(x,^{g_1}b;^{g_2}a)
\frac{1}{\eta_b(g_1, g_1)}\frac{\eta_a(g_2, g_1)}{\eta_a(g_1, g_2)}.
\end{split}
\eeq

\subsection{Relative anomaly}

In order to calculate Eq. \eqref{eq: 3 anomaly indicators app}, we need to have better understanding of the structure of Eq. \eqref{eq: master anomaly indicator app}. For all three anomaly indicators in Eq. \eqref{eq: 3 anomaly indicators app}, the arguments of $\mc{I}_3$ take the form of $g_1=sU_\pi$ and $g_2=sU'_\pi$, where $s$ can be either $C_2$, $T_1T_2C_2$ or $T_1C_2$, which are all order-2 rotations. This observation motivates us to first calculate the relative anomaly between two $\z_2\times SO(3)$ symmetric $SU(2)_k$ topological orders.

Concretely, consider $\z_2\times SO(3)$ symmetric $SU(2)_k$ topological orders with a given anyon permutation pattern. The classification of such SET phases is given by the classification of the $\eta$-symbols, \ie $H^2(\z_2\times SO(3), \mc{A})=H^2(\z_2, \z_2)\oplus H^2(SO(3), \z_2)=\z_2\oplus\z_2$, which can be labeled by $r,\sigma\in\{0, 1\}$, where $r$ and $\sigma$ can be viewed as the generators of $H^2(\z_2, \z_2)$ and $H^2(SO(3), \z_2)$, respectively. Since this $\z_2\times SO(3)$ is a subgroup of the original $p4\times SO(3)$, there is an inclusion map $i: \z_2\times SO(3)\rightarrow p4\times SO(3)$. The $r$ and $\sigma$ can be related to $n_{1,2,3,4}$ in Eq. \eqref{eq:w_parameterization} by the pullback from $H^2(p4\times SO(3), \mc{A})$ to $H^2(\z_2\times SO(3), \mc{A})$ induced by $i$. Specifically, we have
\beq \label{eq: pullback}
r_1=n_2,
\quad
r_2=n_1+n_2,
\quad
r_3=n_2+n_3,
\quad
\sigma=n_4,
\eeq
where $r_{1,2,3}$ are the values of $r$ when the $\z_2$ factor of the $\z_2\times SO(3)$ subgroup is generated by $C_2$, $T_1T_2C_2$ and $T_1C_2$, respectively. The explicit values of $\eta$-symbols for each $\z_2\times SO(3)$ symmetric $SU(2)_k$ topological order can be read off from the $\eta$-symbols of the original $p4\times SO(3)$ symmetric $SU(2)_k$ topological order using this pullback relation Eq. \eqref{eq: pullback}, \ie substituting Eq. \eqref{eq: pullback} into Eq. \eqref{eq:eta_general} results in the $\eta$-symbols of the $\z_2\times SO(3)$ symmetric $SU(2)_k$ topological order.

The relative anomaly here refers to $\frac{\mc{I}_3^{r\sigma}(sU_\pi, sU'_\pi)}{\mc{I}_3^{00}(sU_\pi, sU'_\pi)}$, which measures the difference between the anomaly indicators $\mc{I}_3$ for two $\z_2\times SO(3)$ symmetric $SU(2)_k$ topological orders, where one of them has the $\eta$-symbol being described by $r$ and $\sigma$, and the other has a trivial set of $\eta$-symbols. Here $s$ stands for the generator of $\z_2$ and $U_\pi$ and $U'_\pi$ are two $\pi$ rotations around two orthogonal axes. Note that this relative anomaly should be defined for each anyon permutation pattern. The remarkably simple result is that
\beq \label{eq: relative anomaly}
\frac{\mc{I}_3^{r\sigma}(sU_\pi, sU'_\pi)}{\mc{I}_3^{00}(sU_\pi, sU'_\pi)}=(-1)^{kr\sigma}
\eeq
for all anyon permutation patterns.

The rest of this sub-appendix is devoted to deriving Eq. \eqref{eq: relative anomaly}.

We start from the general result of relative anomaly \cite{Barkeshli2019a}. Suppose that the anomaly is described by an element in $H^4(\z_2\times SO(3), U(1))$. Then Eq. (43) of Ref. \cite{Barkeshli2019a} gives the relative anomaly between two SET phases:
\beq \label{eq: relative anomaly original}
\begin{split}
&\mathcal O_t(g,h,k,l)\\
={}&
R^{{}^{gh}t(k,l),\,t(g,h)}
\,
\eta_{{}^{gh}t(k,l)}^{(1)}(g,h)
\left[
U_g({}^{g}t(hk,l),{}^{g}t(h,k))
\right]^*
U_g({}^{g}t(h,kl),{}^{gh}t(k,l))\\
&\cdot F^{t(ghk,l),\,t(gh,k),\,t(g,h)}
\left[
F^{t(ghk,l),\,t(g,hk),\,{}^gt(h,k)}
\right]^*
F^{t(g,hkl),\,{}^gt(hk,l),\,{}^gt(h,k)}
\left[
F^{t(g,hkl),\,{}^gt(h,kl),\,{}^{gh}t(k,l)}
\right]^*
\\
&\cdot
F^{t(gh,kl),\,t(g,h),\,{}^{gh}t(k,l)}\left[
F^{t(gh,kl),\,{}^{gh}t(k,l),\,t(g,h)}
\right]^*,
\end{split}
\eeq
where $g, h, k, l$ are elements in the symmetry group, and $t(g, h)$ is an Abelian anyon such that the $\eta$-symbols of these two SET phases, $\eta^{(1)}_a(g, h)$ and $\eta^{(2)}_a(g,h)$, are related by
\beq \label{eq: eta relation app}
\eta^{(2)}_{a}(g, h)=\eta^{(1)}_{a}(g, h)M_{a, t(g, h)},
\eeq
with $M_{a, t(g, h)}$ the mutual braiding statistics between the anyon $a$ and the Abelian anyon $t(g, h)$. This $O_t(g, h, k, l)$ is the ratio of representative cocycles of two elements in $H^4(\z_2\times SO(3), U(1))$ that characterize the anomalies of the two SET phases. Note that Eq. \eqref{eq: relative anomaly original} only depends on the Abelian anyons of the system.

Great simplifications occur in our context, because the Abelian anyons in our case are always not permuted by the symmetry, $\eta^{(1)}_a(g, h)=1$ for the SET with a trivial set of $\eta$-symbols, and the $F$-symbols are real, and the $U$-symbols are 1 when they are restricted to Abelian anyons. Then Eq. \eqref{eq: relative anomaly original} becomes
\beq \label{eq: relative anomaly simplified}
\begin{split}
&\mathcal O_t(g,h,k,l)\\
=&
R^{t(k,l),\,t(g,h)}
F^{t(ghk,l),\,t(gh,k),\,t(g,h)}
F^{t(ghk,l),\,t(g,hk),\,t(h,k)}
F^{t(g,hkl),\,t(hk,l),t(h,k)}
F^{t(g,hkl),\,t(h,kl),\,t(k,l)},
\end{split}
\eeq
which is a combination of $F$- and $R$-symbols that only involve the Abelian anyons. For such $F$- and $R$-symbols, one useful result from Eqs. \eqref{eq:F_symbol} and \eqref{eq:R_symbol} is
\beq
(F^{\frac{k}{2}\frac{k}{2}\frac{k}{2}}_{\frac{k}{2}})_{00}=(R^{\frac{k}{2}\frac{k}{2}}_0)^2=(-1)^{k}.
\eeq

According to the discussion around Eq. (54) of Ref. \cite{Ye2023SciPost}, the anomaly indicator $\mc{I}_3$ can be detected by
\beq
\mc{I}_3(g_1, g_2)=\frac{
\mathcal O(g_1,g_2,g_1,g_1)
\mathcal O(g_1,g_1,g_1,g_2)
}{
\mathcal O(g_2,g_1,g_1,g_1)
\mathcal O(g_1,g_1,g_2,g_1)
},
\eeq
where $\mc{O}(g, h, k, l)$ is a representative cocycle of the element in $H^4(\z_2\times\z_2, U(1))$ that characterizes a $\z_2\times\z_2$ symmetric topological phase, with $g_{1,2}$ the generators of $\z_2\times\z_2$. Therefore, considering the $\z_2\times\z_2$ group generated by $sU_\pi$ and $sU'_\pi$, we obtain
\beq
\frac{\mc{I}_3^{r\sigma}(sU_\pi, sU'_\pi)}{\mc{I}_3^{00}(sU_\pi, sU'_\pi)}=\frac{
\mathcal O_t(sU_\pi,sU_\pi',sU_\pi,sU_\pi)
\mathcal O_t(sU_\pi,sU_\pi,sU_\pi,sU_\pi')
}{
\mathcal O_t(sU_\pi',sU_\pi,sU_\pi,sU_\pi)
\mathcal O_t(sU_\pi,sU_\pi,sU_\pi',sU_\pi)
},
\eeq
where $t(g, h)$ is determined from Eq. \eqref{eq: eta relation app}, with $\eta^{(1)}_j(g, h)=1$ and $\eta^{(2)}_j(g, h)=M_{j,t(g, h)}\eta^{(1)}_{j}(g, h)$ obtained from Eq. \eqref{eq:eta_general} using the pullback Eq. \eqref{eq: pullback}.

Finally, by combining all above results, direct calculations lead to Eq. \eqref{eq: relative anomaly}.

\subsection{Anomalies of reference SET phases}

Having the relative anomaly in Eq. \eqref{eq: relative anomaly}, to derive Eq. \eqref{eq: calculated anomaly indicators}, we just need to calculate $\mc{I}_3^{00}(sU_\pi, sU'_\pi)$ for each anyon permutation pattern. Because this is the anomaly indicator for the SET phases with a trivial set of $\eta$-symbols, we can actually get the result without directly calculating Eq. \eqref{eq: master anomaly indicator app}. The final result is
\beq \label{eq: reference anomaly app}
\mc{I}_3^{00}(sU_\pi, sU'_\pi)=1
\eeq
for all $k$, for all anyon permutation patterns, and all three cases, with $s=C_2$, $s=T_1T_2C_2$ and $s=T_1C_2$, respectively. Combining Eqs. \eqref{eq: reference anomaly app} and \eqref{eq: relative anomaly} leads to Eq. \eqref{eq: calculated anomaly indicators}.

The rest of this sub-appendix is devoted to deriving Eq. \eqref{eq: reference anomaly app}.

We start with the case where no anyon is permuted by the symmetry. It is generally true that a bosonic SET phase, which does not have to be the $SU(2)_k$ topological order, with a trivial anyon permutation pattern and trivial $U$- and $\eta$-symbols is anomaly free, so Eq. \eqref{eq: reference anomaly app} holds. This is because, in this case, a canonical \(G\)-crossed braided extension is given by $\mathcal C_G^\times=\mathcal C\boxtimes \operatorname{Vec}_G$,
where the component of \(G\)-degree \(g\) is
\(\mathcal C\boxtimes\operatorname{Vec}_g\). This is the trivial
semidirect-product extension associated with the trivial action of
\(G\) on \(\mathcal C\). Its existence implies that the corresponding
obstruction class in \(H^4(G,U(1))\) vanishes \cite{Etingof2010, Barkeshli2019}.

Next, we turn to the case where the anyons can be permuted. There are three nontrivial anyon permutation patterns, determined by the 4 distinct group homomorphisms:
\beq
\varphi: G=p4\times SO(3)\rightarrow G_{\rm top}=\z_2.
\eeq
We will discuss these three cases in turn.

\begin{enumerate}

\item $\varphi(C_4)=\tilde g$ and $\varphi(T_{1,2})=\tilde{\mathbf{1}}$.

In this case, because $\varphi(C_2)=\varphi(T_1T_2C_2)=\varphi(T_1C_2)=\tilde{\mathbf{1}}$, the symmetry does not permute anyons after restricting to the $\z_2\times SO(3)$ subgroup. So we return to the case with no anyon permutation, and we have shown that Eq. \eqref{eq: reference anomaly app} holds.

\item $\varphi(T_{1,2})=\tilde g$ and $\varphi(C_4)=\tilde{\mathbf{1}}$.

In this case, $\varphi(C_2)=\varphi(T_1T_2C_2)=\tilde{\mathbf{1}}$, so the calculation of $\mc{I}_3(C_2U_\pi, C_2U'_\pi)$ and $\mc{I}_3(T_1T_2C_2U_\pi, T_1T_2C_2U'_\pi)$ reduce to the case with no anyon permutation, which means that Eq. \eqref{eq: reference anomaly app} holds if $s=C_2$ and $s=T_1T_2C_2$.

If $s=T_1C_2$, $\varphi(s)=\tilde g$ and will permute the anyons. Note that the $\z_2\times\z_2$ group generated by $sU_\pi$ and $sU'_\pi$ in this case has a projection map into $G_{\rm top}=\z_2$, and this projection map is just defined by $\varphi$. So the anomaly in this case must be the projection-induced pullback of an element in $H^4(G_{\rm top}, U(1))$. Since $H^4(G_{\rm top}, U(1))=\z_1$, its pullback, which is the anomaly, must vanish. So Eq. \eqref{eq: reference anomaly app} also holds in this case.

\item $\varphi(C_4)=\varphi(T_{1,2})=\tilde g$.

The argument for this case is identical to the argument for the previous case, which shows that Eq. \eqref{eq: reference anomaly app} holds for all three cases, with $s=C_2$, $s=T_1T_2C_2$ and $s=T_1C_2$, respectively.

\end{enumerate}

In summary, Eq. \eqref{eq: reference anomaly app} holds for all $k$, all anyon permutation patterns, and all three cases, with $s=C_2$, $s=T_1T_2C_2$ and $s=T_1C_2$, respectively.

\section{Details of the analysis of the critical Chern-Simons-boson theory}
\label{app:boson_details}
 
This appendix supplies the technical ingredients behind
Sec.~\ref{sec:critical_bosons}: the
classification of $USp(2N_f)$ symmetric quartic interactions (Appendix~\ref{app:quartic_classification}), the framework of the $1/N_f$ expansion (Appendix~\ref{subapp: framework boson}), and the details of the calculation of the scaling dimensions of $\sigma$ (Appendix \ref{subapp: sigma}) and $O_m\equiv\Phi^\dag_{\alpha 1}\Phi_{\alpha 2}$ (Appendix \ref{subapp: calculations of Om}).

%---------------------------------------------------------------------
\subsection{Classification of symmetric local quartic interactions}
\label{app:quartic_classification}
%---------------------------------------------------------------------

We first show that $(\sum_{\alpha, i}\Phi_{\alpha, i}^\dag \Phi_{\alpha, i})^2$ is the only gauge invariant and $USp(2N_f)$ symmetric quartic interaction of the bosons.
 
To do so, we enumerate the quartic operators built from the doubled field
$X_{\alpha,I}$, with $\alpha=1,2$ an $SU(2)$ color index and
$I=1,\dots,2N_f$ a $USp(2N_f)$ flavor index. The generic candidate of such a quartic operator
can be expanded by $(\mathcal G_{\alpha\beta\gamma\delta}\mathcal F_{IJKL})
X_{\alpha I}X_{\beta J}X_{\gamma K}X_{\delta L}$, where $G_{\alpha\beta\gamma\delta}$ and $F_{IJKL}$ are two rank-4 tensors. Below we will determine for which tensors $G_{\alpha\beta\gamma\delta}$ and $F_{IJKL}$ can this quartic operator be gauge invariant and $USp(2N_f)$ invariant.
 
\paragraph{Gauge structures.}
We use the standard notation of angular momentum to label the irreducible representation of the $SU(2)$ gauge group. For example, the fundamental representation is denoted by $\mathbf{\frac{1}{2}}$, and a quartic operator is an outcome of fusing four $\mathbf{\frac{1}{2}}$ representations. The decomposition
$\mathbf{\tfrac12}\otimes\mathbf{\tfrac12}\otimes\mathbf{\tfrac12}\otimes\mathbf{\tfrac12} =
2\!\cdot\!\mathbf{0}\oplus 3\!\cdot\!\mathbf{1}\oplus\mathbf{2}$
produces two independent singlet pairings, \ie a quartic operator $(\mathcal G_{\alpha\beta\gamma\delta}\mathcal F_{IJKL})
X_{\alpha I}X_{\beta J}X_{\gamma K}X_{\delta L}$ is gauge invariant if $\mathcal G_{\alpha\beta\gamma\delta}$ is a linear superposition of the following two tensors:
\begin{align}
\mathcal{G}^{(1)}_{\alpha\beta\gamma\delta} &=
\varepsilon_{\alpha\beta}\varepsilon_{\gamma\delta},\\
\mathcal{G}^{(2)}_{\alpha\beta\gamma\delta} &=
\varepsilon_{\alpha\gamma}\varepsilon_{\beta\delta}.
\end{align}
We note that the case with $\mathcal G_{\alpha\beta\gamma\delta}=\varepsilon_{\alpha\delta}\varepsilon_{\beta\gamma}$ is already included, since by the Schouten identity we have
$\varepsilon_{\alpha\beta}\varepsilon_{\gamma\delta}
+\varepsilon_{\alpha\gamma}\varepsilon_{\delta\beta}
+\varepsilon_{\alpha\delta}\varepsilon_{\beta\gamma}=0$.
 
\paragraph{Flavor structures.}
We view the quartic operator as an outcome of fusing two rank-2 operators. The rank-2 tensor $X_{\alpha I}X_{\beta J}$ decomposes under
$USp(2N_f)$ as symplectic singlet representation $\oplus$ anti-symmetric
traceless representation $\oplus$ symmetric representation. There are three types of $USp(2N_f)$ invariant rank-4 tensors, and they can be obtained by contracting a pair of rank-2 singlets, a pair of rank-2 anti-symmetric traceless tensors, and a pair of rank-2 symmetric tensors, respectively \cite{Yamatsu2015}. We denote the orthonormal bases of anti-symmetric traceless tensors and symmetric tensors by
$\{A^a\},\{S^a\}$, respectively (here a set of rank-2 tensors $\{T_i\}$ are said to be orthonormal if ${\rm Tr}(T_i^\dag T_j)=2N_f\delta_{ij}$, and an anti-symmetric traceless tensor means an anti-symmetric tensor that is orthogonal to $\Omega$). Then the three types of $USp(2N_f)$ singlet rank-4 tensors are:
\begin{align}
\mathcal F^{(1)}_{IJKL} &= \Omega_{IJ}\Omega_{KL}
  &&\text{(from\ two\ singlets)},\\
\mathcal F^{(2)}_{IJKL} &= \sum_a A^a_{IJ}A^a_{MN}\Omega_{MK}\Omega_{NL}
  &&\text{(from\ two\ anti-symmetric\ traceless\ tensors)},\\
\mathcal F^{(3)}_{IJKL} &= \sum_a S^a_{IJ}S^a_{MN}\Omega_{MK}\Omega_{NL}
  &&\text{(from\ two\ symmetric\ tensors)}.
\end{align}
Using the completeness relations
\beq \label{eq:completeness}
\bsp
\tfrac{1}{2N_f}\Omega_{IJ}\Omega_{KL}
+\tfrac{1}{2N_f}\!\sum_a\! A^a_{IJ}A^a_{KL}
&=\tfrac12\bigl(\delta_{IK}\delta_{JL}-\delta_{IL}\delta_{JK}\bigr),\\
\tfrac{1}{2N_f}\!\sum_a\! S^a_{IJ}S^a_{KL}
&=\tfrac12\bigl(\delta_{IK}\delta_{JL}+\delta_{IL}\delta_{JK}\bigr),
\esp
\eeq
it is straightforward to check that $\mathcal F^{(1,2,3)}_{IJKL}$ can also be spanned by $\Omega_{IJ}\Omega_{KL}$, $\Omega_{IK}\Omega_{JL}$ and $\Omega_{IL}\Omega_{JK}$.
 
\paragraph{Reduction to one invariant.}
The above analysis shows that there are six candidate gauge invariant and $USp(2N_f)$ invariant quartic interactions of the bosons, \ie $\mc{E}^{(i, j)}=\mc{G}^{(i)}_{\alpha\beta\gamma\delta}\mc{F}^{(j)}_{IJKL} X_{\alpha I}X_{\beta J}X_{\gamma K}X_{\delta L}$, with $i=1,2$ and $j=1,2,3$. However, the bosonic statistics, the Schouten identity $\varepsilon_{\alpha\beta}\varepsilon_{\gamma\delta}
+\varepsilon_{\alpha\gamma}\varepsilon_{\delta\beta}
+\varepsilon_{\alpha\delta}\varepsilon_{\beta\gamma}=0$ and the completeness relations in Eq. \eqref{eq:completeness} imply that all these 6 types are proportional to $\mc{E}^{(1, 1)}\propto(\sum_{\alpha, i}\Phi_{\alpha, i}^\dag \Phi_{\alpha, i})^2$, which shows that this operator is the only gauge invariant and $USp(2N_f)$ invariant quartic interaction of the bosons.

The above statement can be straightforwardly verified by verifying the following relations.

\begin{enumerate}

    \item $\mc{E}^{(2, j)}=\frac{1}{2}\mc{E}^{(1, j)}$ for $j=1,2$. To verify this relation, we can use the Schouten identity $\varepsilon_{\alpha\gamma}\varepsilon_{\beta\delta}
    =\varepsilon_{\alpha\beta}\varepsilon_{\gamma\delta}
    +\varepsilon_{\alpha\delta}\varepsilon_{\beta\gamma}$ and the first completeness relation in Eq. \eqref{eq:completeness}.

    \item $\mc{E}^{(1,2)}=(N_f-1)\mc{E}^{(1,1)}$. To verify this relation, we can use the first completeness relation in Eq. \eqref{eq:completeness}.

    \item $\mc{E}^{(1, 3)}=0$. This relation follows from the fact that a symmetric tensor is contracted with an antisymmetric tensor.

    \item $\mc{E}^{(2,3)}=\frac{3N_f}{2}\mc{E}^{(1, 1)}$. This relation can be verified using the completeness relations in Eq. \eqref{eq:completeness}.

\end{enumerate}

In summary, we have shown that $(\sum_{\alpha, i}\Phi_{\alpha, i}^\dag \Phi_{\alpha, i})^2$ is the only gauge invariant and $USp(2N_f)$ invariant quartic interaction of the bosons.

\subsection{Framework of $1/N_f$ expansion} \label{subapp: framework boson}

Next, we present the framework for the $1/N_f$ expansion. The main result of this sub-appendix is that we can use the effective action in Eq. \eqref{eq: 1/N_f effective action} to calculate the two-point correlation functions and scaling dimensions of the mass operators. The details of these calculations will be presented in Appendices \ref{subapp: sigma} and \ref{subapp: calculations of Om}.

We start with the action $S=S_{\rm CS}+S_{\rm GF}+S_{\rm gh}+S_{\rm matter}$, with each of these terms given by Eqs. \eqref{eq: CS action bosonic matter}, \eqref{eq: gauge fixing action bosonic matter}, \eqref{eq: ghost action bosonic matter} and \eqref{eq: matter action with HS field}, respectively. For the readers' convenience, these terms are copied below:
\beq
\bsp
S_{\mathrm{CS}}
&= \frac{k}{4\pi}\!\int\!d^3x\left(
  -\frac{i}{2}\varepsilon_{\mu\nu\lambda}A_\mu^{a}\partial_\nu A_\lambda^{a}
  -\frac{i}{6}\varepsilon_{\mu\nu\lambda}f^{abc}
  A_\mu^a A_\nu^b A_\lambda^c\right),\\
S_{\mathrm{GF}}
&= -\frac{N_f}{32\xi}\int d^3x d^3y\frac{\partial_\mu A^a_\mu(x)\partial_\nu A^a_\nu(y)}{2\pi^2|x-y|^2},\\
S_{\mathrm{gh}}
&= \int\!d^3x\left(
  \partial_\mu\bar c^{a}\partial^\mu c^{a}
  + f^{abc}\partial_\mu\bar c^{a} A_\mu^{b}c^{c}\right),\\
S_{\rm matter}
&=
\int\!d^3x\Bigg[
(D_\mu\Phi_{\alpha, i})^\dagger(D^\mu\Phi_{\alpha, i})
  -\sigma\left(\Phi_{\alpha,i}^\dagger\Phi_{\alpha, i}\right)
  \Bigg].
\esp
\eeq
In the above, we have already set the mass term of $\sigma$ to zero in order to access the critical fixed point.

We are interested in the scaling dimensions of $\sigma$ and $O_m\equiv \Phi^\dag_{\alpha, 1}\Phi_{\alpha, 2}$ up to the order of $1/N_f$, which, using Eq. \eqref{eq: scaling dimension from correlator}, can be extracted from the two-point correlation functions of these operators up to the order of $1/N_f$. The above action already contains $\sigma$, so its correlation function can be obtained directly. To obtain the correlation function of $O_m$, we add the following source term to the action:
\beq
S_J=\int d^3x\left(J(x)^* O_m(x)+J(x)O_m(x)^\dag\right),
\eeq
where $J(x)$ is a static probe field, so that the total action is
\beq
S_{\rm total}=S+S_J.
\eeq
Then if the generating functional is $Z[J]\equiv\int D\Phi DA D\sigma e^{-S_{\rm total}}$, the correlation function of $O_m$ is
\beq
\langle O_{m}(x) O^\dag_{m}(0)\rangle = \frac{1}{Z[J]}\left.\frac{\delta^2Z[J]}{\delta J^*(x) \delta J(0)}\right|_{J(x)=0}.
\eeq

To calculate these correlation functions to the order of $1/N_f$, we integrate $\Phi$ out in the path integral and obtain an effective action that naturally has a structure for the $1/N_f$ expansion. Concretely, for later convenience, we introduce the rescaled fields
\beq \label{eq: rescaled fields}
\tilde A_\mu^a(x)=\sqrt{N_f}A^a_\mu(x),
\quad
\tilde\sigma(x)=\sqrt{N_f}\sigma(x).
\eeq
Then the effective action obtained by integrating out $\Phi$ takes the following form:
\beq
S_{\rm eff}=S_{\rm CS}+S_{\rm GF}+S_{\rm gh}+S',
\eeq
where
\beq
\bsp
S_{\rm CS}=&\frac{k}{4\pi N_f}\!\int\!d^3x\left(
-\frac{i}{2}\varepsilon_{\mu\nu\lambda}\tilde A_\mu^{a}\partial_\nu \tilde A_\lambda^{a}
-\frac{i}{6\sqrt{N_f}}\varepsilon_{\mu\nu\lambda}f^{abc}
\tilde A_\mu^a\tilde A_\nu^b\tilde A_\lambda^c\right),\\
S_{\rm GF}=&-\frac{1}{32\xi}\int d^3x d^3y\frac{\partial_\mu \tilde A^a_\mu(x)\partial_\nu \tilde A^a_\nu(y)}{2\pi^2|x-y|^2},\\
S_{\rm gh}=&\int\!d^3x\left(
\partial_\mu\bar c^{a}\partial^\mu c^{a}
+ \frac{f^{abc}}{\sqrt{N_f}}\partial_\mu\bar c^{a}\tilde A_\mu^{b}c^{c}\right),\\
S'=&{\rm Tr}[\log(K)],
\esp
\eeq
In the above, the kernel $K$ is
\beq
K=K_k+K_d+K_J,
\eeq
with
\beq
\bsp
K_k=&-\delta_{ij}\delta_{\alpha\beta}\partial^2,\\
K_d=&\delta_{ij}
\Big[
\frac{i}{\sqrt{N_f}}(\partial_\mu\tilde A^a_\mu)T^a_{\alpha\beta}
+\frac{2i}{\sqrt{N_f}}\tilde A^a_\mu T^a_{\alpha\beta}\partial_\mu
+\frac{1}{N_f}\tilde A^a_\mu \tilde A^b_\mu(T^aT^b)_{\alpha\beta}
-\frac{1}{\sqrt{N_f}}\tilde\sigma\delta_{\alpha\beta}
\Big],\\
K_J=&\delta_{\alpha\beta}(\delta_{i1}\delta_{j2}J^*(x)+\delta_{i2}\delta_{j1}J(x)).
\esp
\eeq

The structure of $K$ allows a systematic expansion of $S'$ in terms of $1/N_f$, by using the following general mathematical relations:
\beq
\bsp
{\rm Tr}[\log(A+B)]
=&{\rm Tr}[\log(A)]+{\rm Tr}[\log(1+A^{-1}B)],\\
{\rm Tr}[\log(1+A)]=&\sum_{n=1}^\infty{\rm Tr}\left(\frac{(-1)^{n+1}A^n}{n}\right),\\
(A+B)^{-1}=&\left(\sum_{n=0}^\infty(-A^{-1}B)^n\right)A^{-1}.
\esp
\eeq
However, since our purpose is to calculate the two-point correlation functions of $\sigma$ and $O_m$ to the order of $1/N_f$, in the expansion of $S'$ we just need to retain the terms that will contribute to these correlation functions to this order, while other terms can be neglected. Keeping only these terms, $S'$ becomes
\beq
\bsp
S'=&\frac{1}{2}\int\frac{d^3p}{(2\pi)^3}\tilde\sigma(p) \left(-\frac{1}{4|p|}\right)\tilde\sigma(-p)+\frac{1}{2}\int\frac{d^3p}{(2\pi)^3}\tilde A^a_\mu(p)
\left[\frac{|p|}{32}\left(\delta_{\mu\nu}-\frac{p_\mu p_\nu}{p^2}\right)\right]
\tilde A^a_\nu(-p)\\
&+S_{\sigma^3}+S_{\sigma^4}+S_{\sigma A^2}+S_{\sigma^2A^2}+S_{J^2}.
\esp
\eeq
With the notation $d_G=\delta^{aa}=3$, $N_c=\delta_{\alpha\alpha}=2$, $T_F=1/2$ and $\bar\delta(\cdot) = (2\pi)^3\delta(\cdot)$, different terms in the above are
\beq \label{eq: S sigma^3}
\bsp
S_{\sigma^3} &= \frac{1}{3!\sqrt{N_f}} \int \dq{k_1}\dq{k_2}\dq{k_3} \bar\delta(k_1+k_2+k_3) \tilde\sigma(k_1) \tilde\sigma(k_2) \tilde\sigma(k_3) \Gamma_{\sigma^3}^0(k_1,k_2,k_3),\\
&\Gamma_{\sigma^3}^0(k_1,k_2,k_3) = -2 N_c I_3(k_1,k_2,k_3), \qquad \\
&I_3(k_1,k_2,k_3) = \int \dq{l} \frac{1}{l^2(l+k_1)^2(l+k_1+k_2)^2},
\esp
\eeq

\beq \label{eq: S sigma^4}
\bsp
S_{\sigma^4} &= \frac{1}{4!N_f}\int \dq{k_1}\dq{k_2}\dq{k_3} \dq{k_4} \bar\delta(k_1+k_2+k_3+k_4) \tilde\sigma(k_1) \tilde\sigma(k_2) \tilde\sigma(k_3) \tilde\sigma(k_4) \Gamma_{\sigma^4}^{0}(k_1,k_2,k_3,k_4),  \\
& \Gamma_{\sigma^4}^{0}(k_1,k_2,k_3,k_4) = -2N_c [I_4(k_1,k_2,k_3,k_4)+I_4(k_1,k_3,k_2,k_4)+I_4(k_1,k_2,k_4,k_3)], \\
&I_4(k_1,k_2,k_3,k_4) = \int \dq{l} \frac{1}{l^2(l+k_1)^2(l+k_1+k_2)^2(l+k_1+k_2+k_3)^2},
\esp
\eeq

\beq \label{eq: S sigma A^2}
\bsp
S_{\sigma A^2} &= \frac{1}{2!\sqrt{N_f}}\int\dq{k}\dq{q_1}\dq{q_2} \bar\delta(k+q_1+q_2) \tilde\sigma(k) \tilde A_\mu^a(q_1) \tilde A_\nu^a(q_2) \Gamma_{\sigma A^2;\mu\nu}^0(k,q_1,q_2), \\
& \Gamma_{\sigma A^2;\mu\nu}^0(k,q_1,q_2) =  -2T_F \left( I_{3;\mu\nu}(q_1,q_2) - \frac{1}{8|k|} \delta_{\mu\nu}\right), \\
& I_{3;\mu\nu}(q_1,q_2) = \int \dq{l} \frac{(2l+q_1)_\mu(2l-q_2)_\nu}{l^2(l+q_1)^2(l-q_2)^2},
\esp
\eeq

\beq \label{eq: S sigma^2 A^2}
\bsp
S_{\sigma^2 A^2} &= \frac{1}{2!2!N_f}\int\dq{k_1}\dq{k_2}\dq{q_1}\dq{q_2}  \bar\delta(k_1+k_2+q_1+q_2) \cdot \tilde\sigma(k_1)\tilde\sigma(k_2)\tilde A_\mu^a(q_1) \tilde A_\nu^a(q_2) \Gamma_{\sigma^2A^2;\mu\nu}^0 (k_1,k_2,q_1,q_2),\\
& \Gamma_{\sigma^2 A^2;\mu\nu}^{0}(k_1,k_2,q_1,q_2) = 2T_F\Big[ 2\delta_{\mu\nu} I_{3}(k_1,k_2,-k_1-k_2) - I_{4;\mu\nu}(k_1,k_2,q_1,q_2) \\
&\qquad\qquad - I_{4;\mu\nu}(k_2,k_1,q_1,q_2) - I_{4;\mu\nu}'(k_1,k_2,q_1,q_2) \Big],\\
& I_{4;\mu\nu}(k_1,k_2,q_1,q_2)  = \int_{p_1} \frac{(2p_1-q_1)_\mu (2p_1-2q_1-q_2)_\nu}{p_1^2(p_1-q_1)^2(p_1-q_1-q_2)^2(p_1+k_2)^2} \\
& I'_{4;\mu\nu}(k_1,k_2,q_1,q_2) =\int_{p_1} \frac{(2p_1-q_1)_\mu (2p_1-q_1-k_2+k_1)_\nu}{p_1^2(p_1-q_1)^2(p_1-q_1-k_2)^2(p_1+k_1)^2},
\esp
\eeq
and
\beq \label{eq: S J^2}
\bsp
S_{J^2} 
=& -N_c\int\frac{d^3p_1 d^3p_2}{(2\pi)^6} \frac{1}{p_1^2 p_2^2} J(p_1-p_2) J^*(p_1-p_2) \\
&- \frac{N_c}{N_f} \int\frac{d^3p_1d^3p_2d^3p_3d^3p_4}{(2\pi)^{12}}
\frac{1}{p_1^2p_2^2p_3^2p_4^2} \tilde\sigma(p_1-p_2) J(p_2-p_3) \tilde\sigma(p_3-p_4) J^*(p_1-p_4) \\
&-\frac{T_F}{N_f}\int\frac{d^3p_1d^3p_2d^3p_3d^3p_4}{(2\pi)^{12}}\frac{(p_1+p_2)_\mu(p_3+p_4)_\nu}{p_1^2p_2^2p_3^2p_4^2} \tilde A_\mu^a (p_1-p_2)\tilde A_\nu^a(p_3-p_4)J(p_2-p_3) J^*(p_1-p_4) \\
&-\frac{2N_c}{N_f} \int\frac{d^3p_1d^3p_2d^3p_3d^3q}{(2\pi)^{12}}\frac{1}{ p_1^2 p_2^2 p_3^2 q^2} J(p_1-p_2) \tilde\sigma(p_2-q)\tilde\sigma(q-p_3) J^*(p_1-p_3) \\
&-\frac{2T_F}{N_f} \int\frac{d^3p_1d^3p_2d^3p_3d^3q}{(2\pi)^{12}}\frac{(p_2+q)_\mu(q+p_3)_\nu}{p_1^2 p_2^2 p_3^2 q^2 } J(p_1-p_2)  \tilde A^a_\mu(p_2-q)\tilde A^a_\nu(q-p_3) J^*(p_1-p_3)
\esp
\eeq

In fact, we can check that the triple-gauge-field coupling in the Chern-Simons action $S_{\rm CS}$ and the ghost action does not contribute at the order of $1/N_f$. Therefore, in summary, to calculate the correlation functions and scaling dimensions of $\sigma$ and $O_m$ to the order of $1/N_f$, it is sufficient to use the following effective action
\beq \label{eq: 1/N_f effective action}
S_{\rm eff}=S_0+S_V,
\eeq
where $S_0$ includes all quadratic terms:
\beq
\bsp
S_0
=&
-\frac{ik}{8\pi N_f}\int d^3x\varepsilon_{\mu\nu\lambda}\tilde A_\mu^a\partial_\nu \tilde A^a_\lambda+\frac{1}{32\xi}\int d^3xd^3y\frac{\partial_\mu\tilde A^a_\mu(x) \partial_\nu\tilde A^a_\nu(y)}{2\pi^2|x-y|^2}\\
&+\frac{1}{2}\int\frac{d^3p}{(2\pi)^3}\tilde A^a_\mu(p)
\left[\frac{|p|}{32}\left(\delta_{\mu\nu}-\frac{p_\mu p_\nu}{p^2}\right)\right]
\tilde A^a_\nu(-p)\\
&+\frac{1}{2}\int\frac{d^3p}{(2\pi)^3}\tilde\sigma(p) \left(-\frac{1}{4|p|}\right)\tilde\sigma(-p),
\esp
\eeq
and $S_V=S_{\sigma^3}+S_{\sigma^4}+S_{\sigma A^2}+S_{\sigma^2A^2}+S_{J^2}$, with $S_{\sigma^3}$, $S_{\sigma^4}$, $S_{\sigma A^2}$, $S_{\sigma^2 A^2}$ and $S_{J^2}$ given by Eqs. \eqref{eq: S sigma^3}, \eqref{eq: S sigma^4}, \eqref{eq: S sigma A^2}, \eqref{eq: S sigma^2 A^2} and \eqref{eq: S J^2}, respectively.

In Eq. \eqref{eq: 1/N_f effective action}, $S_0$ is regarded as order 1 and $S_V$ is suppressed by powers of $1/N_f$. In the following, we will calculate the correlation functions and scaling dimensions of $\sigma$ and $O_m$ to the order of $1/N_f$ in Appendices \ref{subapp: sigma} and \ref{subapp: calculations of Om}. It turns out that different terms in the calculations have diagrammatic interpretations, as summarized in Fig. \ref{fig:boson-diagrams-combined}.

For the calculations in Appendices \ref{subapp: sigma} and \ref{subapp: calculations of Om}, it is useful to first establish some notations. From $S_0$, we obtain the effective propagators of $\tilde\sigma$ and $\tilde A$:
\begin{enumerate}
 	\item The propagator of the $\tilde\sigma$ field is
	\beq \label{eq: sigma propagator}
	\langle\tilde\sigma(p)\tilde\sigma(-p)\rangle_0=-4|p|.
	\eeq
    We will use the notation $\tilde D_{\sigma}(p) = \langle\tilde\sigma(p)\tilde\sigma(-p)\rangle_0$ in the following.
    
	\item The propagator of the gauge field $\tilde A$ is
	\beq \label{eq: gauge propagator}
	\bsp
	\langle \tilde A^a_\mu(p) \tilde A^b_\nu(-p)\rangle_0
	=\frac{32\delta^{ab}}{1+\zeta^2}\left[\left(\frac{\delta_{\mu\nu}}{|p|}-\frac{p_\mu p_\nu}{|p|^3}\right)+\zeta\frac{\epsilon_{\mu\nu\lambda}p_\lambda}{|p|^2}\right]-16\delta^{ab}\xi\frac{p_\mu p_\nu}{|p|^3},
	\esp
	\eeq
	with $\zeta=\frac{8k}{\pi N_f}$. We will use the notation $\tilde D_{\mu\nu}^{ab}(p)= \delta^{ab} \tilde D_{\mu\nu}(p) = \langle \tilde A^a_\mu(p) \tilde A^b_\nu(-p)\rangle_0$ in the following.
\end{enumerate}
We also denote 
\begin{equation}
    G_0(p) = \frac{1}{p^2},
\end{equation}
which is indeed the propagator of the original $\Phi$ field.

In the diagrammatic representation of different terms in the calculations below, we will use a dashed line to represent $\tilde\sigma$, a wavy line to represent $\tilde A$ and a solid line to represent $G_0(p)$.

%%----------------------------------------------------
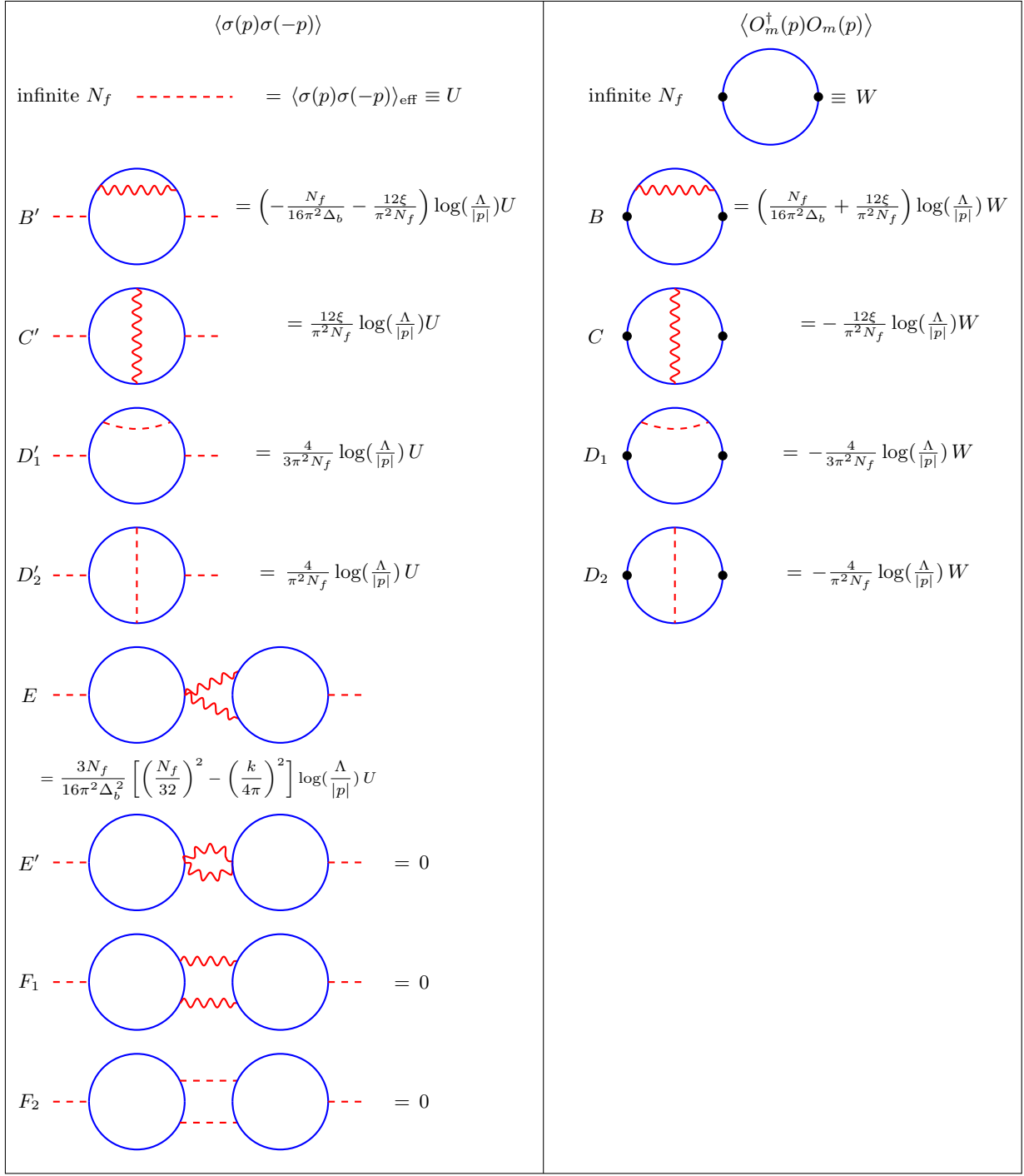
\begin{figure*}[t]
\centering
\begin{tikzpicture}[scale=0.38]

% outer box + divider
\draw[black] (-5.5,4) -- (37,4) -- (37,-45) -- (-5.5,-45) -- cycle;
\draw[black] (17,4) -- (17,-45);
\node at (5.5,3) {\textbf{$\langle\sigma(p)\sigma(-p)\rangle$}};
\node at (28,3) {\textbf{$\bigl\langle O_m^\dag(p)O_m(p)\bigr\rangle$}};

%========== LEFT COLUMN: sigma ==========
% tree
\node at (-3,0) {\small infinite $N_f$};
\draw[dashed,red,thick] (0,0) -- (4,0);
\node at (9.5,0) {$=\,\langle\sigma(p)\sigma(-p)\rangle_{\mathrm{eff}}\equiv U$};

% B'
\node at (-4.5,-5) {\small $B'$};
\draw[blue,thick] ([shift=(180:2cm)]0,-5) arc (180:0:2cm);
\draw[blue,thick] ([shift=(0:2cm)]0,-5) arc (0:-180:2cm);
\draw[dashed,red,thick] (-3.5,-5) to (-2,-5);
\draw[dashed,red,thick] (2,-5) to (3.5,-5);
\draw[photon,red] (-1.66,-3.9) to (1.66,-3.9);
\node at (10,-4.6) {$=\left(-\frac{N_f}{16\pi^2\Delta_b}-\frac{12\xi}{\pi^2N_f}\right)\log(\frac{\Lambda}{|p|})U$};

% C'
\node at (-4.5,-10) {\small $C'$};
\draw[blue,thick] ([shift=(180:2cm)]0,-10) arc (180:0:2cm);
\draw[blue,thick] ([shift=(0:2cm)]0,-10) arc (0:-180:2cm);
\draw[photon,red] ([shift=(90:2cm)]0,-10) to ([shift=(-90:2cm)]0,-10);
\draw[dashed,red,thick] (-3.5,-10) to (-2,-10);
\draw[dashed,red,thick] (2,-10) to (3.5,-10);
\node at (9.5,-9.6) {$=\frac{12\xi}{\pi^2N_f}\log(\frac{\Lambda}{|p|})U$};

% D_1' : sigma on same scalar line
\node at (-4.5,-15) {\small $D_1'$};
\draw[thick,blue] ([shift=(180:2cm)]0,-15) arc (180:135:2cm);
\draw[thick,blue] ([shift=(135:2cm)]0,-15) arc (135:45:2cm);
\draw[thick,blue] ([shift=(45:2cm)]0,-15) arc (45:0:2cm);
\draw[thick,blue] ([shift=(0:2cm)]0,-15) arc (0:-180:2cm);
\draw[thick,red,dashed] (-1.4,-13.6) to[out=-20,in=200] (1.4,-13.6);
\draw[dashed,red,thick] (-3.5,-15) to (-2,-15);
\draw[dashed,red,thick] (2,-15) to (3.5,-15);
\node at (8.5,-15) {$=\,\frac{4}{3\pi^2 N_f}\log(\frac{\Lambda}{|p|})\,U$};

% D_2' : sigma between scalar lines
\node at (-4.5,-20) {\small $D_2'$};
\draw[thick,blue] ([shift=(180:2cm)]0,-20) arc (180:0:2cm);
\draw[thick,blue] ([shift=(0:2cm)]0,-20) arc (0:-180:2cm);
\draw[dashed,red,thick] ([shift=(90:2cm)]0,-20) to ([shift=(-90:2cm)]0,-20);
\draw[dashed,red,thick] (-3.5,-20) to (-2,-20);
\draw[dashed,red,thick] (2,-20) to (3.5,-20);
\node at (8.5,-20) {$=\,\frac{4}{\pi^2 N_f}\log(\frac{\Lambda}{|p|})\,U$};

% E : seagull--jj two-bubble
\node at (-4.5,-25) {\small $E$};
\draw[thick,blue] ([shift=(0:2cm)]0,-25) arc (0:180:2cm);
\draw[thick,blue] ([shift=(-180:2cm)]0,-25) arc (-180:0:2cm);
\draw[photon,red] (2,-25) to ([shift=(150:2cm)]6,-25);
\draw[photon,red] (2,-25) to ([shift=(-150:2cm)]6,-25);
\draw[thick,blue] ([shift=(180:2cm)]6,-25) arc (180:0:2cm);
\draw[thick,blue] ([shift=(0:2cm)]6,-25) arc (0:-180:2cm);
\draw[dashed,red,thick] (-3.5,-25) to (-2,-25);
\draw[dashed,red,thick] (8,-25) to (9.5,-25);

\node[anchor=north, align=center] at (3,-27.35) {\scriptsize $\displaystyle
=\frac{3N_f}{16\pi^2\Delta_b^{\,2}}
\left[
\left(\frac{N_f}{32}\right)^2
-\left(\frac{k}{4\pi}\right)^2
\right]
\log(\frac{\Lambda}{|p|} )\,U
$};

% E' : two-bubble with two AAPhiPhi coupling
\node at (-4.5,-32) {\small $E'$};

% left bubble
\draw[thick,blue] ([shift=(0:2cm)]0,-32) arc (0:180:2cm);
\draw[thick,blue] ([shift=(-180:2cm)]0,-32) arc (-180:0:2cm);

% middle flatter wavy loop
\draw[photon,red]
    (2,-32)
    .. controls (2.45,-31.35) and (3.55,-31.35) ..
    (4,-32);

\draw[photon,red]
    (2,-32)
    .. controls (2.45,-32.65) and (3.55,-32.65) ..
    (4,-32);

% right bubble
\draw[thick,blue] ([shift=(180:2cm)]6,-32) arc (180:0:2cm);
\draw[thick,blue] ([shift=(0:2cm)]6,-32) arc (0:-180:2cm);

% external dashed legs
\draw[dashed,red,thick] (-3.5,-32) to (-2,-32);
\draw[dashed,red,thick] (8,-32) to (9.5,-32);

% result
\node at (11.5,-32) {$=\,0$};

% F_1 : two-bubble with two gauge lines
\node at (-4.5,-37) {\small $F_1$};
\draw[thick,blue] ([shift=(0:2cm)]0,-37) arc (0:180:2cm);
\draw[thick,blue] ([shift=(-180:2cm)]0,-37) arc (-180:0:2cm);
\draw[photon,red] ([shift=(26:2cm)]0,-37) to ([shift=(154:2cm)]6,-37);
\draw[photon,red] ([shift=(-26:2cm)]0,-37) to ([shift=(-154:2cm)]6,-37);
\draw[thick,blue] ([shift=(180:2cm)]6,-37) arc (180:0:2cm);
\draw[thick,blue] ([shift=(0:2cm)]6,-37) arc (0:-180:2cm);
\draw[dashed,red,thick] (-3.5,-37) to (-2,-37);
\draw[dashed,red,thick] (8,-37) to (9.5,-37);
\node at (11.5,-37) {$=\,0$};

% F_2 : two-bubble with two sigma lines
\node at (-4.5,-42) {\small $F_2$};
\draw[thick,blue] ([shift=(0:2cm)]0,-42) arc (0:180:2cm);
\draw[thick,blue] ([shift=(-180:2cm)]0,-42) arc (-180:0:2cm);
\draw[dashed,red,thick] ([shift=(26:2cm)]0,-42) to ([shift=(154:2cm)]6,-42);
\draw[dashed,red,thick] ([shift=(-26:2cm)]0,-42) to ([shift=(-154:2cm)]6,-42);
\draw[thick,blue] ([shift=(180:2cm)]6,-42) arc (180:0:2cm);
\draw[thick,blue] ([shift=(0:2cm)]6,-42) arc (0:-180:2cm);
\draw[dashed,red,thick] (-3.5,-42) to (-2,-42);
\draw[dashed,red,thick] (8,-42) to (9.5,-42);
\node at (11.5,-42) {$=\,0$};

%========== RIGHT COLUMN: adjoint ==========
\begin{scope}[shift={(22.5,0)}]

% tree
\node at (-1.6,0) {\small infinite $N_f$};
\draw[blue,thick] (4,0) circle (2cm);
\draw[fill] (2,0) circle (5pt);
\draw[fill] (6,0) circle (5pt);
\node at (7.5,0) {$\equiv\,W$};

% B
\node at (-3.3,-5) {\small $B$};
\draw[blue,thick] ([shift=(180:2cm)]0,-5) arc (180:0:2cm);
\draw[blue,thick] ([shift=(0:2cm)]0,-5) arc (0:-180:2cm);
\draw[fill] (-2,-5) circle (5pt);
\draw[fill] (2,-5) circle (5pt);
\draw[photon,red] (-1.66,-3.9) to (1.66,-3.9);
\node at (8.2,-4.6) {$=\left(\frac{N_f}{16\pi^2\Delta_b} + \frac{12\xi}{\pi^2N_f}\right)\log(\frac{\Lambda}{|p|})\,W$};

% C
\node at (-3.3,-10) {\small $C$};
\draw[blue,thick] ([shift=(180:2cm)]0,-10) arc (180:0:2cm);
\draw[blue,thick] ([shift=(0:2cm)]0,-10) arc (0:-180:2cm);
\draw[photon,red] ([shift=(90:2cm)]0,-10) to ([shift=(-90:2cm)]0,-10);
\draw[fill] (-2,-10) circle (5pt);
\draw[fill] (2,-10) circle (5pt);
\node at (9,-9.6) {$=-\,\frac{12\xi}{\pi^2N_f}\log(\frac{\Lambda}{|p|})W$};

% D_1
\node at (-3.3,-15) {\small $D_1$};
\draw[thick,blue] ([shift=(180:2cm)]0,-15) arc (180:135:2cm);
\draw[thick,blue] ([shift=(135:2cm)]0,-15) arc (135:45:2cm);
\draw[thick,blue] ([shift=(45:2cm)]0,-15) arc (45:0:2cm);
\draw[thick,blue] ([shift=(0:2cm)]0,-15) arc (0:-180:2cm);
\draw[thick,red,dashed] (-1.4,-13.6) to[out=-20,in=200] (1.4,-13.6);
\draw[fill] (-2,-15) circle (5pt);
\draw[fill] (2,-15) circle (5pt);
\node at (8.5,-15) {$=\,-\frac{4}{3\pi^2 N_f}\log(\frac{\Lambda}{|p|})\,W$};

% D_2
\node at (-3.3,-20) {\small $D_2$};
\draw[thick,blue] ([shift=(180:2cm)]0,-20) arc (180:0:2cm);
\draw[thick,blue] ([shift=(0:2cm)]0,-20) arc (0:-180:2cm);
\draw[dashed,red,thick] ([shift=(90:2cm)]0,-20) to ([shift=(-90:2cm)]0,-20);
\draw[fill] (-2,-20) circle (5pt);
\draw[fill] (2,-20) circle (5pt);
\node at (8.5,-20) {$=\,-\frac{4}{\pi^2 N_f}\log(\frac{\Lambda}{|p|})\,W$};

\end{scope}

\end{tikzpicture}
\caption{$\mathcal{O}(1/N_f)$ Feynman graphs contributing to
$\langle\tilde\sigma(p)\tilde\sigma(-p)\rangle$ (left) and
$\langle O_m^\dag(p)O_m(p)\rangle$
(right). Here blue solid lines represent $G_0(p)=1/p^2$, red dashed lines represent the propagator of $\tilde\sigma$ in Eq. \eqref{eq: sigma propagator} and red wavy lines represent the propagator of the gauge field in Eq.
\eqref{eq: gauge propagator}. Gauge-parameter ($\xi$) dependence cancels between $B$
and $C$ ($B'$ and $C'$). In the above, only the logarithmically divergent parts are retained.}
\label{fig:boson-diagrams-combined}
\end{figure*}

\subsection{Correlation function and scaling dimension of $\sigma$} \label{subapp: sigma}

Using the effective action Eq. \eqref{eq: 1/N_f effective action}, here we calculate the correlation function and scaling dimension of $\sigma$ to the order of $1/N_f$. The results are summarized in Appendix \ref{subsubapp: singlet bosonic mass dimension}. In Appendix \ref{subapp: calculations of Om}, we will calculate the correlation function and scaling dimension of $O_m$.

\subsubsection{Scaling dimension at infinite $N_f$}

The rescaled field $\tilde\sigma=\sqrt{N_f}\sigma$ has the same scaling dimension as $\sigma$, so we can just compute the correlation of $\tilde\sigma$. 
According to Eq. \eqref{eq: sigma propagator}, using Eq. \eqref{eq: scaling dimension from correlator}, at infinite $N_f$ the scaling dimension of $\sigma$ is 
\beq
\Delta_s^{(0)}=\frac{1+3}{2}=2.
\eeq

Next, we calculate the correction to the correlation function and scaling dimension of $\sigma$, up to the order of $1/N_f$.

\subsubsection{ Gauge theory bubble and tadpole}

We first consider the bubble diagram formed by gauge propagators, which, as explained below, consists of graphs $E$, $E'$ and $F_1$ in Fig. \ref{fig:boson-diagrams-combined}.
Its contribution to correlation function of $\tilde\sigma$ is
\begin{equation}
\begin{split}
    &\frac{1}{2}\ew{\tilde\sigma(p)\tilde\sigma(-p) (S_{\sigma A^2})^2 }_0 \\
    =&\frac{1}{2N_f}\cdot(\frac{1}{2})^2\cdot 2^2 \tilde D_\sigma(p)^2 \int\dq{q} \tilde D_{\mu\mu'}^{ab}(q) \tilde D_{\nu\nu'}^{ab}(-p-q) \Gamma_{\sigma A^2;\mu\nu}(p,q,-p-q) \Gamma_{\sigma A^2;\mu'\nu'}(-p,-q,p+q)\\
    =&\frac{\tilde D_\sigma(p)^2(2T_F)^2}{2N_f}\int\frac{d^3q}{(2\pi)^3}\tilde D_{\mu\mu'}^{ab}(q) \tilde D_{\nu\nu'}^{ab}(-p-q)\\
    &\cdot
    \Big[
    I_{3;\mu\nu}(q,-p-q) I_{3;\mu'\nu'}(-q,p+q)
    +
    (\frac{1}{8|p|})^2\delta_{\mu\nu}\delta_{\mu'\nu'}
    -\frac{1}{8|p|}I_{3;\mu\nu}(q,-p-q)\delta_{\mu'\nu'}
    -\frac{1}{8|p|}I_{3;\mu'\nu'}(-q,p+q)\delta_{\mu\nu}
    \Big].
\end{split}
\end{equation}
The factor $2^2$ comes from two ways of contracting external legs and two ways of contracting the internal gauge fields.

The first term corresponds to the diagram $F_1$ in Fig. \ref{fig:boson-diagrams-combined}. Here a graph only counts the Feynman rules involved and does not include any symmetry factor or multiplicity. Indeed, by counting the coefficients, we find its contribution is equal to 
\[
\frac{2}{N_f}T_F^2 d_G \tilde D_\sigma(p)^2\int\dq{q} \tilde D_{\mu\mu'}(q)\tilde D_{\nu\nu'}(-p-q)I_{3;\mu\nu}(q,-p-q) I_{3;\mu'\nu'}(-q,p+q) = 2\times {\rm Graph}\ F_1.
\]
By direct computation, we find that graph $F_1$ does not contribute to logarithmic divergence.

The second term, which corresponds to graph $E'$, actually gives a convergent integral, and does not contribute to the logarithmic divergence, either.

The last two terms have the same contribution, which is
\begin{equation}
    -\frac{4}{N_f} T_F^2 d_G \tilde D_\sigma(p)^2 \int\dq{q} \tilde D_{\mu\mu'}(q) \tilde D_{\mu\nu'}(-p-q) I_{3;\mu'\nu'}(q,-p-q) \frac{1}{8|p|} = 2\times {\rm Graph}\ E
\end{equation}
The graph computation in the manuscript shows that, indeed, the log part of the following expression (with $T_F = 1/2, d_G = 3$) is
\begin{equation}\label{eq: graph_E}
\begin{split}
    {\text{graph }E} &\equiv -\frac{2}{N_f} T_F^2 d_G \tilde D_\sigma(p)^2\int\dq{q} \tilde D_{\mu\mu'}(q) \tilde D_{\mu\nu'}(-p-q) I_{3,\mu'\nu'}(q,-p-q) \frac{1}{8|p|} \\
    &=\frac{3N_f}{32\pi^2\Delta_b^2} \left[(\frac{N_f}{32})^2 - (\frac{k}{4\pi})^2 \right] \log(\frac{\Lambda}{|p|}) \tilde D_\sigma(p) + {\rm finite},
\end{split}
\end{equation}
where an extra negative sign coming from one of $D_\sigma(p)$ cancelled with the sign in the prefactor, and
\beq
\Delta_b=(\frac{N_f}{32})^2+(\frac{k}{4\pi})^2.
\eeq

Hence, we find the log divergence of gauge bubble is
\begin{equation}
    \frac{1}{2}\ew{\tilde\sigma(p)\tilde\sigma(-p) (S_{\sigma A^2})^2}_0\Big|_{\rm NLO;\,Log} = \frac{3N_f}{16\pi^2\Delta_b^2} \left[(\frac{N_f}{32})^2 - (\frac{k}{4\pi})^2 \right] \log(\frac{\Lambda}{|p|}) \tilde D_\sigma(p)
\end{equation}
This result is exhibited in Fig. \ref{fig:boson-diagrams-combined} with a multiplicity 2 included therein. Graphically, this multiplicity comes from the insertion of quartic vertex into either of the two solid line loops in graph $E$.

Next we consider the tadpole diagrams (graphs $B'$ and $C'$):
\begin{equation}\label{eq: gauge_tadpole}
\begin{split}
    &- \ew{\tilde\sigma(p)\tilde\sigma(-p)S_{\sigma^2 A^2}}_0 \\
    =& - \frac{1}{(2!)^2 N_f} \cdot d_G \cdot 2 \tilde D_\sigma(p)^2\int\dq{q} \Gamma_{\sigma^2A^2;\mu\nu}(p,-p,q,-q) \tilde D_{\mu\nu}(q) \\
    =& \frac{3}{2N_f} \tilde D_\sigma(p)^2 \int \dq{q} \tilde D_{\mu\nu}(q) \Big[ -2\delta_{\mu\nu}I_{3}(p,-p,0) + I_{4;\mu\nu}(p,-p,q,-q) + I_{4;\mu\nu}(-p,p,q,-q) + I'_{4;\mu\nu}(p,-p,q,-q)\Big].
\end{split}
\end{equation}

There are four pieces in the bracket and we analyze their contributions one by one. The $\delta_{\mu\nu}I_3(p,-p,0)$ term does not contribute and its corresponding graph is not shown in Fig. \ref{fig:boson-diagrams-combined}, because its divergence is solely from the loop integration of $q$ and can be absorbed into a counterterm. The two terms involving $I_4$ correspond to the two $B'$ graphs in Fig.\ \ref{fig:boson-diagrams-combined}, and they have equal contributions. The last piece, $I'_{4;\mu\nu}(p,-p,q,-q)$, corresponds to graph $C'$ in Fig.\ \ref{fig:boson-diagrams-combined} and contributes only to the gauge dependent part. 

With explicit computations, we find 
\begin{equation}
{\text{graph }B'}= \frac{3}{2N_f}\tilde D_\sigma(p)^2\int \dq{q} \tilde D_{\mu\nu} I_{4;\mu\nu}(p,-p,q,-q) = \left(-\frac{N_f}{32\pi^2\Delta_b} -\frac{6\xi}{\pi^2N_f}\right)\log(\frac{\Lambda}{|p|}) \tilde D_\sigma(p) + {\rm finite}
\end{equation}
and
\begin{equation}
{\text{graph }C'} = \frac{3}{2N_f} \tilde D_\sigma(p)^2\int \dq{q} \tilde D_{\mu\nu}(q) I'_{4;\mu\nu}(p,-p,q,-q) = \frac{12\xi}{\pi^2N_f} \log(\frac{\Lambda}{|p|}) \tilde D_\sigma(p) + {\rm finite}.
\end{equation}
Hence the correct contribution, from Eq.\eqref{eq: gauge_tadpole}, should be (2$\times$graph $B'$ + graph $C'$), which is exhibited in Fig.\ \ref{fig:boson-diagrams-combined},
\begin{equation}
    -\ew{\tilde\sigma(p)\tilde\sigma(-p) S_{\sigma^2 A^2}}_0 \Big|_{\rm NLO; log} = -\frac{N_f}{16\pi^2\Delta_b} \log(\frac{\Lambda}{|p|}) \tilde D_\sigma(p),
\end{equation}
where ``NLO'' stands for next-to-leading order. Notice the gauge dependence, \ie the dependence on $\xi$, has been cancelled.

\subsubsection{$\sigma$ field bubble and tadpole}

We now consider the contributions of insertions with $S_{\sigma^3}$ and $S_{\sigma^4}$.

For $S_{\sigma^3}$, its contribution to the correlation function is
\begin{equation}
\begin{split}
    \frac{1}{2}\ew{\tilde\sigma(p)\tilde\sigma(-p) (S_{\sigma^3})^2}_0 &= \frac{1}{2N_f} (\frac{1}{3!})^2 \cdot 2\cdot 3^2\cdot 2 \tilde D_\sigma(p)^2\int \dq{q} \tilde D_\sigma(q)\tilde D_{\sigma}(p+q) \Gamma_{\sigma^3}^0(p,q,-p-q) \Gamma_{\sigma^3}^0(-p,-q,p+q) \\
    &= \frac{2}{N_f} N_c^2 \tilde D(p)^2\int \dq{q} \tilde D_\sigma(q) \tilde D_{\sigma}(p+q) I_{3}(p,q,-p-q) I_3(-p,-q,p+q) \\
    &= 2\times{\rm graph}\ F_2 
\end{split}
\end{equation}
The graph $F_2$ in Fig.\ \ref{fig:boson-diagrams-combined} does not contribute to the logarithmic divergence.

Next we consider the tadpole diagram from $S_{
\sigma^4}$ and its contribution to the correlation function is 
\begin{equation}
\begin{split}
    -\ew{\tilde\sigma(p)\tilde\sigma(-p)S_{\sigma^4}}_0 &= -\frac{1}{4!}\times 4\times 3 \tilde D_\sigma(p)^2\int\dq{q} \tilde D_{\sigma}(q)\Gamma_{\sigma^4}^0(p,-p,q,-q) \\
    & = \frac{N_c}{N_f} \tilde D_\sigma(p)^2 \int\dq{q} \tilde D_{\sigma}(q) [I_4(p,-p,q,-q)+I_4(p,q,-p,-q) + I_4(p,-p,-q,q)]
\end{split}
\end{equation}
The expressions of three pieces of $I_4$ are treated as
\begin{equation}
\begin{split}
    I_4(p,-p,q,-q) &= \int \dq{l} G_0(l) G_0(l+p) G_0(l) G_0(l+q)\\
    I_4(p,q,-p,-q) &= \int \dq{l} G_0(l) G_0(l+p) G_0(l+p+q) G_0(l+q) \\
    I_4(p,-p,-q,q) &=\int \dq{l} G_0(l) G_0(l+p) G_0(l) G_0(l-q).
\end{split}
\end{equation}
We find the contributions of $I_{4}(p,-p,q,-q)$ and $I_{4}(p,-p,-q,q)$ in the integral are the same due to the fact $\tilde D_\sigma(p)$ is parity even, and each of them contributes the same as graph $D_1'$. The remaining $I_4(p,q,-p,-q)$ contributes the same as graph $D_2'$. Hence, we get
\begin{equation}
\begin{split}
    -\ew{\tilde\sigma(p)\tilde\sigma(-p)S_{\sigma^4}}_0 \Big|_{\rm NLO; log}&= 2\times {\rm Graph}\ D_1' + {\rm Graph}\ D_2' \\
    &= 2\times\frac{2}{3\pi^2N_f}\log(\frac{\Lambda}{|p|})\tilde D_\sigma(p) + \frac{4}{\pi^2N_f}\log(\frac{\Lambda}{|p|}) \tilde D_\sigma(p)\\
    &= \frac{16}{3\pi^2 N_f}\log(\frac{\Lambda}{|p|}) \tilde D_\sigma(p)
\end{split}
\end{equation}

\subsubsection{Summary and extracting anomalous dimension} \label{subsubapp: singlet bosonic mass dimension}

Collecting all $1/N_f$ order contributions together, we find
\begin{equation}\label{eq: tilde_sigma_log}
    \ew{\tilde\sigma(p)\tilde \sigma(-p)}\Big|_{\rm NLO;log} = \left(-\frac{N_f}{16\pi^2\Delta_b}+ \frac{3N_f}{16\pi^2\Delta_b^2}\left[(\frac{N_f}{32})^2 - (\frac{k}{4\pi})^2\right] + \frac{16}{3\pi^2N_f} \right)\log(\frac{\Lambda}{|p|}) \tilde D_\sigma(p),
\end{equation}
where ``NLO'' stands for next-to-leading order and
\beq
\Delta_b=(\frac{N_f}{32})^2+(\frac{k}{4\pi})^2.
\eeq 
Therefore, up to the order of $1/N_f$,
\beq
\langle \tilde\sigma(p)\tilde\sigma(-p)\rangle=-4|p|\left(\frac{\Lambda}{|p|}\right)^{-\frac{N_f}{16\pi^2\Delta_b}+ \frac{3N_f}{16\pi^2\Delta_b^2}\left[(\frac{N_f}{32})^2 - (\frac{k}{4\pi})^2\right] + \frac{16}{3\pi^2N_f} }
\eeq
According to Eq. \eqref{eq: scaling dimension from correlator}, the scaling dimension of $\tilde\sigma$ and $\sigma$ is
\beq
\Delta_\sigma=2+\gamma_\sigma
\eeq
with
\beq
\gamma_\sigma = \frac{N_f}{32\pi^2\Delta_b} - \frac{3N_f}{32\pi^2\Delta_b^2}\left[(\frac{N_f}{32})^2 - (\frac{k}{4\pi})^2\right] - \frac{8}{3\pi^2N_f}.
\eeq

\subsection{Correlation function and scaling dimension of $O_m$} \label{subapp: calculations of Om}

In this sub-appendix, based on the effective action Eq. \eqref{eq: 1/N_f effective action}, we compute the correlation function $\ew{O_m(p)O_m^\dagger(p)}$, and extract the scaling dimension of $O_m$.

By functional derivative,
\begin{equation}
    \ew{O_m(p) O_m^\dagger(p)} = \left.\frac{\delta^2 Z[J]}{\delta J(p)\delta J^*(p)}\right|_{J=J^*=0} = -\ew{ \frac{\delta^2 S_{J^2}}{\delta J(p) \delta J^*(p)}} + O(1/N_f^2)
\end{equation}
we notice that the linear in $J$ terms of effective action vanish under the flavor trace, otherwise we would have the contribution from linear actions. We now use the expressions in Eq.\ \eqref{eq: S J^2} to compute the correlation function.

The leading order contribution is
\begin{equation}
    W(p) = \ew{O_m(p) O_m^\dagger(p)}_0 = N_c \int \dq{q}\frac{1}{q^2}\frac{1}{(p+q)^2} = \frac{1}{4|p|}.
\end{equation}
So the scaling dimension of $O_m$ at infinite $N_f$ is $1$, according to Eq. \eqref{eq: scaling dimension from correlator}.

In the following, we calculate the order $1/N_f$ correction to the correlation function and scaling dimension of $O_m$.

There are two $\sigma\sigma J J^*$ couplings in Eq.\ \eqref{eq: S J^2}. The first gives
\begin{equation}
    \frac{N_c}{N_f} \int\dq{p_1}\dq{q} \frac{1}{p_1^2 (p_1-q)^2 (p_1-q-p)^2(p_1-p)^2 } \tilde D_\sigma(q) = {\rm graph}\ D_2
\end{equation}
The second one give
\begin{equation}
    \frac{2N_c}{N_f}\int\dq{p_1}\dq{q}\frac{1}{p_1^2 (p_1-p)^2(p_1-p)^2 (p_1-p-q)^2} \tilde D_\sigma(q) = 2\times {\rm graph}\ D_1
\end{equation}
Comparing these two expressions with graph $D_1', D_2'$ in the $\sigma$ correlation function, we find the ratios with respect to their corresponding leading order contribution differ by a negative sign. Hence we get
\begin{equation}
\begin{split}
     2\times {\rm graph}\ D_1 &= -\frac{4}{3\pi^2 N_f}\log(\frac{\Lambda}{|p|}) W(p), \\
     {\rm graph}\ D_2 &= -\frac{4}{\pi^2 N_f}\log(\frac{\Lambda}{|p|}) W(p).
\end{split}
\end{equation}

Next we consider the coupling $A A J J^*$ in Eq.\ \eqref{eq: S J^2}. There are also two terms, one gives
\begin{equation}
    \frac{T_F d_G}{N_f} \int \dq{p_1}\dq{q}\frac{(2p_1-q)_\mu (2p_1-q-2p)_\nu}{p_1^2(p_1-q)^2(p_1-q-p)^2(p_1-p)^2} \tilde D_{\mu\nu}(q) = {\rm graph}\ C,
\end{equation}
and the other gives
\begin{equation}
    \frac{2T_F d_G}{N_f}\int\dq{p_1}\dq{l}\frac{(2p_1-2p-l)_\mu (2p_1-2p-l)_\nu}{p_1^2(p_1-p)^2(p_1-p)^2 (p_1-p-l)^2} \tilde D_{\mu\nu}(l) = 2\times{\rm graph}\ B
\end{equation}
Comparing these two expressions with graph $B', C'$ in the $\sigma$ correlation function, we find the ratios with respect to their corresponding leading order contribution differ by a negative sign. Hence we get
\begin{equation}
    \begin{split}
        2\times{\rm graph}\ B &= \left(\frac{N_f}{16\pi^2\Delta_b}+\frac{12\xi}{\pi^2N_f}\right)\log(\frac{\Lambda}{|p|}) W(p), \\
        {\rm graph}\ C &= -\frac{12\xi}{\pi^2N_f}\log(\frac{\Lambda}{|p|})W(p).
    \end{split}
\end{equation}

In summary, the correlation function of $O_m$ up to $1/N_f$ order is
\begin{equation}
    \ew{O_m(p) O_m^\dagger(p)} = \left[1+\left(\frac{N_f}{16\pi^2\Delta_b} - \frac{16}{3\pi^2 N_f}\right)\log(\frac{\Lambda}{|p|})\right] W(p).
\end{equation}
According to Eq. \eqref{eq: scaling dimension from correlator}, we write the scaling dimension of $O_m$ as $\Delta_{O_m} = 1+\gamma_{O_m}$, then 
\begin{equation}
    \gamma_{O_m} = -\frac{N_f}{32\pi^2\Delta_b} +\frac{8}{3\pi^2 N_f}, 
\end{equation}
where
\beq
\Delta_b=(\frac{N_f}{32})^2+(\frac{k}{4\pi})^2.
\eeq

%=====================================================================
%=====================================================================
\section{Details of the analysis of the critical Chern-Simons-fermion theory} \label{app: fermion analysis}

In this appendix, we present the details of the calculations of the scaling dimensions of the operators $\mc{O}_S$ and $\mc{O}_{A_0}$ in Sec. \ref{subsec:bilinear_operators}. As in Appendix \ref{app:boson_details}, we first derive an effective action in Appendix \ref{subapp: fermionic effective action}, and then carry out the calculations of the correlation functions and scaling dimensions in Appendices \ref{subapp: fermion singlet} and \ref{subapp: fermion multiplet}. Here for $\mc{O}_{A_0}$, we will explicitly consider $\mc{O}_{A_0}=\bar\psi_{\alpha 1}\psi_{\alpha 2}$. Up to the order of $\mc{O}(1/N_f)$, we find that the scaling dimensions of these operators being given by Eqs. \eqref{eq:Delta_S_fermion} and \eqref{eq:Delta_A0_fermion}, respectively.

\subsection{Fermion effective action} \label{subapp: fermionic effective action}

The action of the fermionic setup is 
\begin{equation}
    S = S_{\rm CS} + S_{\rm GF} + S_{\rm gh} + S_{\rm matter} + S_{J}
\end{equation}
where $S_{\rm CS}, S_{\rm GF}$ and $S_{\rm gh}$ are exactly the same as those in the bosonic case,
\begin{equation}
    S_{\rm matter} = \int \dq{p_1}\dq{p_2} \bar\psi_j(p_1) \left(i\gamma^\mu p_{1\mu} \bar\delta(p_1-p_2) - i\gamma^\mu T^aA_\mu^a(p_1-p_2)\right) \psi_j(p_2)
\end{equation}
and 
\begin{equation}
    S_{J} = \int \dq{p_1}\dq{p_2} [\bar\psi_1(p_1) J_1(p_1-p_2) \psi_2(p_2) + \bar\psi_2(p_1) J_{1}^*(p_1-p_2) \psi_1(p_2) + \frac{1}{\sqrt{N_f}}J_{0}(p_1-p_2)\bar\psi_j(p_1)\psi_j(p_2)],
\end{equation}
with $J_0$ a real source field and $J_1$ a complex source field, and $\bar\delta(p)=(2\pi)^3\delta(p)$.

In the path integral representation of the generating functional $Z[J_0,J_1,J_1^*]$, we integrate out the fermions and get the effective action
\begin{equation}
    S_{\rm eff} = S_{\rm CS} + S_{\rm GF} + S_{\rm gh} - \Trace\ln[ K + K^J_0+ K^J_1],
\end{equation}
with
\begin{equation}
\begin{split}
    K(p_1,p_2)_{jl, \alpha\beta} &=i \delta_{jl}[\gamma^\mu p_\mu\bar\delta(p_1-p_2)\delta_{\alpha\beta} - T^a_{\alpha\beta}\gamma^\mu A_\mu^a(p_1-p_2)], \\
    K^J_1(p_1,p_2)_{jl,\alpha\beta} &= \delta_{\alpha\beta}[J_1(p_1-p_2)\delta_{j1}\delta_{l2} + J_1^*(-p_1+p_2)\delta_{j2}\delta_{l1}], \\
    K^J_0(p_1,p_2)_{jl,\alpha\beta} &=\frac{\delta_{\alpha\beta}}{\sqrt{N_f}} \delta_{jl} J_0(p_1-p_2).
\end{split}
\end{equation}

Expansion of the trace log in the effective action is
\begin{equation}
    \Trace\ln[K+K^J] = \Trace\ln K + \Trace\ln[1+K^{-1}(K^J_0 + K_1^J)],
\end{equation}
where the trace $\Trace$ includes summations over spinor, flavor and color indices, as well as integration over the momentum spaces.

In order to derive the effective action of gauge field, we need the expansion of the first term $\Trace\ln(K)$ up to quadratic terms. Defining $G_{0f}(p) = (i\gamma^\mu p_\mu)^{-1} = -i\frac{\gamma^\mu p_\mu}{p^2}$ and $V=K-G_{0f}^{-1}$,  we have
\begin{equation}
\begin{split}
    \Trace\ln K &= \Trace\ln(G_{0f}^{-1}) + \Trace\ln (1+G_{0f}V) \\
    \Tr\ln(1+G_{0f}V) &= \Trace(G_{0f}V) - \frac{1}{2} \Trace(G_{0f} V G_{0f} V) +\cdots 
\end{split}
\end{equation}
Direct computations show that the quadratic kernel in $-\Trace\ln(K)$ is 
\begin{equation}
    \frac{1}{2}\int \dq{q} A_\mu^a(q) \Pi_{\mu\nu}^{ab}(q) A_\nu^b(-q),
\end{equation}
with
\begin{equation}
    \Pi_{\mu\nu}^{ab}(q) = N_f T_F \delta^{ab} \int\dq{p} \frac{\tr[\slashed p \gamma^\mu (\slashed p+\slashed q)\gamma^\nu]}{p^2(q+p)^2} = \frac{N_f}{32}|q|(\delta_{\mu\nu} - \frac{q_\mu q_\nu}{q^2})
\end{equation}
where we have used $T_F = 1/2$. Combined with the Chern-Simons and gauge fixing term, we find the effective propagator 
\begin{equation} \label{eq: gauge field propagator fermion}
    \ew{A_\mu^a(p)A_\nu^{b}(-p)} = \frac{32\delta^{ab}}{N_f(1+\zeta^2)}\left[\left(\frac{\delta_{\mu\nu}}{|p|}-\frac{p_\mu p_\nu}{|p|^3}\right)+\zeta\frac{\epsilon_{\mu\nu\lambda}p_\lambda}{|p|^2}\right]-\frac{16\delta^{ab}\xi}{N_f}\frac{p_\mu p_\nu}{|p|^3},
\end{equation}
with $\zeta = \frac{8k}{\pi N_f}$. This expression is indeed exactly the same as the effective propagator in bosonic theory.

%%----------------------------------------------
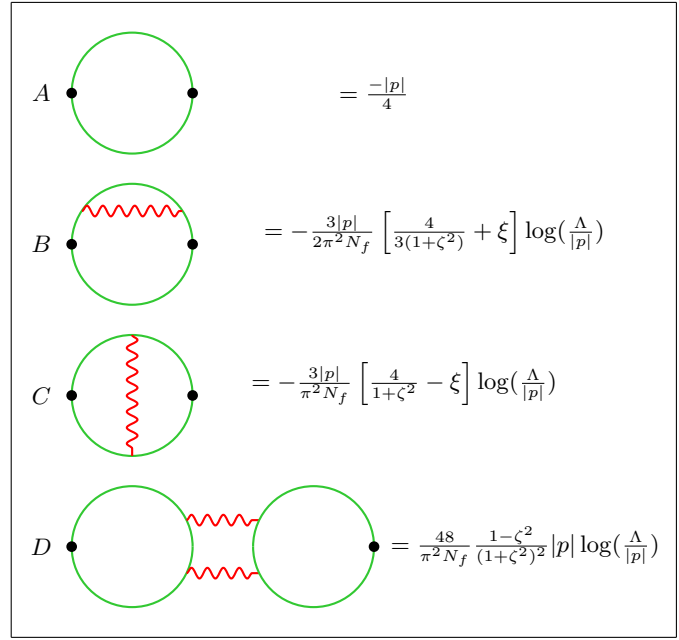
\begin{figure}[t]
\centering
\begin{tikzpicture}[scale=0.4]

\draw[black] (-4,3) to (18,3);
\draw[black] (-4,3) to (-4,-30+2+5+5);
\draw[black] (18,3) to (18,-30+2+5+5);
\draw[black] (-4,-30+2+5+5) to (18,-30+2+5+5);

\draw[thick, mygreen] (0,0) circle (2cm);
\draw[thick,fill] (-2,0) circle (4pt);
\draw[thick,fill] (2,0) circle (4pt);

\node at (8,0) {$=\frac{-|p|}{4}$};

\node at (-3,0)  {$A$};
\node at (-3,-5)  {$B$};
\node at (-3,-10)  {$C$};
\node at (-3,-15)  {$D$};

\draw[mygreen,thick] ([shift=(180:2cm)]0,-5)   arc (180:0:2cm);
\draw[mygreen,thick] ([shift=(0:2cm)]0,-5)   arc (0:-180:2cm);
\draw[photon,red] (-1.66,-3.9) to (1.66,-3.9);
\draw[thick,fill] (-2,-5) circle (4pt);
\draw[thick,fill] (2,-5) circle (4pt);
\node at (10,-5+0.4) {$ =-\frac{3|p|}{2\pi^2N_f}\left[\frac{4}{3(1+\zeta^2)}+\xi\right]\log(\frac{\Lambda}{|p|})$};

\draw[mygreen,thick] ([shift=(180:2cm)]0,-10)   arc (180:0:2cm);
\draw[mygreen,thick] ([shift=(0:2cm)]0,-10)   arc (0:-180:2cm);
\draw[photon, red] (0,-10) ([shift=(90:2cm)]0,-10)  to   ([shift=(-90:2cm)]0,-10);
\draw[thick,fill] (-2,-10) circle (4pt);
\draw[thick,fill] (2,-10) circle (4pt);
 \node at (9,-10+0.4) {$=-\frac{3|p|}{\pi^2N_f}\left[\frac{4}{1+\zeta^2} - \xi\right]\log(\frac{\Lambda}{|p|})$}; 

\draw[thick,mygreen]  ([shift=(0:2cm)]0,-15)  arc (0:180:2cm);
\draw[thick,mygreen]  ([shift=(-180:2cm)]0,-15) arc (-180:0:2cm); 
\draw[photon, red] ([shift=(40-14:2cm)]0,-15)  to   ([shift=(140+14:2cm)]6,-15);
\draw[photon, red] ([shift=(-40+14:2cm)]0,-15)  to   ([shift=(-140-14:2cm)]6,-15);
\draw[thick,mygreen ] ([shift=(180:2cm)]6,-15) arc (180:0:2cm);
\draw[thick,mygreen] ([shift=(0:2cm)]6,-15) arc (0:-180:2cm);
\draw[thick,fill] (-2,-15) circle (4pt);
\draw[thick,fill] (8,-15) circle (4pt);
\node at (13,-15) {$=\frac{48}{\pi^2N_f}\frac{1-\zeta^2}{(1+\zeta^2)^2} |p|\log(\frac{\Lambda}{|p|})$};

  \end{tikzpicture}
  \caption{Feynman diagrams contributing to the two-point functions of the
flavor-singlet mass operator \(O_S\) and the antisymmetric-traceless
mass operator \(O_{A_0}\) in the fermionic matter theory.
Diagram \(A\) is the leading-order contribution, while \(B\) and \(C\)
give the single-trace corrections at order \(O(1/N_f)\).
The diagram \(B\) also has a reflected counterpart with the same
value. Diagram \(D\) is the double-trace contribution and occurs only
for \(O_S\); it vanishes for \(O_{A_0}\) because the corresponding
flavor generator is traceless. Green solid lines represent free
fermion propagators, red wavy lines represent the effective gauge-field
propagator in Eq.~\eqref{eq: gauge field propagator fermion}, and black dots denote
mass-operator insertions. Only the logarithmically divergent parts are
displayed.}
\label{fig:fermion_matter_diagram}
  \end{figure}

%%----------------------------------------------

\subsection{Scaling dimension of singlet operator} \label{subapp: fermion singlet}

In this section we compute the scaling dimension of $O_S=\frac{1}{\sqrt{N_f} }\sum_{i}\bar\psi_i\psi_i$. 

At $N_f=\infty$, it is clear that
\begin{equation}
    \Delta_{O_S}^{(0)} = 2.
\end{equation}
In the following we compute the anomalous dimension at $\mO(1/N_f)$ order.

We expand the source part of $S_{\rm eff}$ to compute the correlators of singlet operators. Concretely, it is
\begin{equation}\label{eq: expansion_probing_field}
    -\Trace\ln[1+K^{-1}(K_0^J + K_1^J)] =-\Trace[K^{-1}(K_0^J+K_1^J)]+ \frac{1}{2}\Trace[K^{-1}(K_0^J + K_1^J)K^{-1}(K_0^J + K_1^J) ] + \cdots,
\end{equation}
and its second order term determines the two-point correlation function $\ew{O_S(p)O_{S}(-p)}$ and $\ew{O_{A_0}(p) O^\dagger_{A_0}(p)}$. There is no operator mixing between the two since they are two different irreducible representations. Hence we can ignore the mixing terms involving both $K_0^J$ and $K_1^J$.

We now consider the terms in $S_{\rm eff}$,
\begin{equation}
    S_{\rm eff}\supset -\Trace[K^{-1}K_0^J] + \frac{1}{2}\Trace[K^{-1}K_0^JK^{-1}K_0^J].
\end{equation}
The functional derivative as
\begin{equation}
    \frac{\delta S_{\rm eff}}{\delta J_0(x)}\Big|_{J_0=0} = -\frac{1}{\sqrt{N_f}}\Trace[K^{-1}],\qquad \frac{\delta^2 S_{\rm eff}}{\delta J_0(x) \delta J_0(y) }\Big|_{J_0=0} = \frac{1}{N_f}\Trace[K^{-1} K^{-1}].
\end{equation}
With the generating functional 
\begin{equation}
    Z[J_0,J_1,J_1^*] = \int\mathcal{D}A e^{-S_{\rm eff}[A,J]},
\end{equation}
the two point correlation function is given by
\begin{equation}
\begin{split}
    \ew{O_S(p) O_{S}(-p)} = \left.\frac{\delta^2 \ln Z[J_0,J_1,J_1^*]}{\delta J_0(p) \delta J_0(-p)}\right|_{J=0} &= -\ew{\frac{\delta^2 S_{\rm eff}}{\delta J_0(p) \delta J_0(-p)  }} + \ew{\frac{\delta S_{\rm eff}}{\delta J_0(p)} \frac{\delta S_{\rm eff}}{\delta J_0(-p)}}^{\rm connected} \\ &= - \int\dq{p_1}\dq{p_2}\ew{\Trace[K^{-1}_{p_2,p_1}\cdot (\mathbf{1}_c)_{p_1,p_1-p} \cdot K^{-1}_{p_1-p,p_2-p}\cdot(\mathbf{1}_c)_{p_2-p,p_2}]} \\
    & + N_f \int\dq{p_1} \dq{p_2} \ew{\Trace[K^{-1}_{p_1,p_1-p}] \Trace[K^{-1}_{p_2,p_2+p}]}
\end{split}
\end{equation}
where the trace now includes color index and spinor index and the expectation $\ew{\cdot}$ is with respect to quadratic part of $S_{\rm eff}$. We call the second line single-trace contribution and the third line double-trace contribution.

With notation $V = K-G_{0f}^{-1}$, we get its inverse expanded as
\begin{equation}
    K^{-1} = (1+G_{0f}V)^{-1} G_{0f} = G_{0f} - G_{0f}VG_{0f} + G_{0f}VG_{0f}VG_{0f} + \cdots.
\end{equation}
We only need to keep up to the second order, because if we use rescaled fields
\begin{equation}
    \tilde A_\mu^a = \sqrt{N_f} A_\mu^a,
\end{equation}
then at order $\mO(1/N_f)$ at most two insertions of $A_\mu^a$ fields contribute. Again, the calculations can be organized using diagrams as in Fig. \ref{fig:fermion_matter_diagram}.

\medskip
\subsubsection{Single-trace contribution}

Let us first consider the single-trace case.
The expression in $\ew{\cdot}$ is
\begin{equation}
\begin{split}
    \Trace[K^{-1}_{p_2,p_1}\cdot (\mathbf{1}_c)_{p_1,p_1-p} \cdot K^{-1}_{p_1-p,p_2-p}\cdot(\mathbf{1}_c)_{p_2-p,p_2}] 
    = &\Trace[G_{0f}(p_1)G_{0f}(p_1-p)] \bar\delta(p_2-p_1) \\
    + &\Trace[(G_{0f}VG_{0f})_{p_2,p_1} (G_{0f}VG_{0f})_{p_1-p,p_2-p}] \\
    + &\Trace[G_{0f}(p_1) (G_{0f}VG_{0f}VG_{0f})_{p_1-p,p_2-p} ]\bar\delta(p_2-p_1) \\
    + &\Trace[G_{0f}(p_1-p) (G_{0f}VG_{0f}VG_{0f})_{p_2,p_1} ]\bar\delta(p_2-p_1) 
\end{split}  
\end{equation}
We analyze these contributions to the correlation function.

\begin{enumerate}
    \item The term $\Trace[G_{0f}(p_1)G_{0f}(p_1-p)] \bar\delta(p_2-p_1)$. This is the leading-order contribution,
    \begin{equation}
    \ew{O_S(p) O_S(-p)}\Big|_{\rm LO} = -(i)^2 N_c \int\dq{p_1}  \frac{ \tr[\gamma_\mu \gamma_\nu] p_{1\mu} (p_1-p)_\nu }{p_1^2(p_1-p)^2} = -N_c \frac{|p|}{8} = -\frac{|p|}{4}.
    \end{equation}
    If we transform it to real space with
    \begin{equation}
        \int \dq{p} e^{ip\cdot x}\frac{|p|}{4} = -\frac{1}{4\pi^2|x|^4},  
    \end{equation}
    we will find the leading order value is 
    \[
    \ew{O_S(x) O_S(0)}\Big|_{\rm LO} = \frac{1}{4\pi^2|x|^4}.
    \]

    \item The term $\Trace[(G_{0f}VG_{0f})_{p_2,p_1} (G_{0f}VG_{0f})_{p_1-p,p_2-p}]$. This corresponds to the diagram with a photon propagator connecting two fermion lines, and it is computed as
    \begin{equation}\label{eq: graph_C_integral}
    \begin{split}
        -(i)^6\Trace[T^a T^b]\int \dq{p_1}\dq{q} \frac{\Trace[(\slashed p_1 + \slashed q) \gamma^\mu \slashed p_1 (\slashed p_1 -\slashed p)\gamma^\nu (\slashed p_1+\slashed q -\slashed p)]}{(p_1+q)^2p_1^2(p_1-p)^2(p_1+q-p)^2} D^{ab}_{\mu\nu}(q) = -\frac{3|p|}{\pi^2N_f}\left[\frac{4}{1+\zeta^2} - \xi\right]\log(\frac{\Lambda}{|p|}) + {\rm finite}
    \end{split}
    \end{equation}
    The evaluation of this integral is slightly complicated, and we use the region analysis to distinguish regions that contribute to log divergence. There are only two such regions: $q\gg p_1, p$ and $p_1\gg q,p$, and both have the same contribution. 

    \item The term $\Trace[G_{0f}(p_1) (G_{0f}VG_{0f}VG_{0f})_{p_1-p,p_2-p} ]\bar\delta(p_2-p_1)$. It corresponds to a graph with a photon propagator connecting a fermion line. Concretely,
    \begin{equation}\label{eq: graph_B_integral}
        -(i)^6 \Trace[T^a T^b] \int\dq{p_1}\dq{q} \frac{\Trace[\slashed p_1(\slashed p_1 - \slashed p) \gamma^\mu (\slashed p_1-\slashed p-\slashed q)\gamma^\nu (\slashed p_1-\slashed p)]}{p_1^2(p_1-p)^2(p_1-q-p)^2(p_1-p)^2} D_{\mu\nu}^{ab}(q)= -\frac{3|p|}{2\pi^2N_f}\left[\frac{4}{3(1+\zeta^2)}+\xi\right]\log(\frac{\Lambda}{|p|})+ {\rm finite}
    \end{equation}
    To compute this integral, one should first consider the $q$ subintegral, which gives a self-energy contribution to fermion line. 

    \item The last term $\Trace[G_{0f}(p_1-p) (G_{0f}VG_{0f}VG_{0f})_{p_2,p_1} ]\bar\delta(p_2-p_1)$ has exactly the same contribution as item 3.
\end{enumerate}
Putting these four next-to-leading order (NLO) contributions together, we get
\begin{equation}
    \ew{O_S(p)O_S(-p)}\Big|^{\rm single\ trace}_{\rm NLO;\log} = -\frac{16|p|}{\pi^2(1+\zeta^2)N_f}\log\frac{\Lambda}{|p|}
\end{equation}

\medskip
\subsubsection{Double-trace contribution}

Next, we consider the double-trace part,
\begin{equation}
    N_f \int\dq{p_1} \dq{p_2} \ew{\Trace[K^{-1}_{p_1,p_1-p}] \Trace[K^{-1}_{p_2,p_2+p}]}.
\end{equation}
Due to the factor $N_f^2$, we need up to four insertions of $V$. Since we only need connected correlator, there is no $O(N_f)$ contribution since a single gauge field propagator connecting two fermion loops vanishes as ${\rm Tr}(T^a)=0$. The first nontrivial contribution arises in $O(1)$ which is next-to-leading order of $\ew{O_S O_S}$. We need exactly four insertions of $V$. There are two ways in the Wick contraction of these $A$ fields. So we get a contribution
\begin{equation}\label{eq: graph_D_integral}
\begin{split}
    \ew{O_S(p) O_S(-p)}\Big|^{\rm double\ trace}_{\rm NLO} &= 2N_f \int\dq{p_1}\dq{p_2} \ew{\Trace[(G_{0f}VG_{0f}VG_{0f})_{p_1,p_1-p}]\cdot \Trace[(G_{0f}VG_{0f}VG_{0f})_{p_2,p_2+p}]} \\
    &= -2N_f \int \dq{p_1}\dq{p_2}\dq{q} \tr_\gamma [ \frac{\slashed p_1 \gamma^\mu (\slashed p_1-\slashed q) \gamma^\nu (\slashed p_1-p)}{p_1^2(p_1-q)^2(p_1-p)^2} ] \tr_\gamma[ \frac{\slashed p_2 \gamma^\rho (\slashed p_2+\slashed q) \gamma^\sigma (\slashed p_2+p)}{p_2^2(p_2+q)^2(p_2+p)^2} ] \\
    &\qquad \times D_{\mu\rho}^{ac}(q)  D_{\nu\sigma}^{bd}(-q+p) \cdot\Trace[T^{a}T^{b}] \Trace[T^c T^d] 
\end{split}
\end{equation}
It turns out the logarithmically divergent part is
\begin{equation}
    \ew{O_S(p)O_S(-p)}\Big|_{\rm NLO;\log}^{\rm double\ trace} = \frac{48}{\pi^2}\frac{1-\zeta^2}{(1+\zeta^2)^2N_f} |p|\log\frac{\Lambda}{|p|}
\end{equation}

\subsubsection{Summary} 

Putting both single-trace and double-trace contributions together, we find
\begin{equation}
    \ew{O_S(p)O_S(-p)}\Big|_{\log} = \ew{O_S(p)O_S(-p)}\Big|_{LO}\cdot\left( \frac{64}{\pi^2N_f(1+\zeta^2)} - \frac{192}{\pi^2N_f}\frac{1-\zeta^2}{(1+\zeta^2)^2}\right)\log(\frac{\Lambda}{|p|}).
\end{equation}

By definition the anomalous dimension $\gamma_S$ is 
\begin{equation}
    \gamma_{S} = \Delta_{O_S} - 2
\end{equation}
and we know 
\begin{equation}
\begin{split}
    \ew{O_S(x)O_S(0)}\sim \frac{1}{|x|^{2\Delta_{O_S}}} \Rightarrow \ew{O_S(p)O_S(-p)} &\sim |p|^{2\Delta_{O_S} -3}\\
    \sim |p| (1+2\gamma_S \ln|p| ),
\end{split}
\end{equation}
hence the anomalous dimension is 
\begin{equation}
\begin{split}
    \gamma_S &= -\frac{32}{\pi^2N_f(1+\zeta^2)} + \frac{96}{\pi^2N_f}\frac{1-\zeta^2}{(1+\zeta^2)^2} \\
    &= -\frac{32N_f}{\pi^2N_f^2 + 64k^2} + \frac{96N_f(N_f^2\pi^2 - 64k^2)}{(\pi^2N_f^2+64k^2)^2}.
\end{split}
\end{equation}
This gives Eq. \eqref{eq:Delta_S_fermion}.

\subsection{Scaling dimension of adjoint operator} \label{subapp: fermion multiplet}

In this section we compute the scaling dimension of $O_{A_0} = \bar\psi_1\psi_2$. 
Its leading order scaling dimension is 
\begin{equation}
    \Delta_{O_{A_0}}^{(0)} = 2.
\end{equation}
We now compute its anomalous dimension at $\mO(1/N_f)$ order.

From the expansion of effective action, Eq.\ \eqref{eq: expansion_probing_field}, the needed terms are
\begin{equation}
    S_{eff} \supset -\Trace[K^{-1}K_1^J] + \frac{1}{2}\Trace[K^{-1} K_1^J K^{-1} K_1^J].
\end{equation}
Recall that 
$$K^J_1(p_1,p_2)_{jl,\alpha\beta} = \delta_{\alpha\beta}[J_1(p_1-p_2)\delta_{j1}\delta_{l2} + J_1^*(-p_1+p_2)\delta_{j2}\delta_{l1}].
$$
Since $K^{-1}$ is diagonal in flavor space, and $K_1^J$ has no diagonal elements, the linear term $-\Trace[K^{-1}K_1^J]$ indeed vanishes. So we are only left with the quadratic term $\frac{1}{2}\Trace[K^{-1}K_1^JK^{-1}K^J_1]$.

Similar to the singlet operator case, the two-point correlation function is given by
\begin{equation}
\begin{split}
    \ew{O_{A_0}(p) O_{A_0}^\dag(p)} &= -\ew{\frac{\delta^2 S_{eff}}{\delta J_1(p)\delta J_1^*(p)}} \\
    &= - \int \dq{p_1}\dq{p_2} \ew{\Trace[K^{-1}]_{p_2,p_1}\cdot(\mathbf{1}_c)_{p_1,p_1-p} \cdot K^{-1}_{p_1-p,p_2-p} \cdot (\mathbf{1}_c)_{p_2-p,p_2} }
\end{split}
\end{equation}
We see from this expression, the value of $\ew{O_{A_0}(p)O_{A_0}^\dag(p)}$ is exactly the same as the single-trace contribution to $\ew{O_S(p)O_S^\dag(p)}$. The double-trace contribution drops out due to the vanishing of linear $J_1$ term in the effective action.
Hence from previous computations, we conclude that
\begin{equation}
    \ew{O_{A_0}(p)O_{A_0}^\dag(p)}\Big|_{LO} = -\frac{|p|}{4}
\end{equation}
and
\begin{equation}
\begin{split}
    \ew{O_{A_0}(p)O_{A_0}^\dag(p)}\Big|_{\log} = \ew{O_{A_0}(p)O_{A_0}^\dag(p)}\Big|_{LO}\cdot\left(\frac{64}{\pi^2N_f(1+\zeta^2)} \right)\log(\frac{\Lambda}{|p|}).
\end{split}
\end{equation}
The anomalous dimension can be found directly,
\begin{equation}
    \gamma_A = -\frac{32}{\pi^2N_f(1+\zeta^2)}= -\frac{32N_f}{\pi^2N_f^2 + 64k^2}.
\end{equation}
This gives Eq. \eqref{eq:Delta_A0_fermion}.

\bibliography{lib}

\begin{thebibliography}{23}%
\makeatletter
\providecommand \@ifxundefined [1]{%
 \@ifx{#1\undefined}
}%
\providecommand \@ifnum [1]{%
 \ifnum #1\expandafter \@firstoftwo
 \else \expandafter \@secondoftwo
 \fi
}%
\providecommand \@ifx [1]{%
 \ifx #1\expandafter \@firstoftwo
 \else \expandafter \@secondoftwo
 \fi
}%
\providecommand \natexlab [1]{#1}%
\providecommand \enquote  [1]{``#1''}%
\providecommand \bibnamefont  [1]{#1}%
\providecommand \bibfnamefont [1]{#1}%
\providecommand \citenamefont [1]{#1}%
\providecommand \href@noop [0]{\@secondoftwo}%
\providecommand \href [0]{\begingroup \@sanitize@url \@href}%
\providecommand \@href[1]{\@@startlink{#1}\@@href}%
\providecommand \@@href[1]{\endgroup#1\@@endlink}%
\providecommand \@sanitize@url [0]{\catcode `\\12\catcode `\$12\catcode
  `\&12\catcode `\#12\catcode `\^12\catcode `\_12\catcode `\%12\relax}%
\providecommand \@@startlink[1]{}%
\providecommand \@@endlink[0]{}%
\providecommand \url  [0]{\begingroup\@sanitize@url \@url }%
\providecommand \@url [1]{\endgroup\@href {#1}{\urlprefix }}%
\providecommand \urlprefix  [0]{URL }%
\providecommand \Eprint [0]{\href }%
\providecommand \doibase [0]{http://dx.doi.org/}%
\providecommand \selectlanguage [0]{\@gobble}%
\providecommand \bibinfo  [0]{\@secondoftwo}%
\providecommand \bibfield  [0]{\@secondoftwo}%
\providecommand \translation [1]{[#1]}%
\providecommand \BibitemOpen [0]{}%
\providecommand \bibitemStop [0]{}%
\providecommand \bibitemNoStop [0]{.\EOS\space}%
\providecommand \EOS [0]{\spacefactor3000\relax}%
\providecommand \BibitemShut  [1]{\csname bibitem#1\endcsname}%
\let\auto@bib@innerbib\@empty
%</preamble>
\bibitem [{\citenamefont {{Zhang}}\ \emph {et~al.}(2024)\citenamefont
  {{Zhang}}, \citenamefont {{Huang}}, \citenamefont {{Wu}}, \citenamefont
  {{Sheng}},\ and\ \citenamefont {{Gong}}}]{Zhang2024}%
  \BibitemOpen
  \bibfield  {author} {\bibinfo {author} {\bibfnamefont {Xiao-Tian}\
  \bibnamefont {{Zhang}}}, \bibinfo {author} {\bibfnamefont {Yixuan}\
  \bibnamefont {{Huang}}}, \bibinfo {author} {\bibfnamefont {Han-Qing}\
  \bibnamefont {{Wu}}}, \bibinfo {author} {\bibfnamefont {D.~N.}\ \bibnamefont
  {{Sheng}}}, \ and\ \bibinfo {author} {\bibfnamefont {Shou-Shu}\ \bibnamefont
  {{Gong}}},\ }\bibfield  {title} {\enquote {\bibinfo {title} {{Chiral spin
  liquid and quantum phase diagram of spin-12 J1-J2-J{\ensuremath{\chi}} model
  on the square lattice}},}\ }\href {\doibase 10.1103/PhysRevB.109.125146}
  {\bibfield  {journal} {\bibinfo  {journal} {\prb}\ }\textbf {\bibinfo
  {volume} {109}},\ \bibinfo {eid} {125146} (\bibinfo {year} {2024})},\ \Eprint
  {http://arxiv.org/abs/2401.07461} {arXiv:2401.07461 [cond-mat.str-el]}
  \BibitemShut {NoStop}%
\bibitem [{\citenamefont {Luo}\ \emph {et~al.}(2023)\citenamefont {Luo},
  \citenamefont {Huang}, \citenamefont {Sheng},\ and\ \citenamefont
  {Zhu}}]{Luo2023}%
  \BibitemOpen
  \bibfield  {author} {\bibinfo {author} {\bibfnamefont {Wei-Wei}\ \bibnamefont
  {Luo}}, \bibinfo {author} {\bibfnamefont {Yixuan}\ \bibnamefont {Huang}},
  \bibinfo {author} {\bibfnamefont {D.~N.}\ \bibnamefont {Sheng}}, \ and\
  \bibinfo {author} {\bibfnamefont {W.}~\bibnamefont {Zhu}},\ }\bibfield
  {title} {\enquote {\bibinfo {title} {Global quantum phase diagram and
  non-abelian chiral spin liquid in a spin-$3/2$ square-lattice
  antiferromagnet},}\ }\href {\doibase 10.1103/PhysRevB.108.035130} {\bibfield
  {journal} {\bibinfo  {journal} {Phys. Rev. B}\ }\textbf {\bibinfo {volume}
  {108}},\ \bibinfo {pages} {035130} (\bibinfo {year} {2023})},\ \Eprint
  {http://arxiv.org/abs/2212.14223} {arXiv:2212.14223 [cond-mat.str-el]}
  \BibitemShut {NoStop}%
\bibitem [{\citenamefont {Zhou}\ and\ \citenamefont {He}(2025)}]{Zhou2025}%
  \BibitemOpen
  \bibfield  {author} {\bibinfo {author} {\bibfnamefont {Zheng}\ \bibnamefont
  {Zhou}}\ and\ \bibinfo {author} {\bibfnamefont {Yin-Chen}\ \bibnamefont
  {He}},\ }\bibfield  {title} {\enquote {\bibinfo {title} {A new series of 3d
  cfts with $\mathrm{Sp}(n)$ global symmetry on fuzzy sphere},}\ }\href
  {\doibase 10.1103/xstj-xvcy} {\bibfield  {journal} {\bibinfo  {journal}
  {Phys. Rev. Lett.}\ }\textbf {\bibinfo {volume} {135}},\ \bibinfo {pages}
  {026504} (\bibinfo {year} {2025})},\ \Eprint
  {http://arxiv.org/abs/2410.00087} {arXiv:2410.00087 [hep-th]} \BibitemShut
  {NoStop}%
\bibitem [{\citenamefont {Ye}\ and\ \citenamefont {Zou}(2024)}]{Ye2023}%
  \BibitemOpen
  \bibfield  {author} {\bibinfo {author} {\bibfnamefont {Weicheng}\
  \bibnamefont {Ye}}\ and\ \bibinfo {author} {\bibfnamefont {Liujun}\
  \bibnamefont {Zou}},\ }\bibfield  {title} {\enquote {\bibinfo {title}
  {Classification of symmetry-enriched topological quantum spin liquids},}\
  }\href {\doibase 10.1103/PhysRevX.14.021053} {\bibfield  {journal} {\bibinfo
  {journal} {Phys. Rev. X}\ }\textbf {\bibinfo {volume} {14}},\ \bibinfo
  {pages} {021053} (\bibinfo {year} {2024})},\ \Eprint
  {http://arxiv.org/abs/2309.15118} {arXiv:2309.15118 [cond-mat.str-el]}
  \BibitemShut {NoStop}%
\bibitem [{\citenamefont {Barkeshli}\ \emph {et~al.}(2019)\citenamefont
  {Barkeshli}, \citenamefont {Bonderson}, \citenamefont {Cheng},\ and\
  \citenamefont {Wang}}]{Barkeshli2019}%
  \BibitemOpen
  \bibfield  {author} {\bibinfo {author} {\bibfnamefont {Maissam}\ \bibnamefont
  {Barkeshli}}, \bibinfo {author} {\bibfnamefont {Parsa}\ \bibnamefont
  {Bonderson}}, \bibinfo {author} {\bibfnamefont {Meng}\ \bibnamefont {Cheng}},
  \ and\ \bibinfo {author} {\bibfnamefont {Zhenghan}\ \bibnamefont {Wang}},\
  }\bibfield  {title} {\enquote {\bibinfo {title} {Symmetry fractionalization,
  defects, and gauging of topological phases},}\ }\href {\doibase
  10.1103/PhysRevB.100.115147} {\bibfield  {journal} {\bibinfo  {journal}
  {Phys. Rev. B}\ }\textbf {\bibinfo {volume} {100}},\ \bibinfo {pages}
  {115147} (\bibinfo {year} {2019})},\ \Eprint {http://arxiv.org/abs/1410.4540}
  {arXiv:1410.4540 [cond-mat.str-el]} \BibitemShut {NoStop}%
\bibitem [{\citenamefont {{Li}}\ and\ \citenamefont {{Zou}}(2026)}]{Li2026}%
  \BibitemOpen
  \bibfield  {author} {\bibinfo {author} {\bibfnamefont {Yingcheng}\
  \bibnamefont {{Li}}}\ and\ \bibinfo {author} {\bibfnamefont {Liujun}\
  \bibnamefont {{Zou}}},\ }\bibfield  {title} {\enquote {\bibinfo {title}
  {{Microscopic universal theory of symmetry-enriched topological quantum spin
  liquids}},}\ }\href {\doibase 10.48550/arXiv.2606.08558} {\bibfield
  {journal} {\bibinfo  {journal} {arXiv e-prints}\ ,\ \bibinfo {eid}
  {arXiv:2606.08558}} (\bibinfo {year} {2026})},\ \Eprint
  {http://arxiv.org/abs/2606.08558} {arXiv:2606.08558 [cond-mat.str-el]}
  \BibitemShut {NoStop}%
\bibitem [{\citenamefont {{Kawagoe}}\ and\ \citenamefont
  {{Levin}}(2020)}]{Kawagoe2019}%
  \BibitemOpen
  \bibfield  {author} {\bibinfo {author} {\bibfnamefont {Kyle}\ \bibnamefont
  {{Kawagoe}}}\ and\ \bibinfo {author} {\bibfnamefont {Michael}\ \bibnamefont
  {{Levin}}},\ }\bibfield  {title} {\enquote {\bibinfo {title} {{Microscopic
  definitions of anyon data}},}\ }\href {\doibase 10.1103/PhysRevB.101.115113}
  {\bibfield  {journal} {\bibinfo  {journal} {\prb}\ }\textbf {\bibinfo
  {volume} {101}},\ \bibinfo {eid} {115113} (\bibinfo {year} {2020})},\ \Eprint
  {http://arxiv.org/abs/1910.11353} {arXiv:1910.11353 [cond-mat.str-el]}
  \BibitemShut {NoStop}%
\bibitem [{\citenamefont {Thorngren}\ and\ \citenamefont
  {Else}(2018)}]{Thorngren2016}%
  \BibitemOpen
  \bibfield  {author} {\bibinfo {author} {\bibfnamefont {Ryan}\ \bibnamefont
  {Thorngren}}\ and\ \bibinfo {author} {\bibfnamefont {Dominic~V.}\
  \bibnamefont {Else}},\ }\bibfield  {title} {\enquote {\bibinfo {title}
  {Gauging spatial symmetries and the classification of topological crystalline
  phases},}\ }\href {\doibase 10.1103/PhysRevX.8.011040} {\bibfield  {journal}
  {\bibinfo  {journal} {Phys. Rev. X}\ }\textbf {\bibinfo {volume} {8}},\
  \bibinfo {pages} {011040} (\bibinfo {year} {2018})},\ \Eprint
  {http://arxiv.org/abs/1612.00846} {arXiv:1612.00846 [cond-mat.str-el]}
  \BibitemShut {NoStop}%
\bibitem [{\citenamefont {Ye}\ \emph {et~al.}(2022)\citenamefont {Ye},
  \citenamefont {Guo}, \citenamefont {He}, \citenamefont {Wang},\ and\
  \citenamefont {Zou}}]{Ye2022}%
  \BibitemOpen
  \bibfield  {author} {\bibinfo {author} {\bibfnamefont {Weicheng}\
  \bibnamefont {Ye}}, \bibinfo {author} {\bibfnamefont {Meng}\ \bibnamefont
  {Guo}}, \bibinfo {author} {\bibfnamefont {Yin-Chen}\ \bibnamefont {He}},
  \bibinfo {author} {\bibfnamefont {Chong}\ \bibnamefont {Wang}}, \ and\
  \bibinfo {author} {\bibfnamefont {Liujun}\ \bibnamefont {Zou}},\ }\bibfield
  {title} {\enquote {\bibinfo {title} {Topological characterization of
  lieb-schultz-mattis constraints and applications to symmetry-enriched quantum
  criticality},}\ }\href {\doibase 10.21468/SciPostPhys.13.3.066} {\bibfield
  {journal} {\bibinfo  {journal} {SciPost Phys.}\ }\textbf {\bibinfo {volume}
  {13}},\ \bibinfo {pages} {066} (\bibinfo {year} {2022})},\ \Eprint
  {http://arxiv.org/abs/2111.12097} {arXiv:2111.12097 [cond-mat.str-el]}
  \BibitemShut {NoStop}%
\bibitem [{\citenamefont {Po}\ \emph {et~al.}(2017)\citenamefont {Po},
  \citenamefont {Watanabe}, \citenamefont {Jian},\ and\ \citenamefont
  {Zaletel}}]{Po2017}%
  \BibitemOpen
  \bibfield  {author} {\bibinfo {author} {\bibfnamefont {Hoi~Chun}\
  \bibnamefont {Po}}, \bibinfo {author} {\bibfnamefont {Haruki}\ \bibnamefont
  {Watanabe}}, \bibinfo {author} {\bibfnamefont {Chao-Ming}\ \bibnamefont
  {Jian}}, \ and\ \bibinfo {author} {\bibfnamefont {Michael~P.}\ \bibnamefont
  {Zaletel}},\ }\bibfield  {title} {\enquote {\bibinfo {title} {Lattice
  homotopy constraints on phases of quantum magnets},}\ }\href {\doibase
  10.1103/PhysRevLett.119.127202} {\bibfield  {journal} {\bibinfo  {journal}
  {Phys. Rev. Lett.}\ }\textbf {\bibinfo {volume} {119}},\ \bibinfo {pages}
  {127202} (\bibinfo {year} {2017})},\ \Eprint
  {http://arxiv.org/abs/1703.06882} {arXiv:1703.06882 [cond-mat.str-el]}
  \BibitemShut {NoStop}%
\bibitem [{\citenamefont {Else}\ and\ \citenamefont
  {Thorngren}(2020)}]{ElseThorngren2020}%
  \BibitemOpen
  \bibfield  {author} {\bibinfo {author} {\bibfnamefont {Dominic~V.}\
  \bibnamefont {Else}}\ and\ \bibinfo {author} {\bibfnamefont {Ryan}\
  \bibnamefont {Thorngren}},\ }\bibfield  {title} {\enquote {\bibinfo {title}
  {Topological theory of lieb-schultz-mattis theorems in quantum spin
  systems},}\ }\href {\doibase 10.1103/PhysRevB.101.224437} {\bibfield
  {journal} {\bibinfo  {journal} {Phys. Rev. B}\ }\textbf {\bibinfo {volume}
  {101}},\ \bibinfo {pages} {224437} (\bibinfo {year} {2020})},\ \Eprint
  {http://arxiv.org/abs/1907.08204} {arXiv:1907.08204 [cond-mat.str-el]}
  \BibitemShut {NoStop}%
\bibitem [{\citenamefont {{Zou}}\ and\ \citenamefont
  {{Cheng}}(2026)}]{Zou2026}%
  \BibitemOpen
  \bibfield  {author} {\bibinfo {author} {\bibfnamefont {Liujun}\ \bibnamefont
  {{Zou}}}\ and\ \bibinfo {author} {\bibfnamefont {Meng}\ \bibnamefont
  {{Cheng}}},\ }\bibfield  {title} {\enquote {\bibinfo {title}
  {{Lieb-Schultz-Mattis Anomalies and Anomaly Matching}},}\ }\href {\doibase
  10.48550/arXiv.2604.00347} {\bibfield  {journal} {\bibinfo  {journal} {arXiv
  e-prints}\ ,\ \bibinfo {eid} {arXiv:2604.00347}} (\bibinfo {year} {2026})},\
  \Eprint {http://arxiv.org/abs/2604.00347} {arXiv:2604.00347
  [cond-mat.str-el]} \BibitemShut {NoStop}%
\bibitem [{\citenamefont {Ye}\ and\ \citenamefont {Zou}(2023)}]{Ye2023SciPost}%
  \BibitemOpen
  \bibfield  {author} {\bibinfo {author} {\bibfnamefont {Weicheng}\
  \bibnamefont {Ye}}\ and\ \bibinfo {author} {\bibfnamefont {Liujun}\
  \bibnamefont {Zou}},\ }\bibfield  {title} {\enquote {\bibinfo {title}
  {Anomaly of $(2+1)$-dimensional symmetry-enriched topological order from
  $(3+1)$-dimensional topological quantum field theory},}\ }\href {\doibase
  10.21468/SciPostPhys.15.1.004} {\bibfield  {journal} {\bibinfo  {journal}
  {SciPost Phys.}\ }\textbf {\bibinfo {volume} {15}},\ \bibinfo {pages} {004}
  (\bibinfo {year} {2023})},\ \Eprint {http://arxiv.org/abs/2210.02444}
  {arXiv:2210.02444 [cond-mat.str-el]} \BibitemShut {NoStop}%
\bibitem [{\citenamefont {Wang}\ \emph {et~al.}(2017)\citenamefont {Wang},
  \citenamefont {Nahum}, \citenamefont {Metlitski}, \citenamefont {Xu},\ and\
  \citenamefont {Senthil}}]{WangNahumMetlitskiXuSenthil2017}%
  \BibitemOpen
  \bibfield  {author} {\bibinfo {author} {\bibfnamefont {Chong}\ \bibnamefont
  {Wang}}, \bibinfo {author} {\bibfnamefont {Adam}\ \bibnamefont {Nahum}},
  \bibinfo {author} {\bibfnamefont {Max~A.}\ \bibnamefont {Metlitski}},
  \bibinfo {author} {\bibfnamefont {Cenke}\ \bibnamefont {Xu}}, \ and\ \bibinfo
  {author} {\bibfnamefont {T.}~\bibnamefont {Senthil}},\ }\bibfield  {title}
  {\enquote {\bibinfo {title} {Deconfined quantum critical points: Symmetries
  and dualities},}\ }\href {\doibase 10.1103/PhysRevX.7.031051} {\bibfield
  {journal} {\bibinfo  {journal} {Phys. Rev. X}\ }\textbf {\bibinfo {volume}
  {7}},\ \bibinfo {pages} {031051} (\bibinfo {year} {2017})},\ \Eprint
  {http://arxiv.org/abs/1703.02426} {arXiv:1703.02426 [cond-mat.str-el]}
  \BibitemShut {NoStop}%
\bibitem [{\citenamefont {{Zou}}\ \emph {et~al.}(2021)\citenamefont {{Zou}},
  \citenamefont {{He}},\ and\ \citenamefont {{Wang}}}]{Zou2021}%
  \BibitemOpen
  \bibfield  {author} {\bibinfo {author} {\bibfnamefont {Liujun}\ \bibnamefont
  {{Zou}}}, \bibinfo {author} {\bibfnamefont {Yin-Chen}\ \bibnamefont {{He}}},
  \ and\ \bibinfo {author} {\bibfnamefont {Chong}\ \bibnamefont {{Wang}}},\
  }\bibfield  {title} {\enquote {\bibinfo {title} {{Stiefel Liquids: Possible
  Non-Lagrangian Quantum Criticality from Intertwined Orders}},}\ }\href
  {\doibase 10.1103/PhysRevX.11.031043} {\bibfield  {journal} {\bibinfo
  {journal} {Physical Review X}\ }\textbf {\bibinfo {volume} {11}},\ \bibinfo
  {eid} {031043} (\bibinfo {year} {2021})},\ \Eprint
  {http://arxiv.org/abs/2101.07805} {arXiv:2101.07805 [cond-mat.str-el]}
  \BibitemShut {NoStop}%
\bibitem [{\citenamefont {{Chester}}\ and\ \citenamefont
  {{Pufu}}(2016)}]{Chester2016}%
  \BibitemOpen
  \bibfield  {author} {\bibinfo {author} {\bibfnamefont {Shai~M.}\ \bibnamefont
  {{Chester}}}\ and\ \bibinfo {author} {\bibfnamefont {Silviu~S.}\ \bibnamefont
  {{Pufu}}},\ }\bibfield  {title} {\enquote {\bibinfo {title} {{Anomalous
  dimensions of scalar operators in QED$_{3}$}},}\ }\href {\doibase
  10.1007/JHEP08(2016)069} {\bibfield  {journal} {\bibinfo  {journal} {Journal
  of High Energy Physics}\ }\textbf {\bibinfo {volume} {2016}},\ \bibinfo {eid}
  {69} (\bibinfo {year} {2016})},\ \Eprint {http://arxiv.org/abs/1603.05582}
  {arXiv:1603.05582 [hep-th]} \BibitemShut {NoStop}%
\bibitem [{\citenamefont {Benvenuti}\ and\ \citenamefont
  {Khachatryan}(2019)}]{BenvenutiKhachatryan2019}%
  \BibitemOpen
  \bibfield  {author} {\bibinfo {author} {\bibfnamefont {Sergio}\ \bibnamefont
  {Benvenuti}}\ and\ \bibinfo {author} {\bibfnamefont {Hrachya}\ \bibnamefont
  {Khachatryan}},\ }\bibfield  {title} {\enquote {\bibinfo {title} {Easy-plane
  qed$_3$'s in the large $n_f$ limit},}\ }\href {\doibase
  10.1007/JHEP05(2019)214} {\bibfield  {journal} {\bibinfo  {journal} {J. High
  Energy Phys.}\ }\textbf {\bibinfo {volume} {2019}},\ \bibinfo {pages} {214}
  (\bibinfo {year} {2019})},\ \Eprint {http://arxiv.org/abs/1902.05767}
  {arXiv:1902.05767 [hep-th]} \BibitemShut {NoStop}%
\bibitem [{\citenamefont {Zou}\ and\ \citenamefont {He}(2020)}]{Zou2018}%
  \BibitemOpen
  \bibfield  {author} {\bibinfo {author} {\bibfnamefont {Liujun}\ \bibnamefont
  {Zou}}\ and\ \bibinfo {author} {\bibfnamefont {Yin-Chen}\ \bibnamefont
  {He}},\ }\bibfield  {title} {\enquote {\bibinfo {title} {Field-induced
  ${\mathrm{qcd}}_{3}$-chern-simons quantum criticalities in kitaev
  materials},}\ }\href {\doibase 10.1103/PhysRevResearch.2.013072} {\bibfield
  {journal} {\bibinfo  {journal} {Phys. Rev. Res.}\ }\textbf {\bibinfo {volume}
  {2}},\ \bibinfo {pages} {013072} (\bibinfo {year} {2020})},\ \Eprint
  {http://arxiv.org/abs/1809.09091} {arXiv:1809.09091 [cond-mat.str-el]}
  \BibitemShut {NoStop}%
\bibitem [{\citenamefont {{Wen}}(2002)}]{Wen2001}%
  \BibitemOpen
  \bibfield  {author} {\bibinfo {author} {\bibfnamefont {Xiao-Gang}\
  \bibnamefont {{Wen}}},\ }\bibfield  {title} {\enquote {\bibinfo {title}
  {{Quantum orders and symmetric spin liquids}},}\ }\href {\doibase
  10.1103/PhysRevB.65.165113} {\bibfield  {journal} {\bibinfo  {journal}
  {\prb}\ }\textbf {\bibinfo {volume} {65}},\ \bibinfo {eid} {165113} (\bibinfo
  {year} {2002})},\ \Eprint {http://arxiv.org/abs/cond-mat/0107071}
  {arXiv:cond-mat/0107071 [cond-mat.str-el]} \BibitemShut {NoStop}%
\bibitem [{\citenamefont {Bonderson}(2007)}]{Bonderson2007}%
  \BibitemOpen
  \bibfield  {author} {\bibinfo {author} {\bibfnamefont {Parsa~Hassan}\
  \bibnamefont {Bonderson}},\ }\emph {\bibinfo {title} {Non-Abelian Anyons and
  Interferometry}},\ \href {\doibase 10.7907/5NDZ-W890} {\bibinfo {type} {Ph.d.
  thesis}},\ \bibinfo  {school} {California Institute of Technology} (\bibinfo
  {year} {2007})\BibitemShut {NoStop}%
\bibitem [{\citenamefont {{Barkeshli}}\ and\ \citenamefont
  {{Cheng}}(2020)}]{Barkeshli2019a}%
  \BibitemOpen
  \bibfield  {author} {\bibinfo {author} {\bibfnamefont {Maissam}\ \bibnamefont
  {{Barkeshli}}}\ and\ \bibinfo {author} {\bibfnamefont {Meng}\ \bibnamefont
  {{Cheng}}},\ }\bibfield  {title} {\enquote {\bibinfo {title} {{Relative
  anomalies in (2+1)D symmetry enriched topological states}},}\ }\href
  {\doibase 10.21468/SciPostPhys.8.2.028} {\bibfield  {journal} {\bibinfo
  {journal} {SciPost Physics}\ }\textbf {\bibinfo {volume} {8}},\ \bibinfo
  {eid} {028} (\bibinfo {year} {2020})},\ \Eprint
  {http://arxiv.org/abs/1906.10691} {arXiv:1906.10691 [cond-mat.str-el]}
  \BibitemShut {NoStop}%
\bibitem [{\citenamefont {Etingof}\ \emph {et~al.}(2010)\citenamefont
  {Etingof}, \citenamefont {Nikshych},\ and\ \citenamefont
  {Ostrik}}]{Etingof2010}%
  \BibitemOpen
  \bibfield  {author} {\bibinfo {author} {\bibfnamefont {Pavel}\ \bibnamefont
  {Etingof}}, \bibinfo {author} {\bibfnamefont {Dmitri}\ \bibnamefont
  {Nikshych}}, \ and\ \bibinfo {author} {\bibfnamefont {Viktor}\ \bibnamefont
  {Ostrik}},\ }\bibfield  {title} {\enquote {\bibinfo {title} {Fusion
  categories and homotopy theory},}\ }\href {\doibase 10.4171/QT/6} {\bibfield
  {journal} {\bibinfo  {journal} {Quantum Topology}\ }\textbf {\bibinfo
  {volume} {1}},\ \bibinfo {pages} {209--273} (\bibinfo {year} {2010})},\
  \Eprint {http://arxiv.org/abs/0909.3140} {arXiv:0909.3140 [math.QA]}
  \BibitemShut {NoStop}%
\bibitem [{\citenamefont {{Yamatsu}}(2015)}]{Yamatsu2015}%
  \BibitemOpen
  \bibfield  {author} {\bibinfo {author} {\bibfnamefont {Naoki}\ \bibnamefont
  {{Yamatsu}}},\ }\bibfield  {title} {\enquote {\bibinfo {title}
  {{Finite-Dimensional Lie Algebras and Their Representations for Unified Model
  Building}},}\ }\href {\doibase 10.48550/arXiv.1511.08771} {\bibfield
  {journal} {\bibinfo  {journal} {arXiv e-prints}\ ,\ \bibinfo {eid}
  {arXiv:1511.08771}} (\bibinfo {year} {2015})},\ \Eprint
  {http://arxiv.org/abs/1511.08771} {arXiv:1511.08771 [hep-ph]} \BibitemShut
  {NoStop}%
\end{thebibliography}%

\end{document}